\documentclass[10pt,twocolumn]{revtex4-1}

\usepackage{graphics} 
\usepackage{amsmath,amssymb} 
\usepackage{dsfont}

\newcommand{\ket}[1]{\ensuremath{|#1\rangle}}

\newcommand{\sref}[1]{Sect.~\ref{#1}}

\newcommand{\Fref}[1]{Figure~\ref{#1}}
\newcommand{\fref}[1]{Fig.~\ref{#1}}

\newcommand{\eref}[1]{Eq.~(\ref{#1})}

\newcommand{\etoeref}[2]{Eqs.~(\ref{#1}) to (\ref{#2})}

\begin{document}
\title[Spatial adiabatic passage]{Spatial adiabatic passage: a review of recent progress}

\author{R.~Menchon-Enrich$^1$, A.~Benseny$^2$, 
V.~Ahufinger$^1$, A.D.~Greentree$^{3}$, Th.~Busch$^2$ and J.~Mompart$^1$}

\affiliation{$^1$ Departament de F\'isica, Universitat Aut\`onoma de Barcelona, E-08193 Bellaterra, Spain}
\affiliation{$^2$ Quantum Systems Unit, OIST Graduate University, Onna, Okinawa 904-0495, Japan}
\affiliation{$^3$ Australian Research Council Centre of Excellence for Nanoscale BioPhotonics, and Chemical and Quantum Physics Group, School of Applied Sciences, RMIT University, Melbourne 3001, Australia}

\begin{abstract}
Adiabatic techniques are known to allow for engineering quantum states with high fidelity. This requirement is currently of large interest, as applications in quantum information require the preparation and manipulation of  quantum states with minimal errors. Here we review recent progress on developing techniques for the preparation of spatial states through adiabatic passage, particularly focusing on three state systems. These techniques can be applied to matter waves in external potentials, such as cold atoms or electrons, and to classical waves in waveguides, such as light or sound.
\end{abstract}

\maketitle

\setcounter{tocdepth}{1}
\tableofcontents

\section{Introduction and motivation}
\label{sec:intro}

The ability to coherently control the spatial degrees of freedom of matter waves is an important ingredient for the emerging field of quantum engineering~\cite{micheli_single_2004, ruschhaupt_atom_2004, stickney_transistorlike_2007, thorn_experimental_2008, price_single-photon_2008, pepino_atomtronic_2009, benseny_atomtronics_2010, zozulya_principles_2013, ryu_experimental_2013, wright_driving_2013, jendrzejewski_resistive_2014}, with applications in matter wave interferometry, quantum metrology, and quantum computation.
For many purposes, quantum transport of trapped matter waves is performed via direct tunneling through the manipulation of the potential barriers that separate the different traps. This direct transport is strongly dependent on the parameter values giving rise, in general, to very sensitive Rabi-type oscillations of the populations of the localized states of the traps. 
Alternatively, an efficient transfer of population between distant traps can be achieved using spatial adiabatic passage (SAP) processes, which consist of the adiabatic following of an energy eigenstate of the system that is spatially modified either in time or space.
This transfer occurs with high fidelity regardless of the selected specific parameter values used to drive the system and their fluctuations.
SAP processes are a particular case of adiabatic following, a concept arising from the adiabatic theorem~\cite{born_beweis_1928, messiah_quantum_1976}: in the absence of level crossings, a system will remain in one of its eigenstates if the system is perturbed slowly enough.
 
The term adiabatic passage started to be commonly used around three decades ago in the field of quantum optics as two main techniques were introduced to adiabatically transfer population between internal atomic/molecular levels.
These were rapid adiabatic passage (RAP)~\cite{allen_optical_1987, vitanov_laser-induced_2001}
and
stimulated Raman adiabatic passage (STIRAP)~ \cite{bergmann_coherent_1998, fewell_coherent_1997, vitanov_laser-induced_2001}.
The RAP technique is implemented in two-level atomic systems interacting with a chirped laser pulse to transfer the population between the two states by adiabatically following one of the two eigenstates of the system.\
The STIRAP technique is performed in a $\Lambda$-type three-level atomic system interacting with two temporally delayed laser pulses in a counterintuitive sequence in order to completely transfer the population between the two atomic ground states by adiabatically following the so-called dark state~\cite{alzetta_experimental_1976, arimondo_nonabsorbing_1976, gray_coherent_1978, alzetta_nonabsorption_1979}.
These techniques have led to many relevant experimentally implemented applications~\cite{shore_pre-history_2013,bergmann_perspective_2015}.

Following the success of techniques such as RAP and STIRAP, adiabatic passage processes were proposed to coherently transport quantum particles between localized states of spatially separated potential wells, leading to the term \textit{spatial} adiabatic passage.
While some proposals were reported in the early 2000s (see for example Ref.~\cite{renzoni_charge_2001}),
the field gathered more interest after the publication of two seminal papers.
The first one, by Eckert \textit{et al.}~\cite{eckert_three-level_2004}, analyzed the spatial adiabatic passage of a single cold atom in a system of three optical traps and the second one, by Greentree \textit{et al.}~\cite{greentree_coherent_2004}, described the transfer of an electron in a system of three quantum dots.
Both proposals resembled the quantum-optical STIRAP technique, and used tunneling as a way to couple the different localized states of spatially separated wells.
However, it was soon realized that SAP can be extended beyond the applications of STIRAP, as it allows for multi-dimensional configurations and many-particle systems.
Therefore, due to the high efficiency and robustness inherited from STIRAP, many applications, such as vibrational state and velocity filtering~\cite{busch_quantum_2007, loiko_filtering_2011, loiko_coherent_2014}, quantum tomography~\cite{loiko_filtering_2011}, interferometry~\cite{jong_interferometry_2010, menchon-enrich_single-atom_2014}, atomtronics~\cite{lu_coherent_2011, benseny_atomtronics_2010}, and the generation of angular momentum states~\cite{menchon-enrich_tunneling-induced_2014} have been proposed.
Note that SAP in a triple-well system has also been referred as matter-wave STIRAP or coherent tunneling by adiabatic passage (CTAP).

In spite of the significant theoretical interest that they have attracted, SAP processes for matter waves have not been experimentally realized yet.
It is worth noting, however, that due to the wave-like nature of the SAP processes, they can be extended to classical wave systems, and have been experimentally demonstrated with light beams in coupled waveguides. 
Since the initial works of Longhi \textit{et al.}~\cite{longhi_adiabatic_2006, longhi_coherent_2007}, SAP for light beams has also been studied in systems of more than three coupled waveguides~\cite{longhi_optical_2006, valle_adiabatic_2008, rangelov_achromatic_2012, ciret_broadband_2013}, through the continuum ~\cite{longhi_transfer_2008, dreisow_adiabatic_2009}, and in the presence of nonlinearities and absorption~\cite{barak_autoresonant_2009, graefe_breakdown_2013}.
It has also lead to applications such as beam splitting~\cite{dreisow_polychromatic_2009, rangelov_achromatic_2012, chung_broadband_2012, hristova_adiabatic_2013}, spectral filtering~\cite{menchon-enrich_light_2013}, interaction free-measurements~\cite{hill_parallel_2011}, quantum gates via long-range coupling~\cite{hope_long-range_2013}, polarization rotation/conversion ~\cite{xiong_integrated_2013}, and photon pair generation~\cite{wu_photon_2014}.
Finally, SAP processes have also been recently proposed for sound propagation in sonic crystals~\cite{menchon-enrich_spatial_2014}, investigating transfer and splitting of sound beams, as well as the use of the system as a phase analyzer.

This review is organized as follows.
In Section II, we describe the concept of adiabatic following of an eigenvector and connect it to the different physical systems that we will study.
We then, in Section III, review the spatial adiabatic passage of matter waves, which includes systems of single atoms, electrons, and Bose--Einstein condensates. We also discuss SAP in two-dimensional systems and address some practical considerations.
This is followed by Section IV, where we discuss dark state adiabatic passage, and Section V, where recent developments of SAP for light and sound waves are summarized.
Finally, in Section VI we conclude.

\section{Adiabatic following of an energy eigenstate}
\label{sec:general}

The concept of a quantum state evolution that adiabatically follows an energy eigenstate was introduced to quantum mechanics by Max Born and Vladimir Fock in 1928 \cite{born_beweis_1928}. They showed that a physical system remains in its instantaneous eigenstate if a given perturbation is acting on it slowly enough and if there is a gap between the eigenvalue and the rest of the Hamiltonian's spectrum.  This insight is known by today as the adiabatic theorem.

To perform an adiabatic evolution it is therefore necessary that the dynamics do not allow transitions between eigenstates.
While in most cases this can be achieved by changing the system parameters slowly enough, it also requires that all eigenstates evolve smoothly and that their first and second derivatives with respect to the temporal or spatial evolution parameter, $s$, are well defined  \cite{messiah_quantum_1976}.
To quantify the adiabaticity of a process one can calculate the probability to excite a state $\ket{\Psi_j(s)}$ while being in $\ket{\Psi_i(s)}$ as~\cite{messiah_quantum_1976}
\begin{equation}
    p_{i\rightarrow j}\leq\max\left|\frac{1}{\omega_{ij}(s)}\langle \Psi_j(s)|\frac{d}{ds}|\Psi_i(s)\rangle\right|^2,
\label{cap0_adiabatic}
\end{equation}
where $\omega_{ij}(s)=\omega_j(s)-\omega_i(s)$ with $\hbar\omega_k(s)$ being the energy of the eigenstate $\ket{\Psi_k(s)}$.  
Since the value of $p_{i\rightarrow j}$ has to be as small as possible for any~$j\neq i$, one can immediately see that the adiabatic following of an eigenstate does not succeed, in principle, when the corresponding energy eigenvalue becomes degenerate with any other eigenvalue of the system. Note however that adiabatic passage in the presence of quasi degenerate eigenstates is still possible if they are not coupled during the dynamics, i.e., when $\left|\langle \Psi_j(s)|\frac{d}{ds}|\Psi_i(s)\rangle\right| \rightarrow 0$ due to, for instance, some particular symmetries of the system.  

SAP is a particular example of adiabatic following of a spatial eigenvector which is modified in either time or space.
As an example for the first case, one can consider
particles trapped in a local minimum of an external potential which is changed as a function of time.
The second case (spatial evolution) usually refers to waveguides, where a localized wavepacket is travelling with finite velocity and encounters changing couplings between different waveguides along the propagation direction. 
It is worth noting that the concept of adiabatically following an energy eigenstate in quantum mechanics has a close analog in classical light and sound propagation, see \sref{sec:classicalSAP} for a detailed discussion.
There, a propagating light/sound wave can adiabatically follow a global spatially varying mode of the waveguide system, if the change of the mode profile is smooth along the propagation direction. 

For SAP to work, three basic conditions have to be fulfilled. The first one is the existence of different and well-defined trapping regions, each of them supporting their own asymptotic eigenvectors, i.e, the trapped states/modes of each region isolated from the rest. The second requirement is the existence of a coupling mechanism between the asymptotic eigenvectors, and the third one is the ability to control the strength of the couplings and/or the eigenvalues of the asymptotic eigenvectors during the process. In a physical system where these three conditions are fulfilled, it is possible to perform a SAP process by following an eigenvector of the full system. 
The advantage of SAP with respect to other transfer techniques relies on its robustness: in the adiabatic regime the transfer between the asymptotic eigenvectors will be almost $100\%$ efficient no matter the total duration of the process or the particular parameter values chosen for the variation of the couplings and the eigenvalues.

\section{Spatial adiabatic passage of matter waves}
\label{sec:SAP}

\subsection{Formalism}
\label{sec:formalism}

Spatial adiabatic passage of matter waves can be achieved in a variety of trapping geometries by simply manipulating the height of the potential barriers and/or the trap distances in an adiabatic fashion.
As two paradigmatic examples of SAP, we first review the formalism for adiabatic transport of matter waves between the outermost traps of a triple-well potential, and then we discuss the double-well case.
The triple well requires the ability to control the tunneling coupling between the traps, while  the double-well case in addition requires the manipulation of the energy bias between the localized eigenstates. 

\subsubsection{SAP in a triple-well system} 
\label{formalism_SAP_three_well}

\begin{figure}
\centerline{ \resizebox{0.99\columnwidth}{!}{\includegraphics{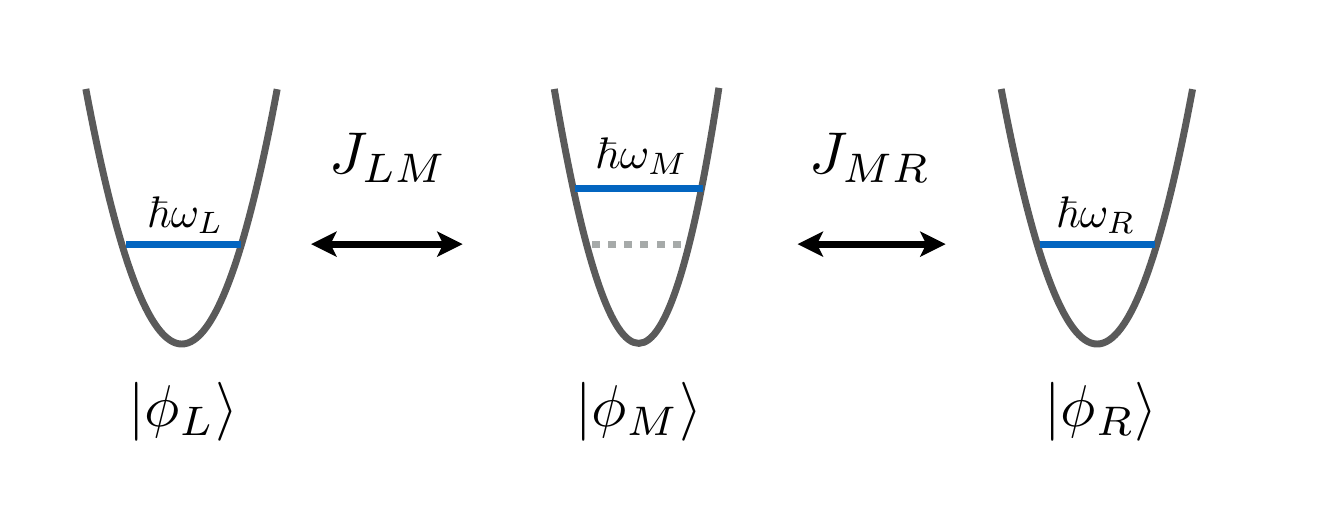} } }
\caption{Sketch of the three-level model approximation used to describe a single quantum particle in a triple-well potential.
}
\label{fig:SAPmodel}
\end{figure}

Consider a single quantum particle of mass $m$ trapped in a triple-well potential, see \fref{fig:SAPmodel}, which we want to move between the outermost potential minima, i.e., from state $\ket{\phi_L}$ to $\ket{\phi_R}$.
Since tunneling will take place along a fixed direction $x$, we can describe the particle's dynamics using the 1D Schr\"odinger equation
\begin{align}
i \hbar\frac{\partial }{\partial t}\psi (x,t)&=
H\psi (x,t) \nonumber \\
&=\left[
- \frac{\hbar^2}{2m}
\frac{\partial^2}{\partial x^2}
+ {V}\left( x,t \right) \right] \psi (x,t),
\label{Yu1xt}
\end{align}
where $V(x,t)$ is the trapping potential.
As we are considering the low energy limit, i.e., only the ground states of each well are involved in the dynamics, we can write the atomic wavefunction as the superposition
\begin{equation}
   \psi (x,t)=\sum_{i=L,M,R} a_i (t) \phi_i(x).
\label{threemode}
\end{equation}
Here $\phi_i(x) = \langle x | \phi_i \rangle$ are the localized ground state wavefunctions of each isolated trap centered at $x_i$ with $i=L,M,R$ accounting for the left, middle and right well, respectively. The probability amplitudes fulfill $\sum_{i} \vert a_i\vert^2=1$ and to guarantee that the ground state wavefunctions satisfy $\int \phi_j^* \phi_i dx = \delta_{ij}$ it is necessary to orthonormalize them by means of, for instance, the Gram--Schmidt \cite{eckert_quantum_2002,loiko_filtering_2011} or the Holstein--Herring methods \cite{holstein_mobilities_1952,herring_critique_1962}.
This lowest-band approximation is good as long as adiabatic evolution is maintained, which in many works we discuss below is supported by numerical integration of the full time-dependent Schr\"odinger equation. 

Inserting ansatz~(\ref{threemode}) into \eref{Yu1xt}, the equation of motion for the probability amplitudes becomes
\begin{equation}
 i \hbar \frac{d}{dt}
\left( \begin{array}{c}
a_L\\
a_M\\
a_R
\end{array} \right) =
H_{3M} \left( \begin{array}{c}
a_L\\
a_M\\
a_R
\end{array}
\right),
\label{threemodeH1}
\end{equation}
with the three-mode Hamiltonian
\begin{equation}
H_{3M} = \hbar 
\left( \begin{array}{ccc}
 \omega_L & - \frac{J_{ML}}{2} & - \frac{J_{RL}}{2} \\
- \frac{J_{LM}}{2} &  \omega_M & - \frac{J_{RM}}{2} \\
- \frac{J_{LR}}{2} & - \frac{J_{MR}}{2} &   \omega_R
\end{array}
\right),
\label{threemodeH2}
\end{equation}
where the energy of the asymptotic vibrational eigenstate $\phi_i$ and the tunnelling rate between states $\phi_i$ and $\phi_j$ are given, respectively, by
\begin{align}
\hbar \omega_i =&  \int \phi_i^* H \phi_i dx, 
\label{onsiteenergy} \\
\hbar J_{ij} =& -2 \int \phi_j^* H \phi_i dx. 
\label{trate}
\end{align}
Without loss of generality we choose the $\phi_i$ eigenfunctions to be real and, therefore, $J_{ij}=J_{ji}$. 

Let us consider now the particular case for which (i) $\omega_L=\omega_R$, i.e., the two outermost wells are in resonance; and (ii) $J_{LR} \sim 0$, i.e., direct tunneling between the outermost traps is negligible. Then
\begin{equation}
H_{3M}= \hbar 
\left( \begin{array}{ccc}
0 & - \frac{J_{LM}}{2} & 0\\
- \frac{J_{LM}}{2} &  \omega_M  & - \frac{J_{MR}}{2} \\
0 & - \frac{J_{MR}}{2} & 0
\end{array}
\right),
\label{threemodeHSAP}
\end{equation}
where, for simplicity, the origin of energies has been taken as $\hbar \omega_L=0$. Diagonalizing \eref{threemodeHSAP} one obtains three eigenstates
\begin{eqnarray}
\ket{+}&=& \sin \Theta \cos \varphi \ket{\phi_L} - \sin \varphi \ket{\phi_M} + \cos \Theta \cos \varphi \ket{\phi_R} , \\
\ket{-}&=& \sin \Theta \sin \varphi \ket{\phi_L} + \cos \varphi \ket{\phi_M} + \cos \Theta \sin \varphi \ket{\phi_R},  \\
\ket{D}&=& \cos \Theta  \ket{\phi_L}-\sin \Theta  \ket{\phi_R},
\label{three_eigenstates}
\end{eqnarray}
with eigenvalues 
\begin{eqnarray}
\lambda_{\pm} &=& \frac{\hbar}{2}\left( \omega_M \pm  \sqrt{\omega_M^2+J_{LM}^2+J_{MR}^2} \right), \\
\lambda_D&=&0 .
\label{three_eigenvalues}
\end{eqnarray}
The mixing angles $\Theta $ and $\varphi$ are
\begin{eqnarray}
\tan \Theta  &=& \frac{J_{LM} }{J_{MR} },  \\
\tan \varphi &=& \frac{\sqrt{J_{LM}^2+J_{MR}^2}}{\sqrt{J_{LM}^2+J_{MR}^2+\omega_M^2}+\omega_M}.
\label{mixing_angles}
\end{eqnarray} 

To adiabatically transfer a particle  in the considered triple-well system from the left to the right trap using SAP, one can follow state $\ket{D}$ in \eref{three_eigenstates}, the so-called spatial dark state, by smoothly varying the tunneling rates such that the mixing angle $\Theta$ evolves from $0$ to $\pi/2$.
This means to favor first the tunneling between the middle and right traps and then the tunneling between the left and middle traps.
This temporal sequence of the tunneling couplings is named the counterintuitive coupling scheme.  
Since the spatial dark state only involves the localized ground states of the outermost wells, the signature of SAP in a triple well is that the middle well is negligibly populated during the whole transport process.

\subsubsection{Comparison to STIRAP}
\label{subsec:differences_SAP_STIRAP}

In quantum-optical STIRAP, two-laser pulses are applied in a counterintuitive temporal sequence to the two adjacent transitions of an atomic or molecular $\Lambda$-type three-level system to transfer the population between the two lower energy internal states~\cite{bergmann_coherent_1998}.
Therefore, SAP in a triple-well potential is the matter wave analogue of STIRAP.
From a physical point of view, SAP deals with the external (localized) degrees of freedom of trapped particles while STIRAP deals with the internal degrees of freedom.
This implies some differences between both techniques:

(i) In STIRAP, to obtain the three-mode Hamiltonian with the two Rabi frequencies playing the role of the tunneling rates, the electric dipole and rotating wave approximations are needed. 
In the SAP case, the fact that the tunneling amplitudes must be smaller than the involved trapping frequencies is equivalent to the rotating wave approximation, and it ensures that, in the adiabatic limit, no transitions to excited vibrational states occur.
Detrimental decoherence mechanisms in STIRAP such as Doppler broadening or photon recoil are not present for SAP although others appear such as trap shaking or finite trapping lifetimes.
The role of some decoherence mechanisms in SAP is discussed in \sref{sec:practical}.

(ii) The signature of both SAP in a triple-well system and STIRAP techniques is that the intermediate state is not being populated during the whole process.
For SAP this seems to indicate that the local continuity equation associated to the Schr\"odinger equation fails. In \sref{sec:TWT} we review this paradoxical issue.

(iii) Being a spatial transfer, SAP can be considered in systems for which it is possible to tunnel-couple all three traps by taking, for instance, three non-aligned wells in two and three-dimensional configurations.
This introduces an additional tunneling rate $J_{LR}$ which allows for potential applications in matter-wave interferometry and for the generation of angular momentum, see \sref{sec:beyond}.
It is worth noting that this extra coupling has been proposed for cyclic STIRAP \cite{unanyan_laser_1997,fleischhauer_coherent_1999} and recently for superadiabatic STIRAP \cite{giannelli_three_2014}. 
In chiral molecules, whose states have ill-definite parity due to the lack of inversion symmetry, the possibility of coupling of all three levels appears naturally \cite{kral_cyclic_2001,kral_two-step_2003}.

(iv) At variance with STIRAP, there is no need to keep the resonance to perform SAP in a triple-well system.
If the resonance condition does not hold during the dynamics, the spatial dark state becomes dressed by the localized state of the middle well.
While in STIRAP the intermediate state is, typically, a fast decaying excited state that one does not want to populate at any time during the dynamics, SAP is free of this problem since the middle localized state is as stable as the localized states of the outermost wells.
However, to achieve significant tunneling rates, it is convenient to keep the resonance condition between the traps.
See \sref{sec:practical} for a detailed discussion on how to keep the resonance condition in practical implementations.

(v) In STIRAP, one can control the relative phase between the pump and Stokes laser pulses.
However, it is not obvious how this can be achieved in SAP
because the tunneling rate between two real-valued localized energy eigenfunctions is also real-valued.
Potential control on the relative phase between the couplings is presently discussed to implement geometrical phases for quantum gates and holonomic quantum computation, see \sref{subsection:geometric_gates} for a more detailed analysis.

(vi)
To ensure the adiabatic transport it is necessary that $J_{MR}(t=0) \gg J_{LM}(t=0)$, and $J_{LM}(t=t_{\max}) \gg J_{MR}(t=t_{\max})$.
In STIRAP, because of the necessity to turn on a laser pulse, it is natural to consider pulses which are symmetric about their midpoint, such as Gaussian pulses~\cite{GRB+88}, even though various pulse schemes have been described~\cite{CH90,LS96}.
In SAP, on the other hand, there is no requirement for the tunneling rates to be initially zero, and many proposals exploring error function pulses \cite{DGH07}, square sinusoidal pulses \cite{rab_spatial_2008}, sinusoidal pulses, linear variation \cite{vaitkus_digital_2013} and so on exist.
The decision of choosing a pulse shape may come down to the availability and convenience of the control, provided that the counterintuitive pulse condition is maintained. 
Interestingly, although smooth variations of the control parameters would seem necessary for adiabatic passage, there are two pulse schemes that show good performances without the need for smoothly varying controls: piecewise adiabatic passage \cite{SMM07,SMS09} and digital adiabatic passage \cite{vaitkus_digital_2013}.

\subsubsection{SAP with three identical harmonic wells}
\label{subsec:harmonic_wells_SAP}

To illustrate the SAP technique let us consider a triple-well potential modeled as three identical truncated harmonic wells \cite{eckert_three-level_2004} .
This academic example allows for keeping the resonance between the three ground states of the triple-well potential and to obtain accurate analytical expressions for the tunneling rates.
Other potentials, such as square wells \cite{opatrny_conditions_2009}, Gaussian \cite{eckert_three_2006} or P\"oschl--Teller \cite{loiko_filtering_2011} are also considered in the literature, and lead to qualitatively similar results.

To control the tunneling rates in time, one can either adjust the well separation or the height of the barriers.
Here we follow the former approach and assume that each well is centered at $x_i(t)$ with $i=L,M,R$ such that the joint trapping potential reads
\begin{equation}
V(x,t)=
\frac{1}{2} m\omega_x^2 \min_i \left\{ \left( x-x_i(t) \right)^2 \right\},
\label{THP}
\end{equation}
with $\omega_x$ being the trapping frequency. The distances between the well centers are given by $d_{LM}=\vert x_L-x_M\vert$ and $d_{MR}=\vert x_R-x_M\vert$. We assume that at the initial time $t=0$ the three wells are negligibly coupled by tunneling and that the single quantum particle is prepared into the localized ground state of the left well, i.e., 
\begin{eqnarray}
\psi(x,t=0)&=&\phi_L(x-x_L(t=0)) \nonumber \\
&=&\frac{\sqrt{\alpha}}{\sqrt[4]{\pi}}e^{-(x-x_L(t=0))^2/2\alpha^2},
\label{initialstate}
\end{eqnarray}
with $\alpha=\sqrt{\hbar/m\omega_x}$ being the width of the ground state of the harmonic potential.

The tunneling rate between the ground vibrational states of two truncated harmonic wells as a function of the separation between the two wells can be straightforwardly calculated as the energy splitting between the lowest symmetric and antisymmetric energy eigenstates of the double-well potential.
Thus, for two piece-wise truncated harmonic wells one obtains \cite{razavy_quantum_2003,menchon-enrich_single-atom_2014}
\begin{equation}
\frac{J( d/\alpha)}{\omega_x}=\frac
{-1+e^{( d/\alpha)^2/4}\left(1+(d/\alpha )\sqrt{\pi} {\rm erfc}( d /2\alpha)/2 \right)
}
{\sqrt{\pi} \alpha \left( e^{(d/\alpha )^2/2}-1 \right)/ d
}.
\label{tunnelingrate}
\end{equation} 
Here ${\rm erfc}(x)=\int_x^\infty e^{-t^2}dt$ is the complementary error function and $d$ the distance between the two well centers.
In the adiabatic limit and in order to avoid excitations to unwanted vibrational states, the distances between the traps must ensure that $J < \omega_x$.

\begin{figure}
\centerline{ \resizebox{0.99\columnwidth}{!}{\includegraphics{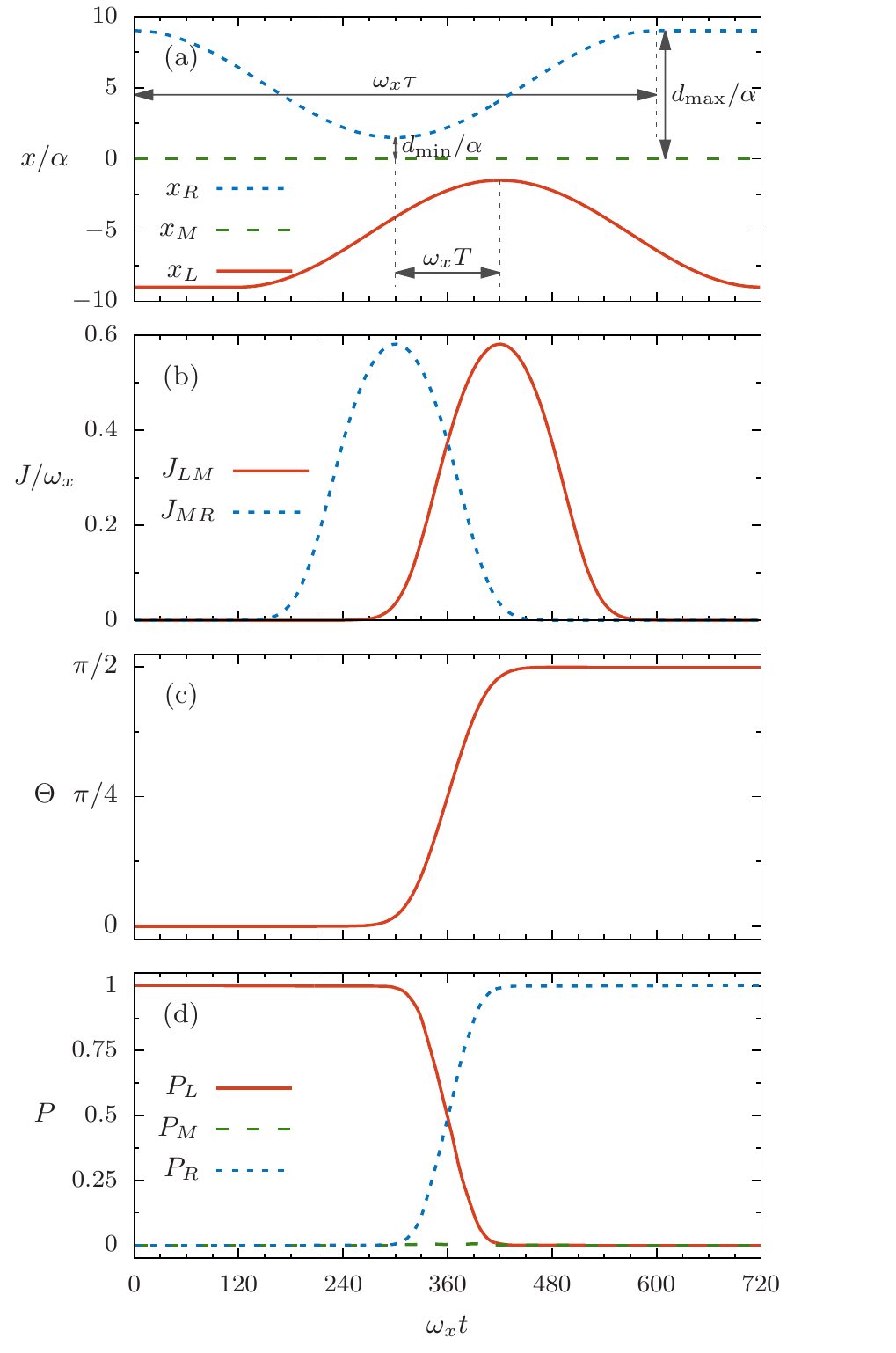} } }
\caption{SAP process for a single quantum particle initially in the left well of a piece-wise truncated triple-well harmonic potential.
Temporal evolution of the (a) well positions, (b) tunneling rates, (c) mixing angle, and (d) populations of the ground state of each well.
Here $d_{\max} = 9 \alpha$, $d_{\min} = 1.5 \alpha $, $\tau = 600 \omega_x^{-1}$, and $T = 120 \omega_x^{-1}$.
}
\label{fig:SAPevolution}
\end{figure}

The counterintutive coupling sequence can then be implemented by the following temporal evolution of the trap centers
\begin{align}
x_{L}(t) &= 
\begin{cases}
-d_{\max} \ & t \leq T , \\
-d_{\min} - (d_{\max} - d_{\min}) \cos^2\left(\frac{\pi (t-T)}{ \tau}\right) \ &t > T ,
\end{cases} \nonumber \\
x_{M}(t) &= 0 , \nonumber \\
x_{R}(t) &= 
\begin{cases}
d_{\min} + (d_{\max} - d_{\min}) \cos^2\left(\frac{\pi t }{ \tau}\right) \ &t \leq \tau , \\
d_{\max} \ & t > \tau .
\end{cases}
\end{align}
The total time is then $t_{\max} = \tau + T$, with $\tau$ being the time it takes to approach and separate the traps and $T$ the delay between the two approaching sequences, see \fref{fig:SAPevolution}(a).
The corresponding tunneling rates calculated from \eref{tunnelingrate} are shown in \fref{fig:SAPevolution}(b), and the resulting mixing angle $\Theta$ can be seen to smoothly evolve from $0$ to $\pi /2$ in \fref{fig:SAPevolution}(c).
This efficiently transports the quantum particle from the left to the right well, see \fref{fig:SAPevolution}(d).

\begin{figure}
\centerline{ \resizebox{0.99\columnwidth}{!}{\includegraphics{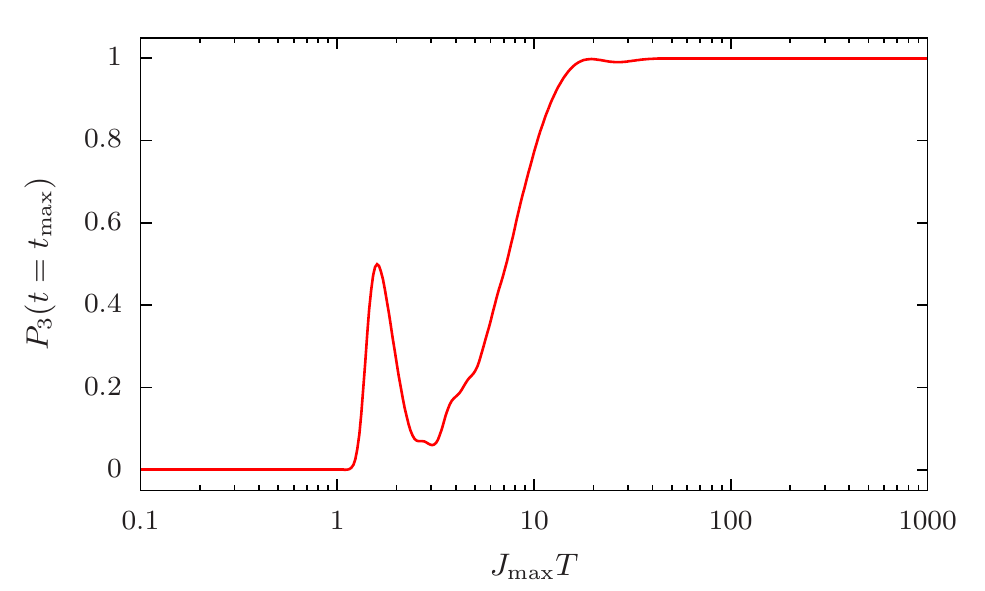} } }
\caption{Population at the right well at the end of the SAP protocol for a particle initially located at the left well as a function of the adiabaticity of the process.
Here $d_{\max} = 9 \alpha$, $d_{\min} = 1.5 \alpha $, $\tau = 5T$, and $J_{\max} = J(d_{\min})$.
}
\label{fig:SAPtemporaldelay}
\end{figure}

For the three-mode case, it is possible to obtain a simple `global' condition which guarantees the adiabaticity of the process.
For tunneling rates with Gaussian-like temporal profiles of width $\sigma$ and peak values $J_{\max}$, separated by an optimal temporal delay $T \sim \sqrt{2}\sigma$, this condition is given by~\cite{bergmann_coherent_1998}
\begin{equation}
J_{\max} T > 10.
\label{adiabaticitycondition}
\end{equation}
The final population in the right well as a function of $J_{\max} T$ is shown in \fref{fig:SAPtemporaldelay}, demonstrating the high fidelity and the robustness of the SAP process for adiabatic time scales.
In the non-adiabatic limit, i.e., for short durations of the process, two factors are inhibiting full transfer: Rabi oscillations between localized states and excitations to higher vibrational bands.

Finally, it is important to highlight that the validity of the three mode approximation to account for the adiabatic transport of a single quantum particle in a triple-well potential has been verified by the direct integration of the Schr\"odinger equation in one \cite{eckert_three-level_2004,cole_spatial_2008}, two \cite{menchon-enrich_single-atom_2014,menchon-enrich_tunneling-induced_2014} and three dimensions \cite{rab_spatial_2008}.
Accurate quantitative agreement was not only obtained when SAP was applied to the adiabatic transport between the localized ground states, but also between excited states \cite{loiko_filtering_2011}.
In addition, as will be discussed in \sref{sec:BEC}, SAP can be applied over a significant parameter range to the non-linear Gross--Pitaevskii equation to deal, in the mean field approximation, with the adiabatic transport of a Bose--Einstein condensate.

\subsubsection{SAP in a double-well system}

Adiabatic passage techniques can also be implemented in double-well potentials to transfer a single particle from the left to the right trap.
The two-mode Hamiltonian in terms of the left, $\ket{\phi_L}$, and right, $\ket{\phi_R}$, localized states is
\begin{equation}
H_{2M} = \hbar
\left( \begin{array}{cc}
0 & - \frac{J}{2} \\
- \frac{J}{2} & \omega 
\end{array}
\right),
\label{twomodeH}
\end{equation}
where $\hbar\omega=\hbar\omega_R-\hbar\omega_L$ is the energy bias and $J$ the tunneling rate. The diagonalization of \eref{twomodeH} gives the two eigenstates
\begin{eqnarray}
\ket{+}=\sin \theta \ket{\phi_L} - \cos \theta \ket{\phi_R} \\
\ket{-}=\cos \theta \ket{\phi_L} + \sin \theta \ket{\phi_R} 
\end{eqnarray}
with eigenvalues $\lambda_{\pm} = \hbar \left( \omega \pm \sqrt{J^2+\omega^2} \right)/2$, and a mixing angle given by $\tan 2\theta = J/\omega$.
Analogously to the quantum-optical RAP technique \cite{vitanov_laser-induced_2001}, it is therefore possible to transfer a single particle initially located in the left trap to the right one by adiabatically following either $\ket{+}$ or $\ket{-}$. 
Alternatively it is straightforward to show that the time evolution of the probability amplitudes $a_L$ and $a_R$ for the quantum particle to be in the left and right traps, respectively, can be mapped to a three-variable model governed by the following equations of motion~\cite{vitanov_stimulated_2006,ottaviani_adiabatic_2010}
\begin{equation}
\frac{d}{dt}  
\left( \begin{array}{c}
U\\
V\\
W
\end{array}
\right)
=
\hbar
\left( \begin{array}{ccc}
0 &  -\omega & 0 \\
\omega & 0 &  J \\
0 & -J &  0
\end{array}
\right)
\left( \begin{array}{c}
U\\
V\\
W
\end{array}
\right).
\label{twomode}
\end{equation}
Here $U=2 {\rm Re}\{ a_L a_R^*\}$, $V=2 {\rm Im} \{ a_L a_R^*\}$ and $W=\vert a_R \vert^2-\vert a_L \vert^2$ are, respectively, three real variables that correspond to two times the real and imaginary part of the spatial coherence and the population imbalance. The conservation of the norm implies that $U^2+V^2+W^2=1$.
The matrix on the r.h.s. of \eref{twomode} is an odd real-valued skew symmetric matrix with one zero eigenvalue, $\lambda_d=0$, and two other ones $ \lambda_\pm = \pm \hbar\sqrt{\omega^2+J^2}$.
The eigenvector associated to the zero eigenvalue, named dark eigenvector, is given by
\begin{equation}
d(\tilde{\theta})= W \cos \tilde{\theta}  + U \sin \tilde{\theta} ,
\label{darkvariable}
\end{equation}
where $\tan \tilde{\theta} = J/\omega$.
The dark eigenvector is stationary and decoupled from the rest of the eigenvectors.
Thus, if the quantum particle is initially located in the left well ($W =-1$, $U=V=0$) and $\tilde{\theta}$ is smoothly varied from 0 to $\pi/2$, the wavefunction splits equally between the two wells ($W=V=0$, $U=-1$).
On the other hand, the particle can be adiabatically transferred from the left ($W=-1$) to the right ($W=1$) well by smoothly varying $\tilde{\theta}$ from $0$ to $\pi$. 
Finally, it is possible to inhibit the transport of a particle initially located in the left trap if the mixing angle evolves from $0$ to any arbitrary value to eventually reach $0$ again.

\subsection{Single atoms}
\label{sec:atoms}

\subsubsection{Three-level atom optics}

\begin{figure}
\centerline{ \resizebox{0.95\columnwidth}{!}{\includegraphics{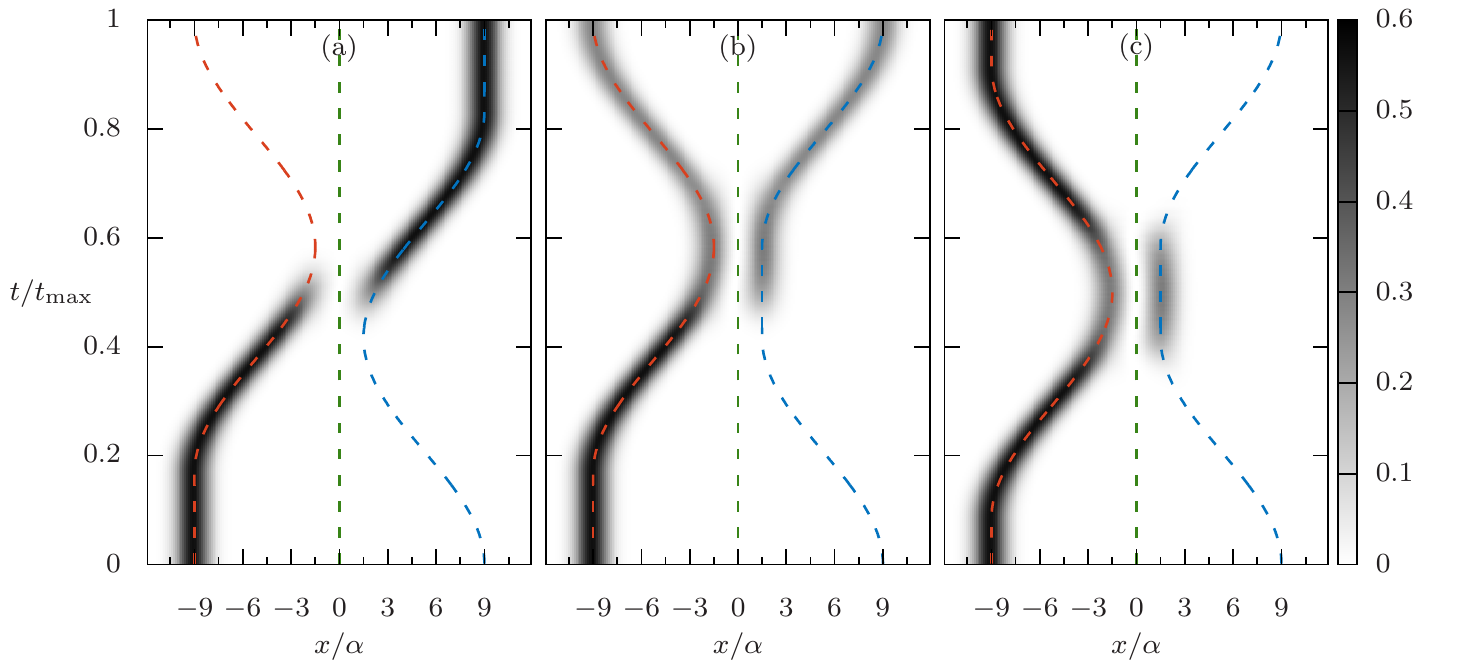}}}
\caption{Trap positions (dashed lines) and atomic densities for TLAO techniques analogous to (a) STIRAP, (b) CPT, and (c) EIT processes (see text).
The atom is initially in the left trap centered at $x = -9 \alpha$.
}
\label{fig:TLAO}
\end{figure}

Controlling the state of single quantum particles is a challenging task and a topic of significant present activity in the fields of quantum computation, quantum metrology, and quantum simulation.
While there is a panoply of techniques to control the internal states of atoms,
there is a need for the development of novel techniques to manipulate the external degrees of freedom of matter waves in optical and magnetic traps.
To this aim, the transport of single atoms via SAP was proposed by Eckert {\it et al.}~\cite{eckert_three-level_2004} in a set of techniques named three-level atom optics (TLAO).
The name suggests an analogy between quantum-optical processes in electronic three-level systems and tunneling-based processes for cold atoms in triple-well potentials.
The analogue of STIRAP in TLAO has already been presented in \sref{sec:formalism}.
Two more analogies were presented in this initial work~\cite{eckert_three-level_2004}: coherent population trapping (CPT), which allows to create a delocalized dark state between the outermost traps by finishing the evolution of the mixing angle $\Theta$ at $\pi/4$,
and electromagnetically-induced transparency (EIT),  which inhibits the transfer of the atom from the left to the right trap by strongly coupling the middle and right traps.
The positioning sequences for the individual traps and the temporal evolution of the atomic density for these three TLAO techniques are shown in \fref{fig:TLAO}.

A follow-up work~\cite{eckert_three_2006} focused on the robustness of the transfer, studied the transfer with two atoms in the triple-well system, introduced Gaussian potentials and also showed the possibility of applying SAP techniques to cold atoms propagating in optical waveguides.
Examples of these processes are discussed later in this section.

\subsubsection{Population in the central well}
\label{sec:TWT}

As discussed above, for the transport of an atom between the two outermost traps in a triple-well following a dark state of the form of \eref{three_eigenstates}, the middle trap remains unpopulated.
In real space this corresponds to the spatial dark state possessing a node in the middle trap.
However, for finite times the dynamical state corresponds to the dark state being weakly dressed by other eigenstates, leading to a finite population in the central trap.
The more adiabatic the process is performed, the better the following of the dark state, and thus, the smaller this population will be.
This therefore allows for the paradoxical possibility that the transport of the atom between the outer traps can be achieved with a negligible population in the middle trap, see \fref{fig:TLAO}(a), which appears to contradict the quantum continuity equation.

This behaviour is not particular to systems of ultracold atoms in harmonic traps, but appears in any system where SAP can be implemented for three spatially separated wells following a spatial dark state of the form of \eref{three_eigenstates}.
For instance, in Ref.~\cite{rab_spatial_2008} this effect was discussed for an atomic Bose--Einstein condensate in a triple well created by adding two Gaussian barriers to a harmonic trap.
SAP was then performed by lowering and raising these barriers in a counterintuitive fashion.
An analytical study of the conditions for a vanishing central-well population was carried out in Ref.~\cite{opatrny_conditions_2009} using three square wells separated by delta-function or square potential barriers. 
The authors found that in order to maintain a negligible occupation of the central trap, the depths of the outer wells need to be varied while the barrier heights are changed time-dependently.

\begin{figure}
\centerline{ \resizebox{0.99\columnwidth}{!}{\includegraphics{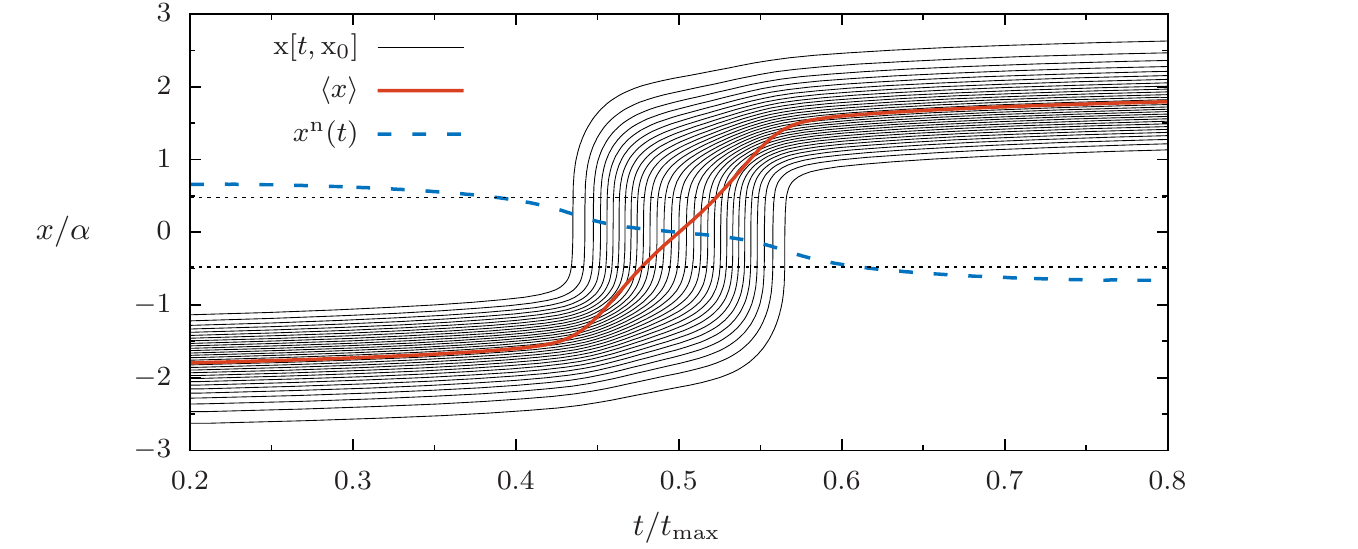}}}
\caption{
Set of 30 Bohmian trajectories (black lines) associated with the dynamics of a Bose--Einstein condensate during a SAP process between the outermost traps of a triple well~\cite{benseny_need_2012}.
The distribution of initial positions for the trajectories ($\textrm{x}_0$) follows the probability distribution of the ground state of the left trap.
Also represented are the atom mean position (red thick solid line), the quasinode position (blue dashed line), and the positions of the barriers (dotted lines). 
$\alpha$ is the width of the ground state of the harmonic trap.
}
\label{fig:trajs}
\end{figure}

However, how can the atom move between the outer wells without populating the middle trap?
To clarify this apparently paradoxical behaviour, this process was studied in Refs.~\cite{benseny_need_2012} and \cite{benseny_atomtronics:_2012} by means of Bohmian mechanics~\cite{bohm_suggested_1952-1,bohm_suggested_1952-2,benseny_applied_2014}.
Bohmian mechanics is a formalism equivalent to standard quantum mechanics in terms of predictions, and provides a good visualization of continuity because of its use of quantum trajectories.
A particle's velocity along a trajectory is given by
\begin{equation}
\label{eq:bohmv}
\vec{v}_\textrm{traj}(\vec{r},t) = \frac{\vec{j}(\vec{r},t)}{\rho(\vec{r},t)} , 
\end{equation}
where $\rho(\vec{r},t) = |\psi(\vec{r},t)|^2$ is the probability density and
\begin{equation}
\vec{j}(\vec{r},t) = \frac{i \hbar}{2 m} \Big [ \psi(\vec{r},t) \nabla \psi^*(\vec{r},t) - \psi^*(\vec{r},t) \nabla \psi(\vec{r},t) \Big] 
\end{equation}
is its associated probability density current.
For the SAP dynamics following the set up suggested in Ref.~\cite{rab_spatial_2008} and described above, the trajectories are shown in \fref{fig:trajs} and can be seen to follow the wavefunction from the left region to the right one, transiting through the middle trap.
However, as the dark state has a node in the middle region, indicated by the green dashed line in \fref{fig:trajs}, the atomic density has a very low value at that position.
Therefore, as can be seen in the figure, the velocity increases dramatically (cf. \eref{eq:bohmv}) around this quasinode, reaching values which are orders of magnitude larger than the mean velocity of the wave packet.
This means that, in the Bohmian picture, the atom transits through the central trap at a high velocity in order to keep the population low.

Slowing down the SAP process will therefore increase the Bohmian velocities~\cite{benseny_need_2012}, because a better following of the dark state leads to a smaller population in the quasinode.
There is thus a regime for a (finite) total time where the trajectories will approach and eventually surpass the speed of light.
Since with a correct relativistic treatment, Bohmian trajectories for massive particles cannot surpass the speed of light~\cite{leavens_are_1998,struyve_uniqueness_2004}, superluminal trajectories are an irrefutable indication of the application of the Schr\"odinger equation in a regime where it is not valid. 
It is remarkable that relativistic corrections are needed to properly address the transport and avoid superluminal propagation in a system where the associated average velocities are orders of magnitude smaller than the speed of light~\cite{benseny_need_2012}.
Using a relativistic formalism to study SAP, however, is still outstanding.

The vanishing of the central trap population also allows an atom to be transported between the outer wells even if a second atom is present in the central well.
In Ref.~\cite{gajdacz_transparent_2011} the transport was shown to be unaffected by the presence of a second atom of a different species that was considered to be deeply trapped, and thus remained essentially static in the central well.
This proposal is also of interest because it is based on an optical superlattice with a unit cell consisting of three wells.
Optical superlattices are the result of the combination of multiple laser beams with different frequencies/orientation and, at variance with regular lattices, consist of unit cells which can contain multiple traps.
Superlattices give the opportunity to perform the same experiment multiple times simultaneously, once in each cell, which allows to easily scale up the size of more complex cold atomic systems.

\subsubsection{Quantum state preparation} \label{quantum_state_preparation}

Besides transport, SAP techniques have also been studied for the preparation of more complex quantum states from simpler ones.
Such techniques can be very helpful to control and manipulate single neutral atoms for potential applications in quantum information processing.
For instance, a configuration proposed in Ref.~\cite{busch_quantum_2007} can be used to create a symmetric or antisymmetric superposition between the ground states of a double well.
In this system of three traps, the rightmost trap is a double well consisting of two harmonic potentials.
The state localized in the leftmost trap can be resonant with either the symmetric or antisymmetric combination of the ground states in the right trap wells, and SAP can be used to completely transfer the atom to that state.
Which of these states is resonant can be chosen by detuning the trapping frequencies of the right-most wells either up to match the symmetric state or down to match the antisymmetric one.

\begin{figure}
\centerline{ \resizebox{0.99\columnwidth}{!}{\includegraphics{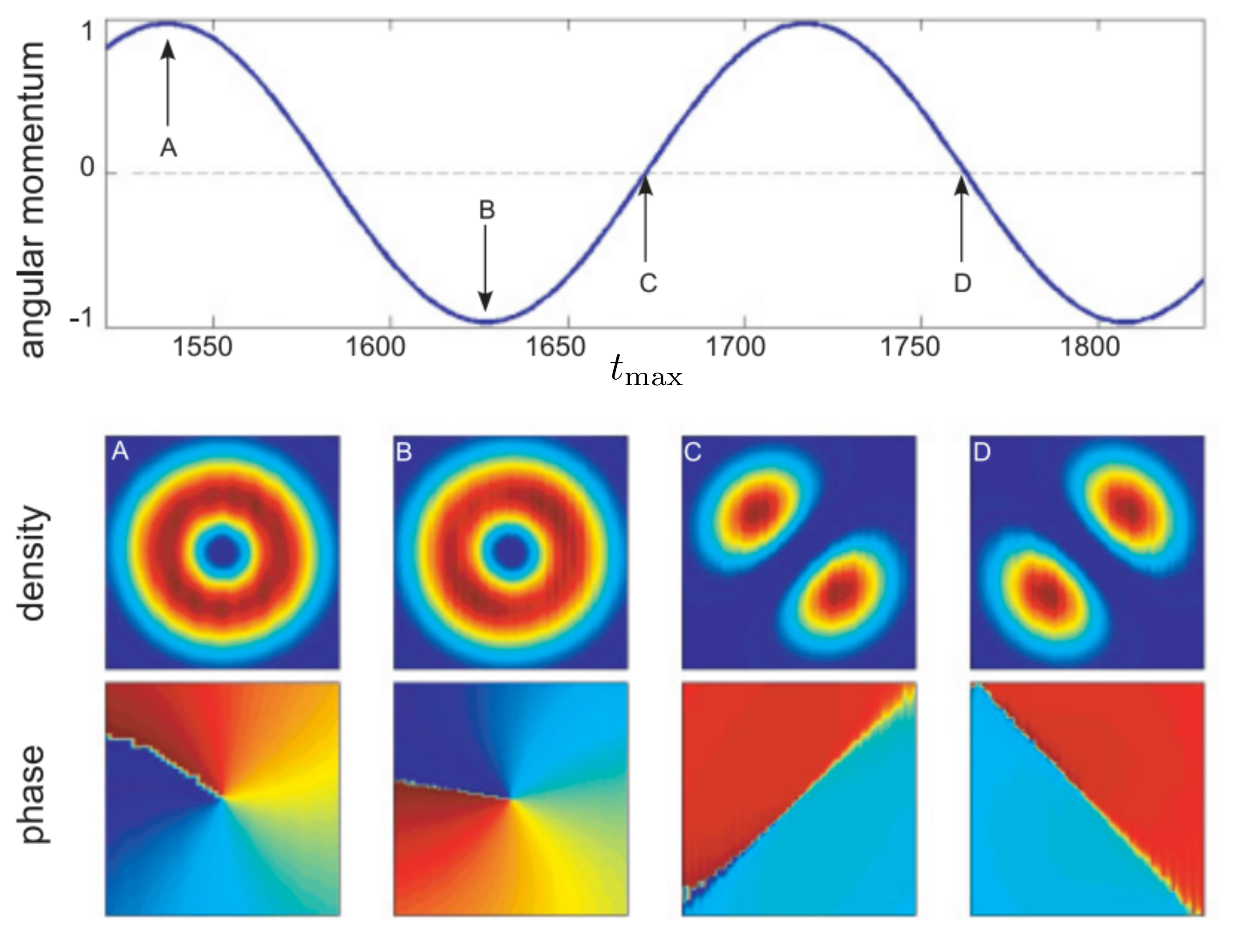}}}
\caption{Angular momentum as a function of the overall duration of the process (upper plot) and the final states in the rightmost trap at points A, B, C, and D (lower plots). All simulations are in the adiabatic regime where almost 100\% transfer is achieved. Time is in units of the inverse of the trapping frequency and angular momentum in units of $\hbar$. From \cite{mcendoo_phase_2010}.}
\label{fig:mcendoo}
\end{figure}

The influence of the SAP dynamics on the phase of a quantum state was studied in Ref.~\cite{mcendoo_phase_2010}, and was shown to allow to control the angular momentum of an atom by transporting it through three 2D harmonic traps.
The initial state is chosen to carry a single unit of angular momentum, and therefore has a phase distribution that increases by $2\pi$ for a closed loop around the centre of the state.
While applying the counterintuitive SAP sequence of couplings leads to a complete transfer of the atom from the initial trap to the most distant one, the angular momentum of the final state oscillates continuously between clockwise and counterclockwise depending on the overall duration of the process, see \fref{fig:mcendoo}.
This shows that SAP is not robust with respect to the conservation of the phase~\cite{mcendoo_phase_2010}.
However, the dependence of the final angular momentum on the overall time of the process is deterministic, therefore it can be used to obtain the full spectrum of angular momentum superposition states, which can have applications for quantum information processing.
Other SAP schemes for manipulating angular momentum when the traps are in a triangular configuration are discussed in \sref{sec:beyond}.

\subsubsection{State filtering}
\label{sec:atom-filter}

\begin{figure}
\centerline{\resizebox{0.70\columnwidth}{!}{\includegraphics{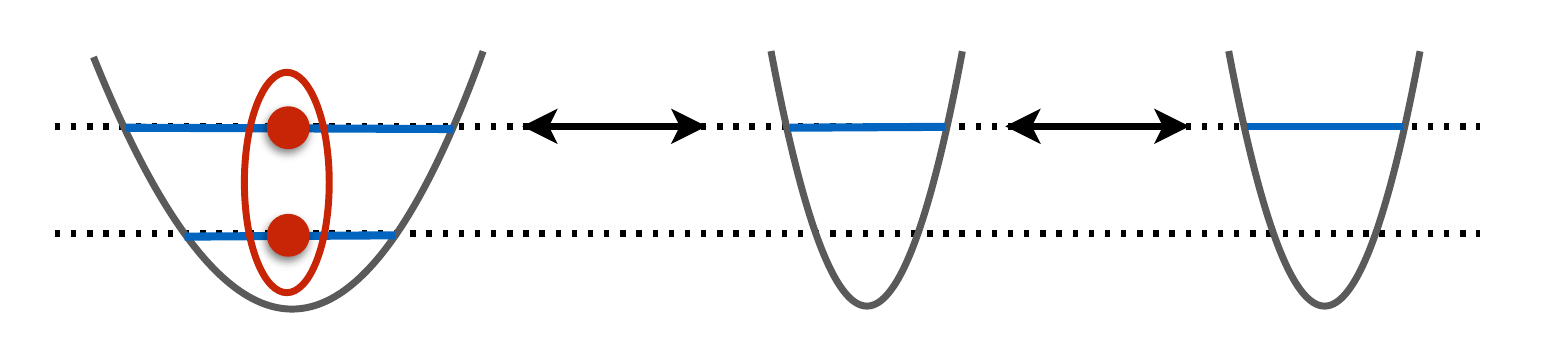}}}
\caption{
Schematic of the state filtering proposal in Ref.~\cite{busch_quantum_2007}.
The middle and right wells have trapping frequencies three times larger than that of the left trap.
The initial state can be a superposition of the ground and first excited states of the left trap.
}
\label{fig:Busch07b}
\end{figure}

The control over the shape of the trapping potentials allows to create filtering mechanisms for vibrational states by engineering the energy spectrum of the system.
A simple SAP mechanism in a triple-well can be devised by making the first vibrational state of the leftmost trap resonant with the ground states of the middle and right traps, see \fref{fig:Busch07b}.
For harmonic traps in 1D, this means making the trapping frequency of the middle and right traps three times larger than that of the left trap.
In this situation, SAP transfers the population of the first excited state of the left trap to the ground state of the right trap, while the population of the ground state of the left trap remains unaffected.

\begin{figure}
\centerline{ \resizebox{0.8\columnwidth}{!}{\includegraphics{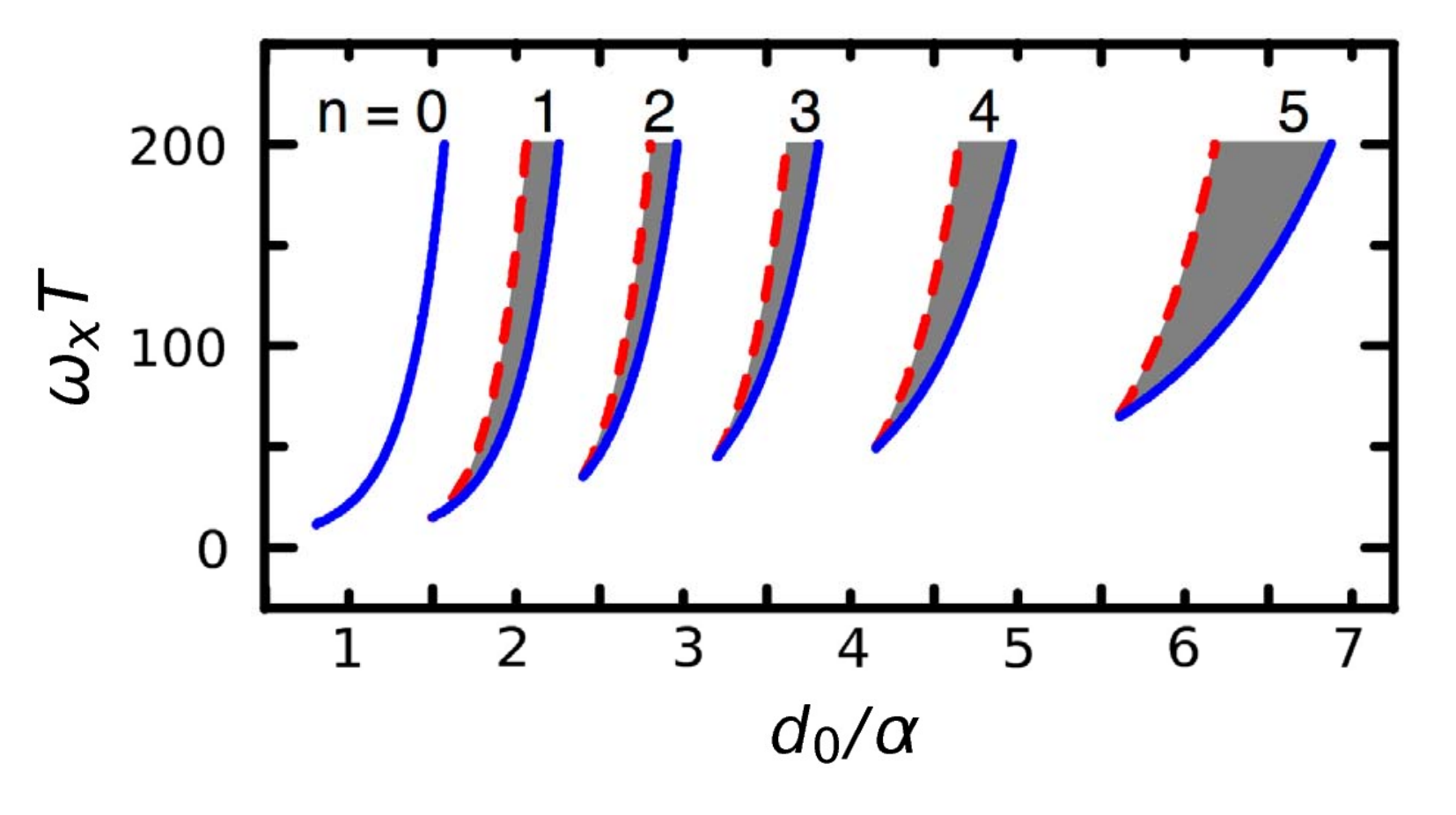}}}
\caption{
State filtering via SAP using the procedure from Ref.~\cite{loiko_filtering_2011}.
Gray areas indicate regions in the parameter plane $(d_0/\alpha, \omega_x T)$ where filtering of the $n$-th vibrational level can be achieved for P\"oschl--Teller potentials with a depth of $42 \hbar \omega_x$.
These areas are limited to the right by $J_{n}^{\rm max}T=10$, where the process is not adiabatic enough for level $n$, (solid blue curves)
and to the left by $J_{n-1}^{\rm max}T=1$, where the process transfers some population within level $n-1$ (dashed red curve).
The maximum couplings $J_n^{\rm max}$ depend on the minimum approach distance between the traps $d_0$.
From Ref.~\cite{loiko_filtering_2011}.
}
\label{fig:yuriadiabaticity}
\end{figure}

Another filtering mechanism can be created by making use of the fact that for two identical potential traps at a fixed distance, the tunneling rate increases with the vibrational state of the traps~\cite{loiko_filtering_2011}.
As a consequence, the adiabaticity condition for SAP transport in a triple well is state-dependent.

Assuming that the energy separation between the different vibrational states of each trap is large enough to avoid cross tunneling among different vibrational states and that there is no significant coupling between the two outermost traps, the Hamiltonian of the system can be separated as
\begin{align}
H=H_0\oplus H_1 \oplus \dots \oplus H_n \oplus \dots 
\end{align}
Here, $H_n$ is the Hamiltonian of the subsystem of the states in the $n$-th vibrational level, which is analogous to the one in \eref{threemodeHSAP} and each has a spatial dark state which allows to transport an atom between the two outermost traps.
However, each subsystem has a different tunneling rate with $J_{n} > J_{n^\prime}$ for $n>n^\prime$, leading to a different adiabaticity condition, see \eref{adiabaticitycondition}.
Therefore, for any given $n$ it is possible to find a parameter regime in which the process is adiabatic for levels $n$ and higher (and SAP transfers the atom to the right trap), but where the tunneling is too weak for those vibrational states lower than $n$ (and SAP leaves them in the left trap), see \fref{fig:yuriadiabaticity}.
This filtering process can be used, for instance, for the preparation of vibrational states on demand and to perform quantum tomography of the initial population of vibrational-states~\cite{loiko_filtering_2011}.
However, it cannot measure the relative phase between the different levels.
In Ref.~\cite{loiko_filtering_2011} this system was studied using P\"oschl--Teller potentials, which approximate experimental Gaussian traps much better than the usually used truncated harmonic potentials and for which analytical expressions for the energy eigenvalues and eigenstates exist.

\subsubsection{Hole transport}\label{Sect:Holes}

\begin{figure}
\centerline{ \resizebox{0.99\columnwidth}{!}{\includegraphics{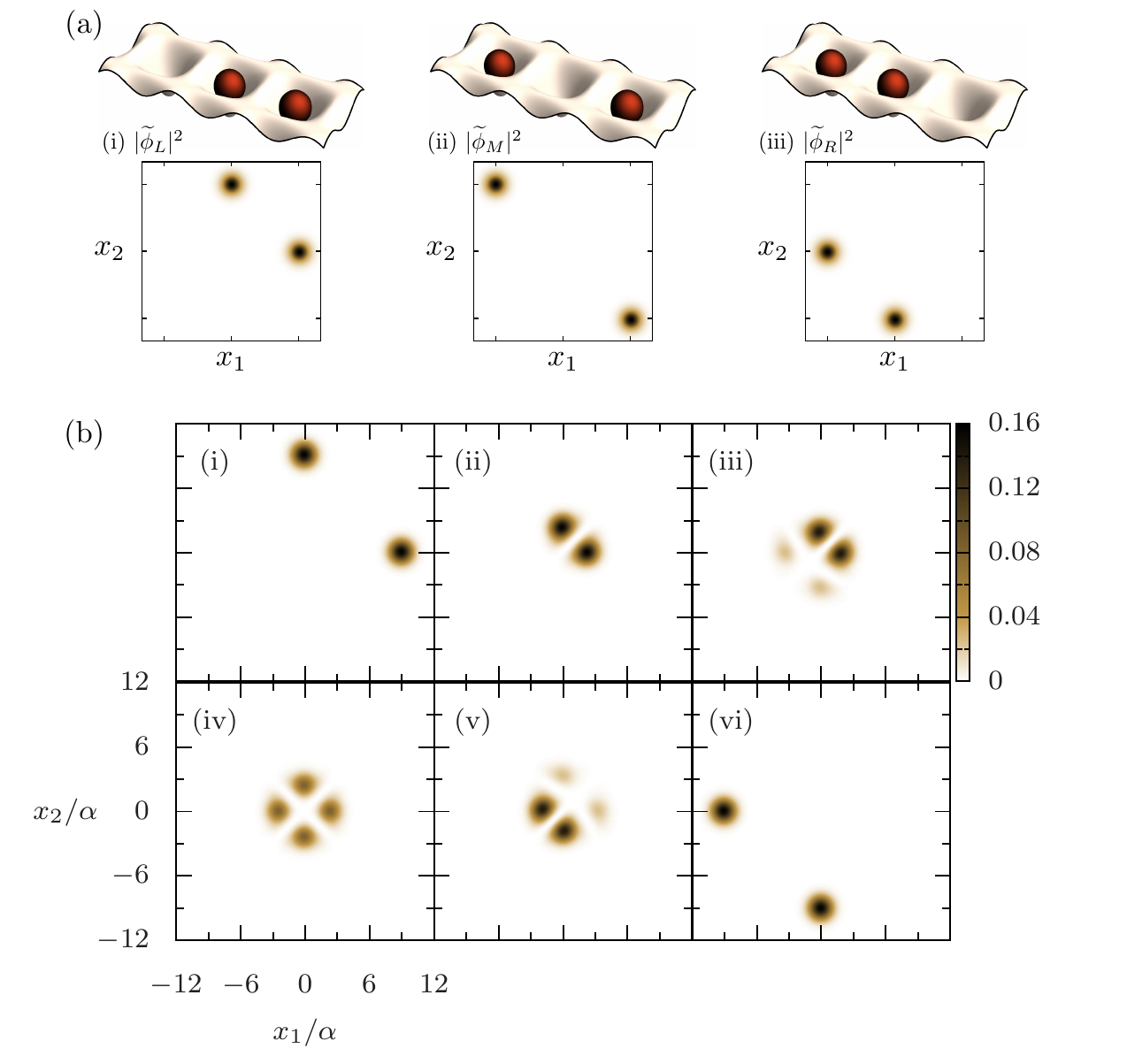}}}
\caption{Transport of a hole in a triple well potential.
(a) Sketch of the states considered with two atoms in three traps, and their representation on the configuration space.
The two spots correspond to the (anti)symmetrization of the wavefunction.
(b) Snapshots of the two-fermion joint probability distribution $| \psi(x_1, x_2, t)|^2$ at different times during the evolution.
The initial and final states correspond, respectively, to $\ket{\widetilde{\phi}_L}$ and $\ket{\widetilde{\phi}_R}$.
The trap approaching scheme is shown in \fref{fig:TLAO}(a).
}
\label{fig:hole}
\end{figure}

The SAP dynamics can also be used for the transport of empty sites, i.e., holes~\cite{benseny_atomtronics_2010}.
For the hole description
to be valid, each trap must contain, at most, one atom in its vibrational ground state at all times, 
which can be achieved by either considering Pauli's exclusion principle (for spin-polarized identical fermions) or by introducing a large enough interaction between atoms.
In the case of two atoms in three identical traps, one can then construct a three-level model for the hole state, see \fref{fig:hole}(a), which supports a spatial dark state for the hole and therefore allows for SAP dynamics.

The efficiency of the process was numerically determined for both spin-polarized fermions and interacting bosons
by solving the two-particle Schr\"odinger equation
with the trapping potential given in \eref{THP} and an interaction potential of the form
\begin{align}
\label{eq.contactUhole}
U(x_1,x_2) = 2 \hbar a_s \omega_p \delta(x_1 - x_2) .
\end{align}
Here $a_s$ is the $s$-wave scattering length of the interaction and $\omega_p$ is the trapping frequency in the transverse direction.

The dynamics of the SAP transport of a fermionic hole in a triple well are shown in \fref{fig:hole}(b).
The initial state corresponds to the hole being in the left trap (atoms in the middle and right traps, see \fref{fig:hole}(b-i)) and the final state corresponds to the hole being in the right trap (atoms in the left and middle traps, see \fref{fig:hole}(b-vi)).
During the entire evolution, the diagonal of the configuration space has a node due to the antisymmetrization of the wavefunction (or the strong contact interaction for bosons).
The fact that the counterdiagonal also has a vanishing population is a signature of the SAP process because it means that state $\ket{\widetilde{\phi}_M}$, i.e., hole in the middle trap (\fref{fig:hole}(a-ii)) is not being populated.

The control over the dynamics offered by the interaction between the atoms and their spin state makes this two-atom system more versatile than the single atom counterpart.
Taking advantage of these two control parameters, the hole SAP transport was proposed to implement two atomtronic devices:
a diode, where the hole transport succeeds in only one direction, and a transistor, where the transport efficiency depends on the spin state of the atoms~\cite{benseny_atomtronics_2010}.
Furthermore, the system's Hamiltonian can be generalized for both fermions and hardcore bosons by means of a Hubbard model considering the hole as an effective particle.
This allows, in a similar manner to multilevel STIRAP/SAP~\cite{shore_multilevel_1991,jong_coherent_2009}, to implement hole transport in arrays of $n$ (odd) traps containing $n-1$ atoms.

\subsubsection{Multiple particle transport}

Another system that shows particle-induced nonlinearities in SAP is the Bose--Hubbard model \cite{bradly_coherent_2012}.  For the case of $N$ particles across 3 wells, using the canonical SAP geometry, it is possible to express the allowed states in a triangular diagram, with the initial state $\ket{N,0,0}$, final state $\ket{0,0,N}$, and the state $\ket{0,N,0}$ at the apex.  Particles hopping from the left to centre wells are represented by diagonal lines going from bottom left to top right, while particles hopping from the centre to the right well are represented by lines from the top left to the bottom right.  By using such a representation, Bradly \textit{et al.} were able to show that the lowest two levels of the non-interacting Bose--Hubbard SAP process were equivalent to the alternating SAP process \cite{jong_coherent_2009},
and hence could be treated following the results given in \eref{eq:35} in \sref{sec:electron_transport} (below).
The results showed the usual hallmarks of SAP, namely particles moving from left to right wells without transient occupation of the central well.

Including particle-particle interactions into the model introduced energy gaps to the intermediate states, which in turn led to an increase in the time required for high-fidelity transport, and central well detuning could be used to mitigate the effects of particle-particle interaction to some extent.
Since the interaction raises the overall energy of the system, additional care needs to be taken in order not to excite transitions to higher-lying bands.

\subsubsection{Coupled matter waveguides}
\label{sec:matterwaveguides}

\begin{figure}
\centerline{ \resizebox{0.95\columnwidth}{!}{\includegraphics{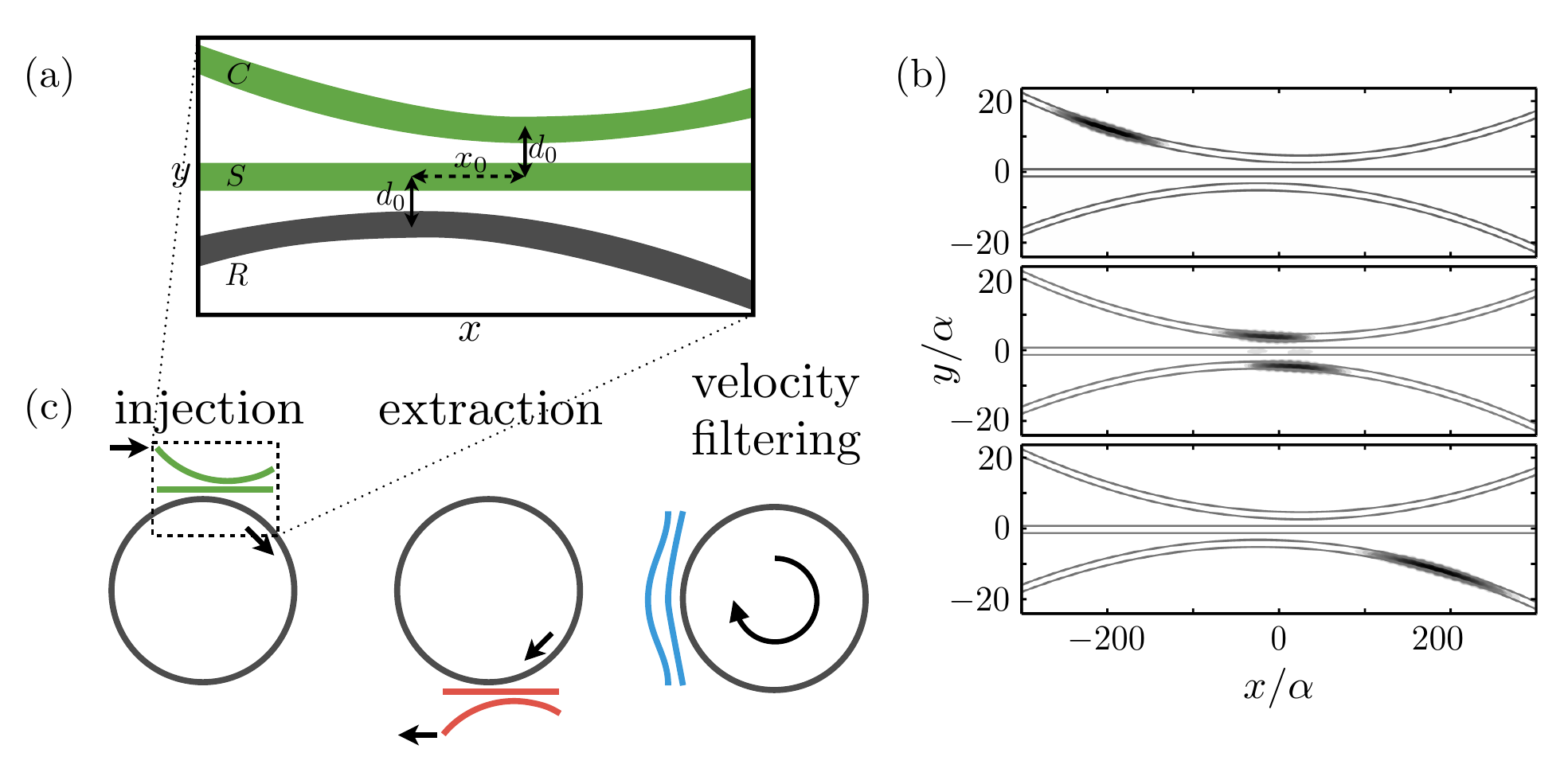}}}
\caption{
(a) Schematic representation of three coupled waveguides $C$, $S$ and $R$ (for curved, straight, and ring) to achieve SAP transport between $C$ and $R$.
$d_0$ corresponds to the minimum $y$ separation between adjacent waveguides while $x_0$ denotes the distance in the $x$ direction between the two positions of minimum separation between the waveguides.
(b) 2D numerical simulations showing the atomic probability distribution at three different consecutive times during the SAP for a $^{87}$Rb atom.
Here, $v_x=0.3\alpha\omega_\bot $, $d_0=3.9\alpha$, $x_0=50\alpha$, and the radius of the ring and curved waveguides is $r=3000\alpha$, with $\omega_{\bot}$ being the transverse trapping frequency of the three waveguides.
(c) Waveguide configurations to (left) inject neutral atoms into, (center) extract from, and (right) velocity filter in a ring.
Adapted from~\cite{loiko_coherent_2014}.}
\label{fig:CSR}
\end{figure}

Up to now we have seen that the external wavefunction of trapped ultracold atoms can be manipulated by a temporal variation of the coupling between potential traps.
Analogously, one can engineer the couplings in space between matter waveguides by appropriately designing a fixed guiding structure, which allows to manipulate the propagation of an atomic wave packet.
In particular, one can implement SAP techniques in a three-waveguide system~\cite{eckert_three_2006} to transport an atomic wave packet between the two outer waveguides by engineering the couplings with position-dependent distances as shown in \fref{fig:CSR}(a).

Considering that the separation between waveguides in the transverse direction, $y$, varies slowly along the longitudinal direction, $x$, the velocity of the atom along any of the waveguides can be approximated as its projection onto the $x$-axis, $v_x$.
This ensures that the atomic longitudinal motion can not excite the transversal modes and it can therefore be decoupled from the transversal dynamics.
Thus, the system can be effectively reduced in the $y$-direction to a 1D triple well potential, analogous to the system presented in \sref{formalism_SAP_three_well}.
The validity of this assumption has been checked by the direct numerical integration of the time-dependent 2D Schr\"odinger equation with the waveguides modeled as truncated harmonic potentials in the transverse direction~\cite{loiko_coherent_2014}.
\Fref{fig:CSR}(b) shows three consecutive snapshots of the 2D atomic probability distribution where, with the SAP counterintuitive approaching sequence, the atom is completely transferred from the transverse vibrational ground state of the $C$ waveguide to the same state of the $R$ waveguide.
Realistic models close to present experiments for the implementation of this technique have been proposed for cold atoms in atom-chip waveguides~\cite{osullivan_using_2010,morgan_coherent_2013},
or radiofrequency traps (for both cold atoms and BECs) which allow for a better control of the asymptotic energy states of each waveguide~\cite{morgan_coherent_2011}, see \sref{sec:practical}.

The SAP process depicted in \fref{fig:CSR}(a,b) constitutes an injection protocol of a single cold neutral atom into a ring trap if we assume that the $R$ waveguide is part of a ring, see \fref{fig:CSR}(c-left).
It is easy to realize that extraction from the ring can be achieved by exchanging the role of the curved and ring waveguides, as depicted in \fref{fig:CSR}(c-center)~\cite{loiko_coherent_2014}.
The two waveguides coupled to the ring can be switched on or off at will by simply turning on or off the laser field that generates them.
As a consequence, these processes can be applied selectively when needed and with higher robustness and efficiency than in the case of simply spatially overlapping the ring and the input/output waveguides \cite{kreutzmann_coherence_2004}.

It is also possible to implement a velocity filter if one takes into account that the adiabaticity of the process is controlled by the atom's injection velocity.
Then, the adiabaticity condition, \eref{adiabaticitycondition}, depends on $v_x$ through the length of the interaction region, $x_0=v_x T$, which allows for the definition of a threshold longitudinal velocity~\cite{loiko_coherent_2014}
\begin{align}
v_x^{th} = \frac{J_n^{\rm max} x_0 }{10},
\end{align}
where $J_n^{\rm max}$ is the maximum tunneling rate for the $n$-th vibrational state.
For injection velocities below $v_x^{th}$, the process is adiabatic and the transfer succeeds, but atoms with higher velocities end up spread over the three waveguides.
While the switch between these two behaviours is smooth, $v_x^{th}$ gives an estimate of the region of parameters for which SAP is efficiently performed.  
The velocity filter can then be implemented with the structure shown in \fref{fig:CSR}(c-right), 
designed to perform a \textit{double} SAP process: from the ring to the external waveguide and back to the ring.
In this situation, slow atoms are able to adiabatically follow the spatial dark state and return to the ring trap with high fidelity, while faster atoms that do not fulfill the adiabaticity condition spread among the three waveguides.

\subsection{Electrons}
\label{sec:electrons}

The challenge of scalable quantum computing using spins \cite{LD98,Kan98} or charges \cite{HDW+04} in semiconductors introduced the need for long-range quantum transport.  This need was either to satisfy the DiVincenzo criteria around flying qubits \cite{DiV00}, or to address gate density issues with donor in silicon approaches to quantum computing \cite{COI+03}.  Electronic SAP provides one potential avenue to long-range transport in a quantum circuit and examples are discussed below.  However the possibility of \textit{engineering} the placement of sites, and hence tailoring the Hilbert space accessible to the electrons, provides other opportunities.  Although it is not discussed in detail here, there is also related work on SAP in superconducting systems \cite{SB04,SBF06}

\subsubsection{Transport}
\label{sec:electron_transport}
The idea of adiabatically transferring electrons in an electrical circuit appears to have originated around 2000/2001 \cite{TRM2000,RTA2000,BRB01}.  These early works considered a STIRAP-like process to transfer a charge through a double quantum dot system, and as such differ qualitatively from the SAP schemes that we are concentrating on systems that do not utilize electromagnetic driving.  Nevertheless, this scheme forms an important bridge between STIRAP and SAP, and it is therefore instructive to review it in some detail.  

\begin{figure}
\centerline{ \resizebox{0.95\columnwidth}{!}{\includegraphics{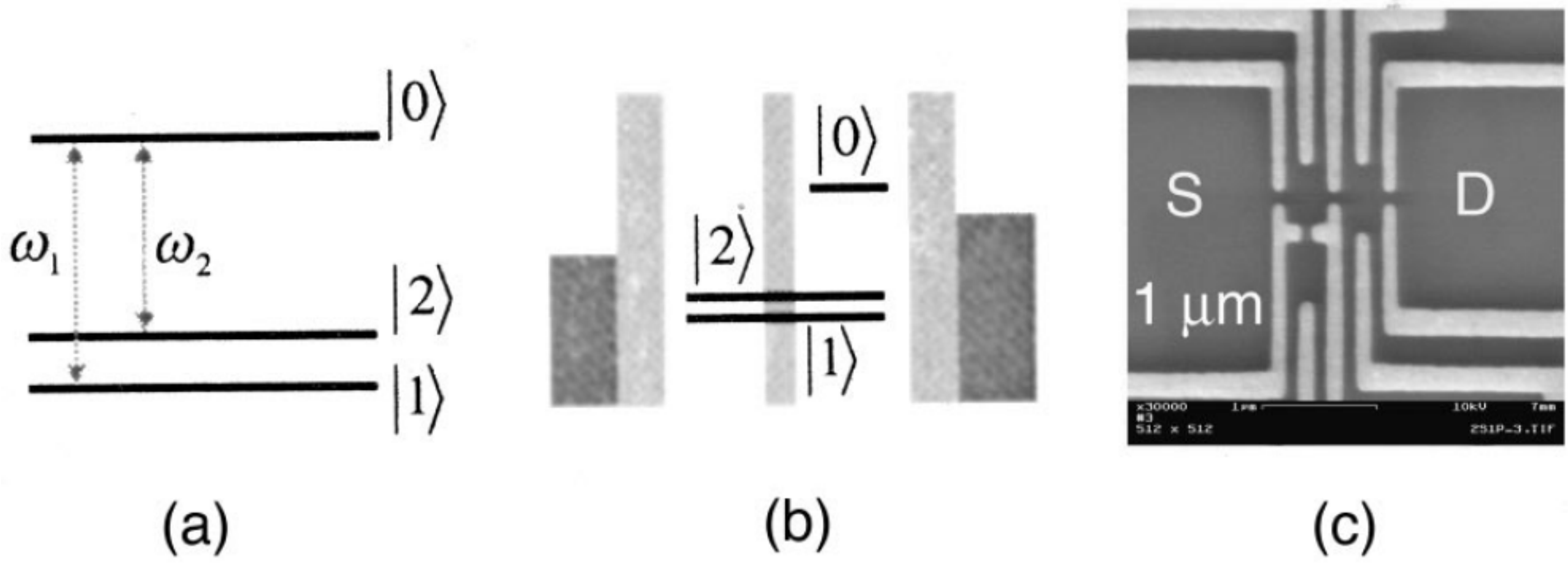}}}
\caption{(a), (b) Schematic of three-level $\Lambda$ system with radio-frequency drive fields. The two ground states are the symmetric and anti-symmetric superpositions of the electron in each of the two quantum dots. Both states are coupled to a shared excited state, $\ket{0}$ via resonant radio-frequency drive.  (c) Image of a potential quantum dot implementation. Reprinted from Ref.~\cite{BRB01}.}
\label{fig:Electronfig1}
\end{figure}

Spatial STIRAP considers an engineered double quantum dot, as shown in \fref{fig:Electronfig1}, in which the electron is moved between the two molecular states. These two molecular states are labelled $\ket{1}$ and $\ket{2}$, and each can be coupled to an excited state $\ket{0}$.  Only the excited state is coupled to the source-drain leads, so that $\ket{1}$ and $\ket{2}$ are long-lived.  Resonant radio-frequency fields are applied to induce STIRAP dynamics, which leads to adiabatic population control.  However, this scheme does not provide net population transfer for complete transport, as the molecular states have equal populations in each of the dots, although it is straightforward to consider schemes where full population transfer would be realized.  Because the excited state is coupled to source and drain leads, the proposal is analogous to STIRAP via ionizing states \cite{PH07}.

In contrast to STIRAP, SAP schemes involve spatially distinct start and end points, and the electronic proposals involve localized sites (e.g. dopants or quantum dots) with control of inter-site coupling via modulation of the tunnelling barrier, usually via electrostatic surface gates.  The other important distinction between electronic SAP and SAP of neutral particles is the role of decoherence, and this will be discussed below.  The first proposal that explored such transport was that of Ref.~\cite{greentree_coherent_2004}, which considered long-range transport of an electron through either a chain of dopant ions (in particular, phosphorus in silicon) or quantum dots.  

While the three-site transport mirrors other SAP proposals, the importance of Ref.~\cite{greentree_coherent_2004} was the extension to chains of more than three sites via the straddling scheme \cite{MT97}, which has applications in phosphorus-in-silicon quantum computing.  This scheme allows for long-range transport to be achieved without the necessity of applying gate control to the dopant sites within the chain, but only to the end-of-chain dopants.  In this way, the overall gate density can be reduced, leading to a proposal for a scalable architecture utilizing adiabatic electronic transport for spin-based quantum computation \cite{HGF+06}.  Adiabaticity arguments show that the time for transport across a chain of $n$ sites scales with $\sqrt{n}$ to leading order \cite{GCH+05}.  Extending electronic transport via the straddling scheme has also been discussed \cite{petrosyan_coherent_2006,jong_coherent_2009}.

Another major reason for exploring SAP for charge-based transport in phosphorus-in-silicon quantum architectures is related to the timescales for control and tunnelling.  For a proposed inter-dopant spacing of about 20~nm, the hopping time is expected to be of order of 100~ps \cite{HGF+06}.  For non-adiabatic control of population on such timescales, a bandwidth at least an order of magnitude faster than this is required, a task highly non-trivial for classical, cryogenic control electronics.  Conversely, using adiabatic passage requires control pulses with a bandwidth at least an order of magnitude or two smaller than the hopping time.  In this way, the full bandwidth implied by strong coupling between the dopants can be utilized without control at this bandwidth, provided the decoherence rates permit the longer adiabatic timescales (see below).

Hydrogenic arguments have been used for determining the relevant energy scales and disorder effects \cite{greentree_coherent_2004,HGF+06,VGA+10}.  More sophisticated treatments have explored the microscopic properties of phosphorus in silicon for adiabatic passage more rigorously using the NEMO3D tight binding code \cite{rahman_atomistic_2009}, including disorder \cite{rahman_coherent_2010}.  The main result from the NEMO3D simulations is the existence of extra molecular states for realistic phosphorus atoms, which are in general not detrimental to SAP.

Quantum dot systems provide considerable opportunities for transport protocols.  In addition to the idealized proposal of \cite{greentree_coherent_2004}, more realistic transport in a triple square well system was analyzed in Refs.~\cite{cole_spatial_2008,huneke_steady-state_2013}.  SAP in exchange-only coupled quantum dots has also been proposed \cite{FMM+15}.  
Whilst electronic SAP is still yet to be demonstrated in any system, recent developments in coupled quantum dot systems suggest that this milestone will soon be achieved.  The first designed triple dots in the one electron domain were shown by Schr\"{o}er \textit{et al.} \cite{SGG+07}, with further developments in Refs.~\cite{BBR+2013,PTS+2015}.

As mentioned above, the key challenge to electronic SAP is to maintain the coherence of the system over a timescale long compared with the transport time.  A naive estimate of the error rate yields the standard result that the transport error is simply the product of the total time with the decoherence rate, provided that the adiabatic limit is satisfied. A slightly more sophisticated calculation gives the error just less than this product, as the population is not in an equal superposition throughout the protocol \cite{HGF+06}.  However, such calculations ignore some more realistic aspects of decoherence in the solid state, which highlight the fact that the role of practical decoherence is interesting and non-trivial.

To go beyond phenomenological models for decoherence requires a microscopic treatment.  The first such treatment was given by Kamleitner \textit{et al.} who considered both spatially-registered and non-Markovian noises \cite{kamleitner_adiabatic_2008}.  Subsequent work by Rech and Kehrein \cite{rech_effect_2011} considered measurement backaction and Vogt \textit{et al.} considered the related problem of STIRAP in the presence of two-level fluctuators \cite{vogt_influence_2012}.  In all of these cases, the analyses show that although decoherence is important when applied to the ends of the chain, SAP is relatively immune to the effect of decoherence, whatever the cause (measuring devices or a fluctuating environment), that is applied solely to the central `bus' states.  

The relative robustness from decoherence arises from the suppression of population in the bus states, and is strongest in the case of the straddling scheme.  For a multi-state chain with $2n+1$ sites, labelled in increasing order from left ($\ket{1}$) to right ($\ket{2n+1}$), the general solution for the dark state is \cite{petrosyan_coherent_2006,jong_coherent_2009}
\begin{align}
\ket{D_0} =& \frac{1}{\sqrt{N}}\left[\prod_{i = 1}^nJ_{2i}\ket{1} + \cdots \right. \nonumber \\
& +(-1)^j\prod_{i=j}^nJ_{2i}\prod_{i=1}^{j}J_{2i-1}\ket{2j+1}+\cdots \nonumber \\
&\left.+(-1)^n\prod_{i=1}^nJ_{2i-1}\ket{2n+1}\right], \label{eq:D0arb}
\end{align}
where we have introduced $J_i$ as the (possibly time-varying) tunnel matrix element between site $\ket{i}$ and $\ket{i+1}$, and the normalization 
\begin{align}
N = \prod_{i = 1}^nJ_{2i}^2 + \cdots + \prod_{i=j}^nJ_{2i}^2\prod_{i = 1}^jJ_{2i-1}^2+ \cdots + \prod_{i=1}^n J_{2i-1}^2.
\end{align}
Note that this result guarantees that all of the even numbered sites will have precisely zero population in the adiabatic limit.  To understand the robustness to decoherence, one can consider the case of the straddling scheme, where $J_1, J_{2n} \ll J_{j}$ with $1 < j < 2n$.  In this limit, and assuming $J_{j} = J_S$, \eref{eq:D0arb} reduces to the simple result \cite{MT97,greentree_coherent_2004}
\begin{align}
\ket{D_0} =& \frac{J_1\ket{1} +(-1)^nJ_{2n}\ket{2n+1}}{\sqrt{J_1^2 + J_{2n}^2}} + \nonumber \\
&\frac{J_1J_{2n}}{J_S\sqrt{J_1^2 + J_{2n}^2}}\left[\sum_{i = 1}^{n-1}(-1)^i\ket{2i+1}\right], \label{eq:35}
\end{align}
for $n>1$.  As can be seen, the form of this straddling dark state is the same as that for conventional three-state SAP, except for the correction term, which is suppressed by the $1/J_S$ factor.  Now if we consider just the case of isolated `decohering centres' (for example localized measurement devices or two-level fluctuators coupled to the moving charge), which are considered one of the main sources of decoherence in the solid state, then the only parts of $\ket{D_0}$ that can be effectively measured are the ends of the chain, because these are the only parts of the system with appreciable population.  This has significant importance for the design of high-fidelity quantum wires in the solid state, since the fragility of a SAP chain does not increase linearly with the length of the chain.  Note, however, that the increased chain length implies a $\sqrt{n}$ increase in transport time, and hence the decoherence environment of the ends of the chain needs to be proportionally longer \cite{GCH+05}.

\subsubsection{Gates, branches and electronic interferometers}
\label{subsection:geometric_gates}

Engineered quantum systems provide the opportunity to design interesting topologies, and in the following we review a few of the suggestions for multi-qubit gate operations, branched and interferometric configurations.
The branched topology approach~\cite{greentree_quantum-information_2006,WC15}, which is also referred to as mutliple-recipient adiabatic passage, MRAP, is a simple extension of conventional SAP, with the consideration of multiple end destinations for the excitation.  These end points connect to a shared `bus', and due to symmetry, an excitation can be distributed to an equally weighted superposition state of the end points.  In its simplest form, where the `bus' comprises a single site, and there are two final states, this scheme is very similar to the Unanyan, Shore and Bergmann (USB) approach to geometric gates \cite{unanyan_laser-driven_1999} (see also \cite{KR02,FSF03}), which we discuss below.

\begin{figure}
\centerline{ \resizebox{0.5\columnwidth}{!}{\includegraphics{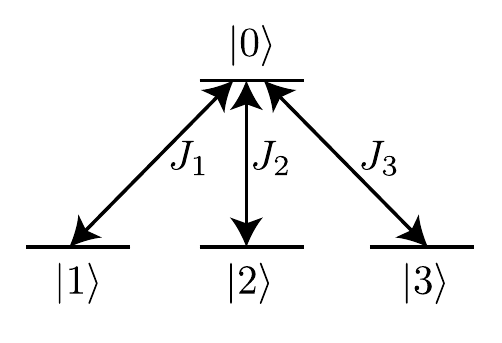} } }
\caption{Sketch of a four-state system with one shared excited state, $\ket{0}$, and three ground states $\ket{1}$, $\ket{2}$ and $\ket{3}$.  The time-varying couplings between state $\ket{i}$ and $\ket{0}$ are $J_i$. 
}
\label{fig:FormalismUSB}
\end{figure}

The USB approach was introduced for realizing robust unitary gates for laser-driven tripod atoms, shown in \fref{fig:FormalismUSB}.
The motivation for this scheme is that in laser driven systems, although absolute intensities, and hence Rabi frequencies, are not always easy to control, relative intensities can be maintained with high accuracy, as can predefined phase shifts.
changeThis overall motivation does not always translate across to SAP systems, for example in evanescently coupled waveguides where errors in position are expected to be uncorrelated.  Nevertheless, this scheme is still explored for adiabatic gates.

The four-state USB Hamiltonian can be written in matrix form with basis ordering $\ket{0}$, $\ket{1}$, $\ket{2}$, $\ket{3}$, where $\ket{0}$ is the shared excited state, as
\begin{align}
H = - \frac{\hbar}{2} \left(\begin{array}{cccc}
0 & J_1 & J_2 & J_3 \\
J_1 & 0 & 0 & 0 \\
J_2 & 0 & 0 & 0 \\
J_3 & 0 & 0 & 0 
\end{array}\right),
\end{align}
and where it is assumed that the Rabi frequencies $J_i$ are real.
The USB scheme assumes that the system is initially in state $\ket{\psi} = \alpha \ket{1}  + \beta\ket{2}$, and it seeks an adiabatic pulsing scheme for the $J_i$ that will perform a geometric gate operation.
When $J_1$ and $J_2$ are non-zero, $\ket{\psi}$ can be expressed in the canonical dark/bright basis,
\begin{align}
\ket{D} = \frac{J_2 \ket{1} - J_1 \ket{2}}{\sqrt{J_1^2 + J_2^2}}, \quad \ket{B} = \frac{J_1\ket{1} + J_2 \ket{2}}{\sqrt{J_1^2 + J_2^2}},
\end{align}
where $\ket{D}$ is the dark and $\ket{B}$ is the bright state.
Note that these states only depend on the ratio $\gamma = J_2/J_1$,
which is kept constant during the USB scheme.
The standard approach for understanding problems involving bright and dark states is to rewrite the Hamiltonian in this basis~\cite{RLA99}. 
The USB scheme now works by employing a pulse sequence with $J_3$ applied first, and $J_1$ and $J_2$ together later, i.e. $J_3(0) \gg J_1(0), J_2(0)$, and $J_3(t_{\max}) \ll J_1(t_{\max}), J_2(t_{\max})$.
This will adiabatically move the population in $\ket{B}$ to $\ket{3}$, while the population in $\ket{D}$ will not evolve.
To achieve a non-trivial gate operation on the qubit subspace $\ket{1}$ and $\ket{2}$, the evolution is then reversed with the addition of an extra $\pi$ phase shift.
This is achieved in the optical implementation of the USB scheme by a non adiabatic phase shift in the Rabi frequency of the field effecting $J_3$.
In SAP, implementing this phase shift is more difficult, and seemingly restricted to purely real variation.
Hope \textit{et al.}~\cite{HNM+15} proposed varying coupling to a `slab' mode (discussed further in \sref{LightSound:Beyond3LA}).
Another option would be to follow a return pathway via an extra two intermediate states, to pick up an additional minus sign, as can be seen from \eref{eq:D0arb} below, although this does increase the complexity of the scheme.

The net result of the forward and backwards SAP with the symmetry breaking $\pi$ phase shift is a rotation gate acting on the qubit space defined by $\ket{1}$, $\ket{2}$, that is \cite{HNM+15} 
\begin{align}
\mathcal{G} = \frac{1}{1 + \gamma^2}\left(\begin{array}{cc} \gamma^2 - 1 & -2\gamma \\ -2 \gamma & 1 - \gamma^2 \end{array}\right).
\end{align}
Because the $J_i$ are real, this gate can only perform rotations in the X-Z plane of the Bloch sphere, and hence arbitrary gate operations cannot be achieved using this method.
The USB scheme can be extended to consider more general Morris--Shore type operations \cite{MS83,RVS06}.

\begin{figure}
\centerline{ \resizebox{0.95\columnwidth}{!}{\includegraphics{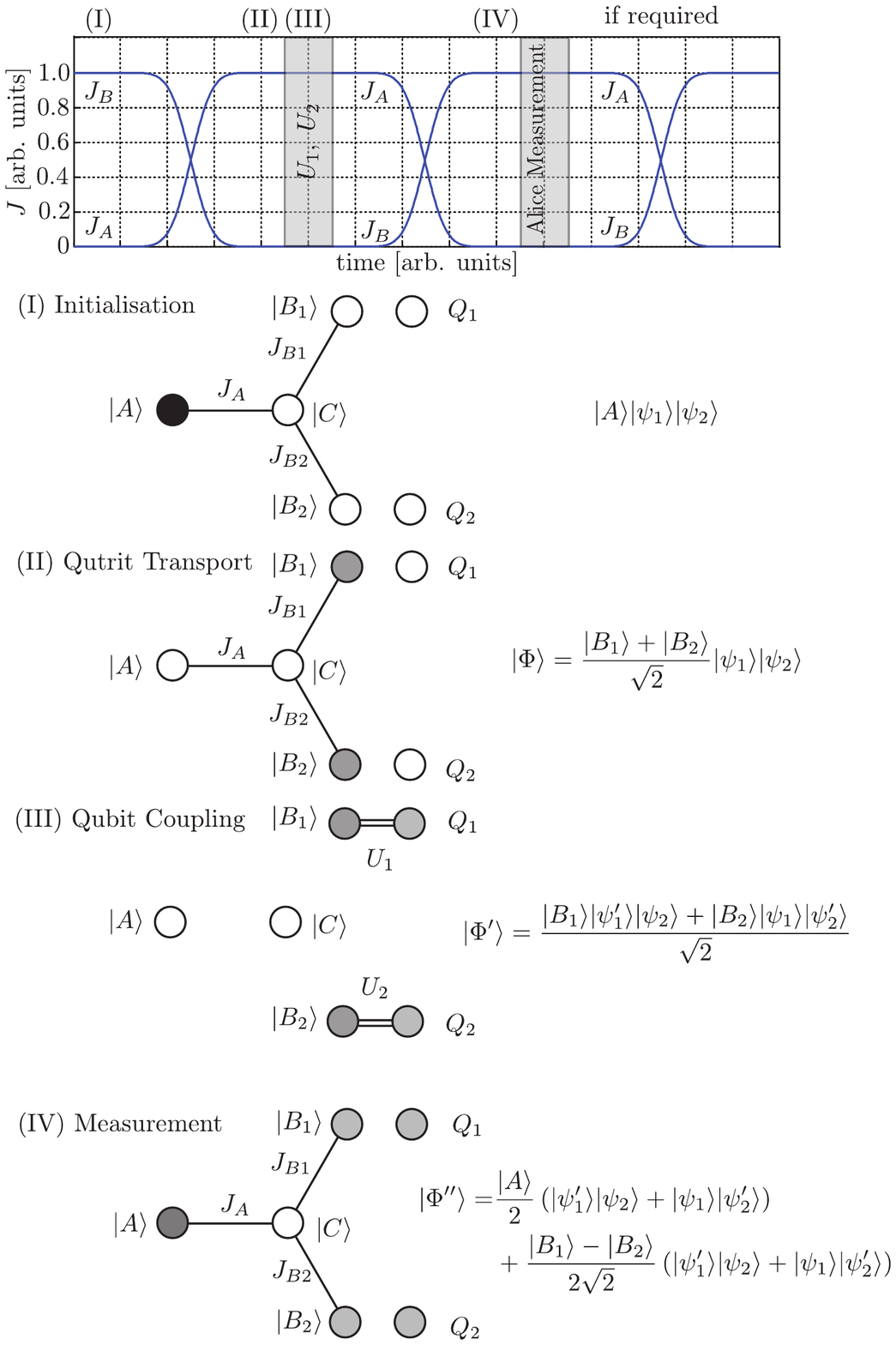}}}
\caption{Scheme for operator measurements between two qubits, via a transported particle. The top panel shows the pulse configuration for the tunnel matrix elements, labelled by the steps in the protocol shown below.  (I) The system is initialised with the transported particle in $\ket{A}$, and the two qubits in some separable state $\ket{\psi_1}\ket{\psi_2}$. (II) SAP is used to transport the particle to an equally weighted superposition $(1/\sqrt{2})(\ket{B_1} + \ket{B_2})$.  (III) Two-qubit controlled unitaries are performed between the particle and the qubits.  (IV) The controlled unitary operations break the time-reversal symmetry of the SAP.  Nevertheless, by performing a binary measurement of the particle present/not present at $\ket{A}$, it is possible to collapse the qubits to a non-trivial entangled state. Reprinted from Ref.~\cite{DGH07}
}
\label{fig:ElectronInfFreeBus}
\end{figure}

Intriguingly, MRAP allows for a symmetry-breaking approach~\cite{DGH07}, qualitatively different from what has been explored in the USB scheme.
Starting from the simple example where a superposition state can be created by effectively performing SAP with two endpoints, rather than one, one immediately realizes than the reverse process trivially returns the system to its initial state.  The symmetry between the forward and backward paths, however, is broken in the USB approach by introducing an additional $\pi$-phase shift, which allows for the nontrivial gate rotation.  However a mechanism for breaking symmetry between the pathways can also be provided by two-qubit interactions, see \fref{fig:ElectronInfFreeBus}.  

The system comprises a particle, for example an electron, which can exist in one of four possible states: initial state $\ket{A}$ (for Alice), final states $\ket{B_1}$ and $\ket{B_2}$ (for Bob 1 and Bob 2), and central state $\ket{C}$.
The Bobs also have qubits, $Q_1$ and $Q_2$, and the idea is to use the transported electron as a means of mediating a two-qubit operator measurement between the qubits.

At the start of the protocol, the system is initialised with the electron in state $\ket{A}$, and the qubits $Q_1$ and $Q_2$ are in the states $\ket{\psi_1}$ and $\ket{\psi_2}$ respectively. Adiabatic passage through $\ket{C}$ is then used create a superposition state at sites $\ket{B_1}$ and $\ket{B_2}$, via
\begin{align}
\ket{D} = \frac{2J_B\ket{A} - J_A\left(\ket{B_1}+\ket{B_2}\right)}{\sqrt{4 J_B^2 + 2 J_A^2}},
\end{align}
where we have introduced $J_A$ as the tunnel matrix element between $\ket{A}$ and $\ket{C}$, and $J_B = J_{B1} = J_{B_2}$ is the tunnel matrix element between $\ket{C}$ and $\ket{B_1}$, and $\ket{C}$ and $\ket{B_2}$.  This brings the system into state
\begin{align}
\ket{\Phi} = \frac{1}{\sqrt{2}}\left(\ket{B_1} + \ket{B_2}\right)\ket{\psi_1}\ket{\psi_2},
\end{align}
which is fully separable.

Because the electron is in a superposition of states, the application of two-qubit gates between the electron and the qubits will in general entangle the qubits.  If we assume that the two-qubit gate takes the form of a controlled operation (e.g. CNOT) where the electron is the control and the qubit the target ($\ket{\psi}\to\ket{\psi'}$), the system transfers to
\begin{align}
\ket{\Phi'} = \frac{1}{\sqrt{2}}\left(\ket{B_1}\ket{\psi'_1}\ket{\psi_2} + \ket{B_2}\ket{\psi_1}\ket{\psi'_2}\right).
\end{align}
While this state appears similar to $\ket{\Phi}$, it is no longer separable, and therefore the reversal of the SAP process does not lead to the electron trivially returning to $\ket{A}$. Instead, it leaves the system in the state
\begin{align}
\ket{\Phi''} = &\frac{1}{2}\ket{A}\left(\ket{\psi'_1}\ket{\psi_2} + \ket{\psi_1}\ket{\psi'_2}\right) \nonumber \\
	&+ \frac{1}{2\sqrt{2}}\left(\ket{B_1}-\ket{B_2}\right)\left(\ket{\psi'_1}\ket{\psi_2} - \ket{\psi_1}\ket{\psi'_2}\right).
\end{align}
The protocol then proceeds by performing a measurement to determine if the electron is at $\ket{A}$.  If the electron is found at $\ket{A}$, then the qubits are projected into the entangled state $\left(\ket{\psi'_1}\ket{\psi_2} + \ket{\psi_1}\ket{\psi'_2}\right)/\sqrt{2}$.  If the electron is not at $\ket{A}$, then a local rotation of the phase of the electron at (for example) $\ket{B_2}$ is performed, and then the electron can be returned to $\ket{A}$ via a reversal of the initial SAP process.  In this case, the qubits are projected to the orthogonal (but known) entangled state $
\left(\ket{\psi'_1}\ket{\psi_2} - \ket{\psi_1}\ket{\psi'_2}\right)/\sqrt{2}$.  This approach can be used as a primitive for realizing more general operator-measurements, and hence may be useful for quantum error correction protocols.

A more direct approach to realising two-qubit gates has been proposed by Kestner and das Sarma \cite{kestner_proposed_2011}.  In this protocol they consider a triangular triple dot configuration with two spins acting as qubits.  Their aim is to effect a CNOT interaction between the qubits.  This is achieved by a combination of SAP pulses, with spin-dependent tunnelling and local spin flips.  The key advantage of the Kestner and das Sarma proposal is that due to the use of SAP, they predict significant robustness against fluctuations in the control parameters, and low frequency environmental noise. 

A concept for interferometer-like behavior in a quantum dot network using a a Mach--Zehnder style configuration was studied in Ref.~\cite{jong_interferometry_2010}.  The authors considered  a central `ring' of four connected quantum dots, with initial and final dots connected to the outside of the ring.  The system shows an interesting interplay between adiabatic and non-adiabatic features as the degeneracy (introduced by an additional pair of control gates) between the two arms of the interferometer is increased. For anti-symmetric detunings of the arms, this effects a kind of adiabatic electrostatic Aharanov--Bohm loop \cite{Boy73}, superimposed on robust adiabatic transport for all other detuning situations.

\subsection{Bose--Einstein condensates}
\label{sec:BEC}

Spatial adiabatic passage is not limited to the transport of single quantum particles and it has been considered, for instance, for Bose--Einstein condensates (BECs).
A BEC is obtained when a dilute gas of identical bosons is cooled down to quantum degeneracy, and it exhibits a non-linear behavior due to interparticle interactions.
In the limit of zero temperature and within the mean-field approximation, the BEC dynamics in a one-dimensional geometry can be described by a wavefunction $\psi(x,t)$ that obeys the 1D Gross--Pitaevskii equation (GPE) \cite{pitaevskii_Bose_2003}
\begin{equation}
i \hbar\frac{\partial }{\partial t}\psi (x,t)= 
\left[
- \frac{\hbar^2}{2m}
\frac{\partial^2}{\partial x^2}  + V\left( x,t \right) 
+g_{\rm 1D} \vert \psi(x,t) \vert^2
\right] \psi (x,t) ,
\end{equation}
where $V(x,t)$ is the trapping potential and
$g_{1\rm D}=2N\hbar \sqrt{\omega_y \omega_z} a_s$ is the 1D non-linear interaction constant.
The total number of atoms in the BEC is given by $N$, $\omega_y$ ($\omega_z$) is the trapping frequency for the $y$ ($z$) harmonic confinement, and $a_s$ is the $s$-wave scattering length.
The normalization of the wavefunction is chosen to be $\int_{-\infty}^{+\infty} \vert \psi(x,t) \vert^2 dx=1$.

\begin{figure}
\centerline{ \resizebox{0.9\columnwidth}{!}{\includegraphics{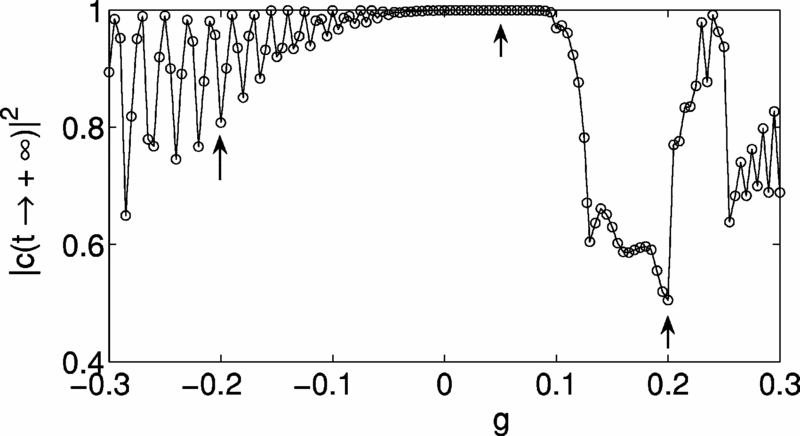}}}
\caption{Numerically calculated transfer efficiency of a BEC in a triple well potential for $\Delta=0.1$ via SAP. From \cite{graefe_mean-field_2006}.}
\label{fig:BEC_graefe}
\end{figure} 

The three-mode Hamiltonian for the BEC in a triple well potential can be derived in a similar manner as in \sref{formalism_SAP_three_well} by writing down the BEC wavefunction as a superposition of the localized eigenstates for the isolated wells with the probability amplitudes  $a_i=\sqrt{N_i/N}e^{i\phi_i}$, where $i=L,M,R$.
Here, $N_i$ and $\phi_i$ are the number of atoms and the phase of the BEC in each well.
Making use of the definitions~(\ref{onsiteenergy}) and (\ref{trate}) one then obtains the BEC three-mode Hamiltonian which is analogous to the three mode Hamiltonian~(\ref{threemodeH2}),  with the diagonal terms $\hbar \omega_i $ replaced by $\hbar \omega_i+ g \vert a_i \vert^2$, where $g$ is the atomic self-interaction energy~\cite{ottaviani_adiabatic_2010}.

Note that, as a consequence of the non-linear contributions in the diagonal of BEC three-mode Hamiltonian, the localized states in each trap are in general not resonant during the SAP sequence. Moreover, the nonlinearity leads to additional non-linear energy eigenstates, level crossing scenarios, and bifurcations that can break up the adiabatic following of the spatial dark state.
Graefe \textit{et al.} \cite{graefe_mean-field_2006} showed that by imposing $\omega_R=\omega_L$, a complete SAP transfer of the BEC between the outermost wells of a triple-well potential can be achieved for $g \Delta \geq 0$ and $\vert g  \vert < \vert \Delta \vert$, 
where $\Delta\equiv\omega_M-\omega_L$ (see \fref{fig:BEC_graefe}).
Additionally, as discussed in \cite{morgan_coherent_2011}, time-dependent trapping frequencies can be considered to compensate for the time-dependent energy shifts that the non-linear interaction produces in each well.
In the context of the three-mode Hamiltonian, the SAP sequence has also been discussed in a cyclic triple well potential \cite{nesterenko_stirap_2009}.
The numerical integration of the GPE was used to study the fidelities for the SAP process in different parameter regimes \cite{rab_spatial_2008} and to show that nonlinearities can improve the sensitivity of matter-wave interferometers based on SAP \cite{rab_interferometry_2012}.

For adiabatic passage of a BEC within the two mode approximation \cite{nesterenko_adiabatic_2009, ottaviani_adiabatic_2010,nesterenko_adiabatic_2010}, 
one obtains the same Hamiltonian as for the single particle case, see \eref{twomode}, but now replacing $\omega$ by $\omega + g  W$. Thus, there exists a dark variable $d (\theta)= W \cos \theta + U \sin \theta $ with a  mixing angle $\theta$ that now reads $\tan \theta = J / (\omega + g  W )$.
In Ref.~\cite{ottaviani_adiabatic_2010}, this model is used to investigate the robust splitting (varying $\theta$ from $0$ to $\pi/2$), transport (varying $\theta$ from $0$ to $\pi$), or inhibiting transport (varying $\theta$ from $0$ to $0$ through arbitrary intermediate values of $\theta$) of a BEC initially located in the left site of a double-well potential.
Temporal control of the mixing angle $\theta $ can now be achieved via a temporal variation of either the energy bias or the nonlinear interaction with respect to the tunneling rate.
The dynamics were investigated from a nonlinear systems point of view, by deriving the stationary solutions, performing a linear stability analysis, and discussing possible bifurcation scenarios.
In \cite{nesterenko_adiabatic_2009,nesterenko_adiabatic_2010} protocols generalizing the Landau--Zener and Rosen--Zener schemes were presented, stressing the role of the nonlinearity for transport in double-well potentials.

\subsection{SAP in two dimensions}
\label{sec:beyond}

In the following, we will review recent results related to SAP in two-dimensional systems.
Considering more spatial dimensions can lead to new scenarios where additional couplings can be obtained.

In fact, in the triple-well case, the simultaneous coupling of all traps during the SAP procedure can be achieved by designing triangular trapping geometries. 
In quantum optics, to couple the two ground states of a $\Lambda$ system requires advanced electric or magnetic couplings \cite{giannelli_three_2014} or the use of chiral molecules \cite{kral_cyclic_2001,kral_two-step_2003}.

\subsubsection{Triangular trap configuration}
\label{triangular}

\begin{figure}
\centerline{ \resizebox{0.6\columnwidth}{!}{\includegraphics{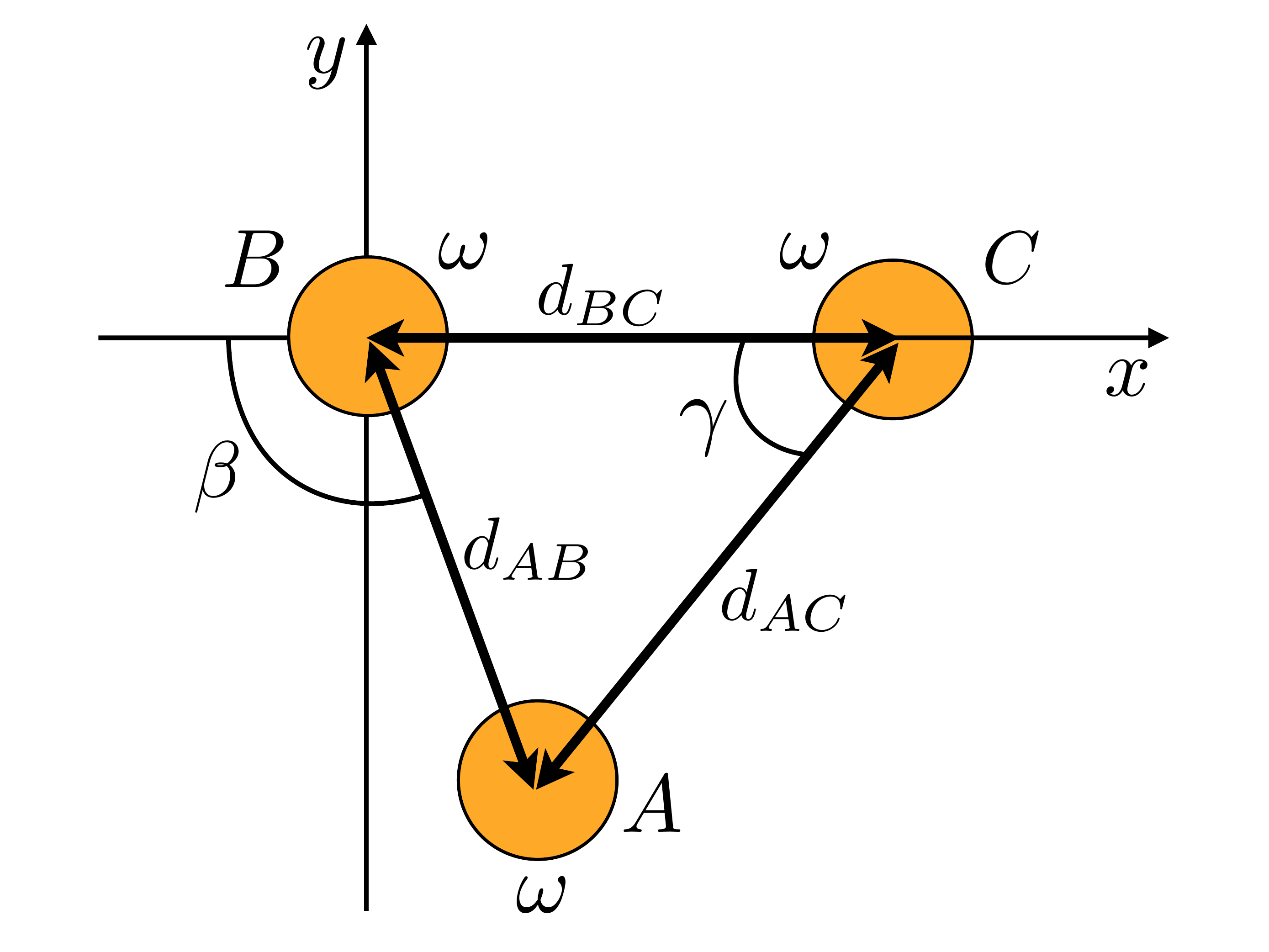}}}
\caption{Schematic representation of the system of three harmonic traps, $A$, $B$ and $C$ with equal trapping frequencies, $\omega$ in a triangular configuration.
The distance between the centers of traps $i$ and $j$ is $d_{ij}$ with $i,j=A,B,C$. From \cite{menchon-enrich_single-atom_2014}.}
\label{fig1}
\end{figure}

The simplest two-dimensional system that can be considered is formed by three harmonic potentials (labeled $A$, $B$ and $C$) with equal trapping frequencies forming a triangle~\cite{menchon-enrich_single-atom_2014}, as schematically shown in \fref{fig1}. 
The tunneling rates between traps $i$ and $j$ are denoted by $J_{ij}$, and their explicit dependence on the distance $d_{ij}$ is given by \eref{tunnelingrate} in \sref{formalism_SAP_three_well}.
Assuming that the dynamics of the system are restricted to the space spanned by the  localized ground states of the three traps, $\left\{ \psi_{A} (t),\psi_{B} (t),\psi_{C} (t) \right\}$
(see \sref{formalism_SAP_three_well}),
the Hamiltonian that governs the particle's evolution can be written as 
\begin{equation}
H = -\hbar\left( \begin{array}{ccc}
0 & \frac{J_{AB}}{2} & \frac{J_{AC}}{2}\\
\frac{J_{AB}}{2} & 0 & \frac{J_{BC}}{2} \\
\frac{J_{AC}}{2} & \frac{J_{BC}}{2} & 0 \\
\end{array} \right).
\label{ham_eq}
\end{equation}

Diagonalizing this Hamiltonian gives the energy eigenvalues
\begin{equation}
E_k=2\sqrt{-\frac{p}{3}}\cos\left[\frac{1}{3}\arccos\left(\frac{3q}{2p}\sqrt{\frac{-3}{p}}\right)+k\frac{2\pi}{3}\right],
\label{eigenval}
\end{equation}
where $k=1,2,3$ and 
$p=-\hbar^2(J_{AB}^2+J_{BC}^2+J_{AC}^2)/4$, and  $q=\hbar^3 J_{AB}J_{BC}J_{AC}/4$.
The corresponding eigenstates read
\begin{equation}
\Psi_{k}=\frac{1}{N}\left(a_k\psi_A+b_k\psi_B-c_k\psi_C\right),
\label{estats}
\end{equation}
with
\begin{equation}
a_k=J_{BC}-\frac{2E_{k}J_{AC}}{\hbar J_{AB}},
\end{equation}
\begin{equation}
b_k=J_{AC}-\frac{2E_{k}J_{BC}}{\hbar J_{AB}},
\end{equation}
\begin{equation}
c_k=J_{AB}-\frac{4E_{k}^2}{\hbar^2J_{AB}},
\end{equation}
and
\begin{equation}
N=\sqrt{a_k^2+b_k^2+c_k^2}.
\end{equation}
For $J_{AC}=0$, which means $q=E_2=b_2=0$, the system recovers the same expressions as for the 1D SAP case, see \sref{formalism_SAP_three_well}.
 
\begin{figure}
\centerline{\resizebox{0.99\columnwidth}{!}{\includegraphics{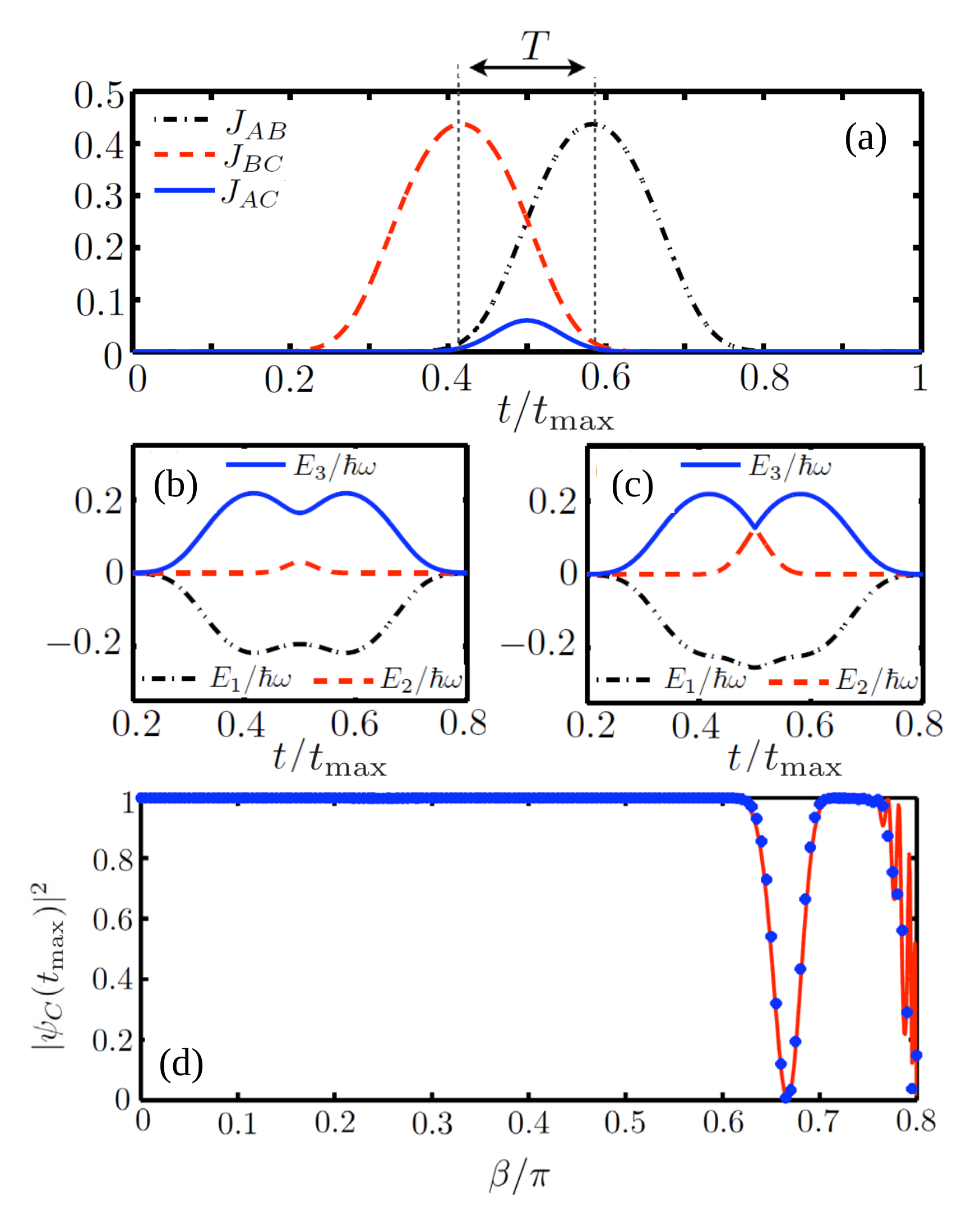}}}
\caption{(a) Temporal evolution of the couplings $J_{AB}$, $J_{BC}$ and $J_{AC}$ during the SAP process for $\beta=\pi/2$. Energy eigenvalues as a function of time for (b) $\beta=\pi/2$ and (c) $\beta=2\pi/3$. (d) Final population in trap $C$ as a function of $\beta$ for a total time of the process $t_{\max}=5000\omega^{-1}$. The red curve represents the numerical integration of the Hamiltonian~(\ref{ham_eq}) and the blue dots the results of the numerical integration of the 2D Schr\"odinger equation. In all the cases, $T=0.2 t_{\max}$. From \cite{menchon-enrich_single-atom_2014}.
}
\label{fig:triangleplots}
\end{figure}

Starting with a single atom located in the vibrational ground state of trap $A$, and keeping trap $B$ fixed, the SAP sequence consists in initially approaching and separating traps $C$ and $B$.
Later on, and with a certain temporal delay $T$, traps $A$ and $B$ are approached and separated, keeping the angle $\beta$ fixed.
This variation of the distances between traps leads to the familiar counterintuitive sequence of couplings and also introduces a coupling between traps $A$ and $C$, see \fref{fig:triangleplots}(a). 

From the analytical expression of the eigenstates~(\ref{estats}) one can see that at $t=0$ the particle is in state $\Psi_{2}(t=0) = \psi_A$.
Following this eigenstate, see \fref{fig:triangleplots}(b), at the end of the adiabatic process, the atom will be in state $\Psi_2(t=t_{\max})=-\psi_C$, and full transfer is achieved.
As one can see from \fref{fig:triangleplots}(d), this process is successful in the interval $0 \leq \beta \lesssim \beta_{th} = 2\pi/3$.
For $\beta=\beta_{th}$, the three tunneling rates become identical at a certain time during the dynamics, which corresponds to the appearance of a level crossing in the spectrum, see \fref{fig:triangleplots}(c).
In this case, it is no longer possible to adiabatically follow $\Psi_2$, and the system is instead transferred to $\Psi_3$.
At the end of the process, the particle state will therefore be in an equal superposition of the ground states of traps $A$ and $B$.
For $\beta > \beta_{th}$ the level crossing is again avoided but for large $\beta$ the transfer of the atom between traps $A$ and $C$ fails again, because the coupling strength $J_{AC}$ is significant during the first stage of the process and leads to unwanted Rabi-like oscillations.
This behaviour has been confirmed by numerically integrating both the Hamiltonian~(\ref{ham_eq}) and the 2D Schr\"odinger equation, obtaining a very good agreement as shown in \fref{fig:triangleplots}(d).

The coherent splitting of the atomic wavefunction between the traps $A$ and $B$ occurring for $\beta=2\pi/3$ in the triangular configuration of traps can be used to design an interferometer to measure spatial field inhomogeneities.
After the splitting, a relative phase $\varphi$ is picked up by the parts of the wavefunction in traps $A$ and $B$, leading to a state of the system of the form $(\psi_{A}-e^{i\varphi}\psi_{B})/\sqrt{2}$, which can be decomposed in a superposition of $\Psi_1$ and $\Psi_3$.
Reversing the temporal evolution of the SAP couplings, the contribution of $\Psi_3$ will be transferred to $\Psi_2$ at the level crossing and end up in trap $A$, while the contribution of $\Psi_1$ will evolve backwards and at the end of the process will be in a superposition of traps $B$ and $C$.
By then measuring the population of the three traps, one can infer the relative phase before the recombination.
Simulating this process, very good agreement between a numerically imprinted phase and the one obtained from the populations has been obtained, demonstrating the excellent performance of this system as a matter-wave interferometer~\cite{menchon-enrich_single-atom_2014}. 

\subsubsection{Generation of angular momentum}

In addition to the already discussed possibilities for quantum state preparation through SAP processes in 1D (see \sref{quantum_state_preparation}), SAP in 2D can also be used to create and control angular momentum~\cite{menchon-enrich_tunneling-induced_2014}.
For this, one can consider a geometrical arrangement of three 2D harmonic traps, analogous to the one depicted in \fref{fig1},
but with the trapping frequency of trap $C$ chosen as half that of traps $A$ and $B$.
In this configuration, the ground energy levels in traps $A$ and $B$ are resonant with the first excited level in trap $C$, allowing for a large tunnel coupling between them.
Since the first excited energy level of trap $C$ is doubly degenerate, it supports an angular momentum carrying state through a superposition of the two energy eigenstates $\psi^{C}_{1,0}(x,y)$ and $\psi^{C}_{0,1}(x,y)$ in the chosen $x$--$y$ reference frame.
In particular, maximum angular momentum, $\langle L_z\rangle=\pm\hbar$, occurs when these two degenerate states are equally populated and have a phase difference of $\pi/2$.

In the considered triangular configuration, tunneling into the first excited states of trap $C$ is described by the rates $J^{AC}_{1,0}$, $J^{AC}_{0,1}$, and $J^{BC}$~\cite{menchon-enrich_tunneling-induced_2014}.
Therefore, the dynamics of the system in the basis of the asymptotic states of the traps, $\{\psi^{A}_{0,0}, \psi^{B}_{0,0}, \psi^{C}_{1,0},\psi^{C}_{0,1}\}$, can be described by the $4\times 4$ Hamiltonian
\begin{equation}
H = - \frac{\hbar}{2}\left( \begin{array}{cccc}
0 & J^{AB} & J^{AC}_{1,0} & J^{AC}_{0,1}\\
J^{AB} & 0 & J^{BC} & 0\\
J^{AC}_{1,0} & J^{BC} & 0 & 0\\
J^{AC}_{0,1} & 0 & 0 & 0
\end{array} \right).
\label{japham}
\end{equation}

\begin{figure}
\centerline{ \resizebox{0.99\columnwidth}{!}{\includegraphics{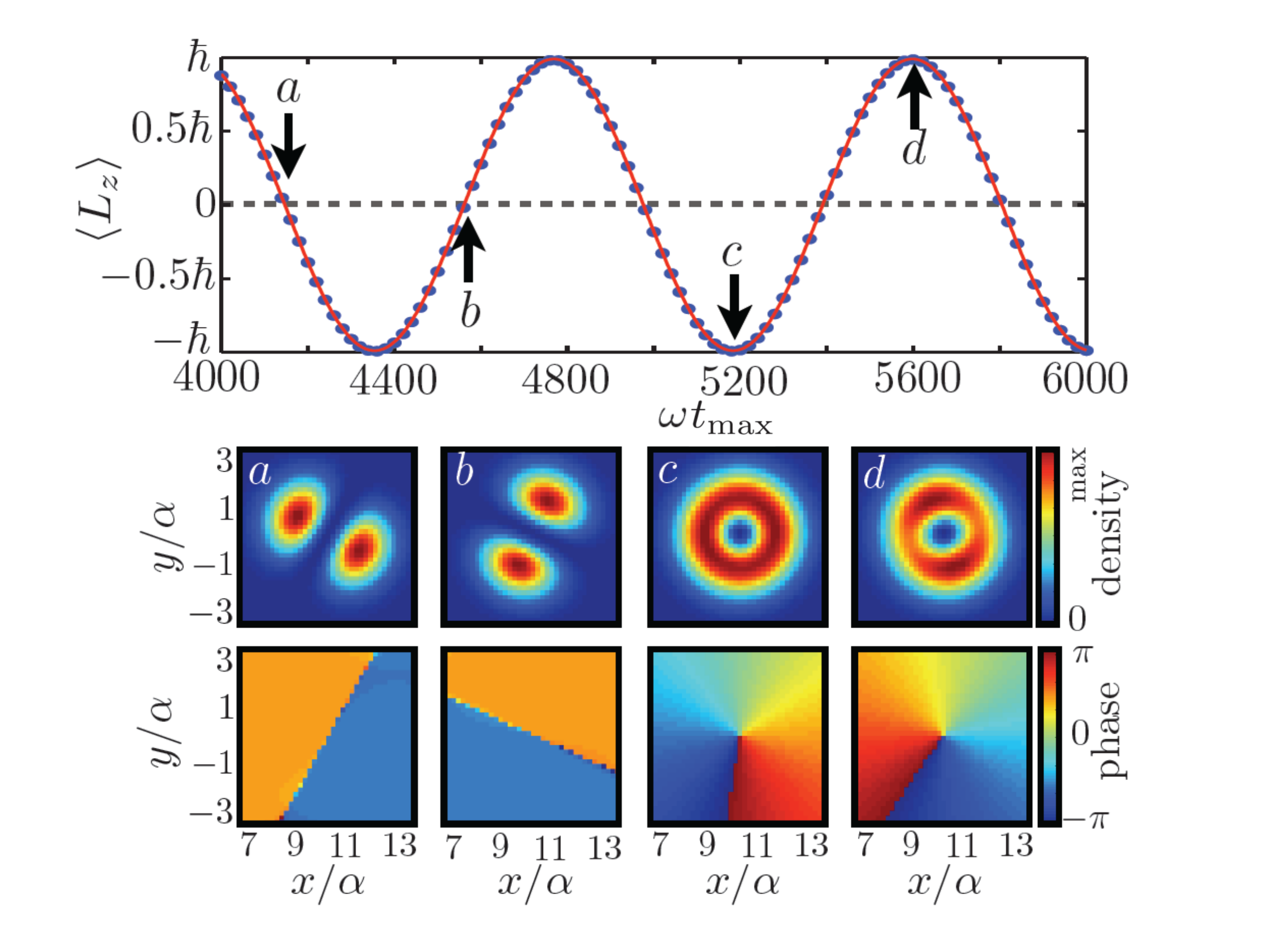}}}
\caption{Generated angular momentum as a function of the total time of the process $t_{\max}$ (upper plot),  obtained by numerically integrating the 2D Schr\"odinger equation.
Densities and phases for the final states in trap $C$ for four different final total times $a$, $b$, $c$ and $d$ (lower plots).
The parameter values are $\beta=0.55\pi$ and $T=0.2 t_{\max}$, and $d_{BC}$ and $d_{AB}$ vary from $10\alpha$ to $3.5\alpha$, and from $9\alpha$ to $2.5\alpha$, respectively.
From \cite{menchon-enrich_tunneling-induced_2014}.}
\label{japfig5}
\end{figure}

The generation of angular momentum occurs during the SAP transfer of the particle from trap $A$ to trap $C$, with the same trap movement as described in \sref{triangular}.
The initial state of the system (particle in trap $A$) can be written as a superposition of two of the eigenstates of the system.
Thus, if the process is adiabatic and level crossings are absent, this superposition of eigenstates is followed all through the process, leading to a final state that only involves the asymptotic states of trap $C$, $\psi^{C}_{1,0}(x,y)$ and $\psi^{C}_{0,1}(x,y)$.
In fact, the two asymptotic states of trap $C$ are equally populated at the end of the process, with a phase difference, $\varphi(t_{\max})$, proportional to the total time of the process $t_{\max}$.
The expectation value of the angular momentum can be found to be $\langle L_z(t_{\max})\rangle=\hbar\sin[\varphi(t_{\max})]$, which allows to control the angular momentum of the system by choosing an appropriate value for $t_{\max}$, see \fref{japfig5}.

\subsubsection{Two-dimensional lattices}

SAP can also be applied between distant sites of 2D rectangular and triangular lattices by dynamically controlling the tunneling rates.
For this, one maps the motion of a particle in the 2D lattice to the Fock space dynamics of a second-quantized Hamiltonian for appropriate bosonic fields~\cite{longhi_coherent_2014}.
To illustrate the procedure, one can consider a $3\times3$ rectangular lattice  in the tight binding and nearest-neighbor approximations.
In this case, the temporal evolution equations for the probability amplitudes for finding the particle in each of the sites can be derived from the time-dependent Schr\"odinger equation with a quadratic Hamiltonian.
This requires the introduction of six coupled bosonic oscillators consisting of products of annihilation and creation operators of independent bosonic modes and of a state vector which is a linear superposition of nine states resulting from the application of two different creation bosonic operators onto the vacuum.
The Heisenberg equations of motion for the six bosonic field operators are formally analogous to two sets of three-level equations for SAP, provided that the operators are replaced by c-numbers~\cite{longhi_coherent_2014}.
Thus, the adiabatic evolution of the operators leads to adiabatic passage in Fock space.
This analysis can also be extended to rectangular lattices (with an odd number of sites in each direction) and to triangular lattices \cite{longhi_coherent_2014}.

\subsection{Practical considerations}
\label{sec:practical}

Implementing SAP techniques requires identification of experimental systems that have localized asymptotic states between which a controllable time-dependent coupling exists. 
To achieve this, one typically considers trapping potentials in which the low energy part of the spectrum is experimentally accessible.
Changing either the distance between adjacent traps or the height of the separating barrier between them allows to create a time-dependent tunneling coupling.
The first possibility is usually considered in microtrap systems for atoms or ions~\cite{eckert_three-level_2004}, whereas the latter one is more suitable for electrons in quantum dots~\cite{greentree_coherent_2004}.

A second requirement is that the system must posses a state, such as the spatial dark state, that allows transfer. From the three-state model it is clear that this is fulfilled if the asymptotic states for each potential at any point during the process are in resonance. However the exact diagonalisation of the Schr\"odinger equation shows that high fidelity transport can still be achieved in the presence of a small detunings between the traps. If these detunings are too large, however, effective tunneling between the potentials is prevented and the transport will break down.

From an experimental point of view the second requirement means that the trapping potentials should not change significantly over the whole process.
Changing the distance between microtraps, however, leads to certain overlap between neighboring traps which can alter the shape of the individual potentials.
This can lead to level crossings in the spectrum that can make following of the dark state harder, or detunings that become too large for the process to be efficient.
Ensuring that this does not happen usually involves significant experimental resources or restrictions on the parameter space~\cite{eckert_three-level_2004,eckert_three_2006,rab_spatial_2008}.

Finally, the whole SAP process has to be carried out adiabatically, so that the system does not leave the lowest band or the dark state at any point during the evolution.
This requires precise dynamical control as the time scales for the process need to be chosen such that they fulfill an adiabaticity condition, like the one given in 
Eq.~\eqref{adiabaticitycondition}, and are shorter than the life- and coherence times of the system.

\subsubsection{Optical traps}

One of the first experimental systems suggested for observing SAP were optical potentials generated through a microlens array~\cite{eckert_three-level_2004}.
By illuminating the lenses with red-detuned laser light, ultracold atoms can be trapped in the focus above the lens, with typical trapping frequencies for $^{87}$Rb atoms on the order of $10^5-10^6$ s$^{-1}$ in the transverse directions and $10^4-10^5$ s$^{-1}$ along the laser beam direction \cite{birkl_atom_2001,dumke_micro-optical_2002}.
This means that the traps can be adiabatically approached in the millisecond range or even faster by using optimization techniques. 

\begin{figure}
\centerline{ \resizebox{0.99\columnwidth}{!}{\includegraphics{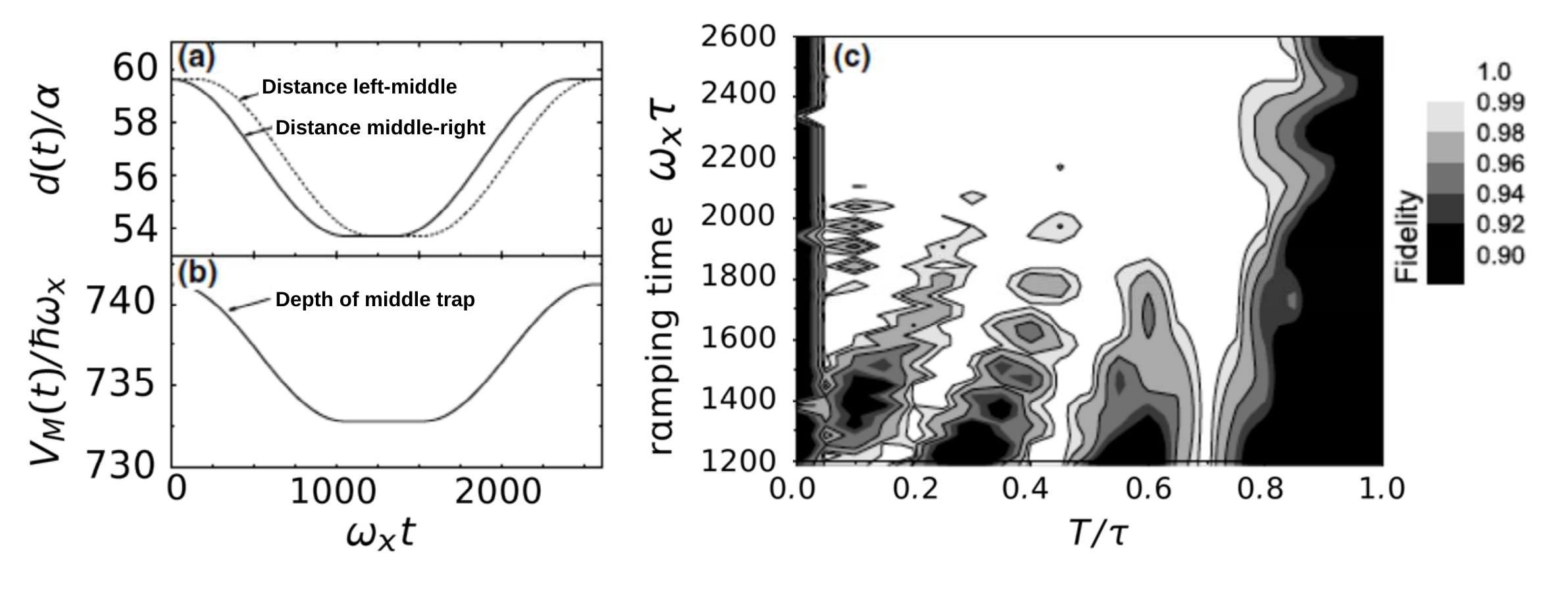}}}
\caption{SAP transport between three Gaussian-shaped traps. Variation of the (a) trap distances and (b) the depth of the center trap. (c) Fidelity, i.e., final
population of the right trap, as the approaching time $\tau$ and the delay time $T$ are varied.
From \cite{eckert_three_2006}, notation adjusted to match text.}
\label{Fig:DipoleTraps}
\end{figure}

The distances between the individual traps can be adjusted by using separate laser beams illuminating each lens.
However, this also leads to an increased overlap of the optical intensities and therefore to a significant distortion of the resulting potential.
The resonances between the individual traps are then no longer guaranteed and one needs to consider the use of additional compensation techniques to restore them.
For example, a simple and robust manner to do this in a system of three Gaussian-shaped dipole traps is to dynamically adjust the depth of the center trap~\cite{eckert_three_2006}, see \fref{Fig:DipoleTraps}.

\subsubsection{Atom chips}

As discussed in \sref{sec:matterwaveguides}, SAP processes can also be implemented for particles moving in a waveguides system by changing the distances between three different waveguides in the direction of propagation.
Particle waveguides can be created using optical or magnetical potentials and in  \cite{morgan_coherent_2013} a system of three waveguides on an atom chip was investigated.
Atom chips are versatile experimental tools that are by today used extensively in experiments with ultracold atoms.
A small current flowing through nanofabricated wires on the substrate produces a magnetic field gradient in such a way that cold atoms can be trapped very close to the surface.
Because the layout of the nanowires can be chosen during the chip production process and the currents are small, they allow for highly stable potential generation \cite{AtomChipReview}.

\begin{figure}
\centerline{ \resizebox{0.99\columnwidth}{!}{\includegraphics{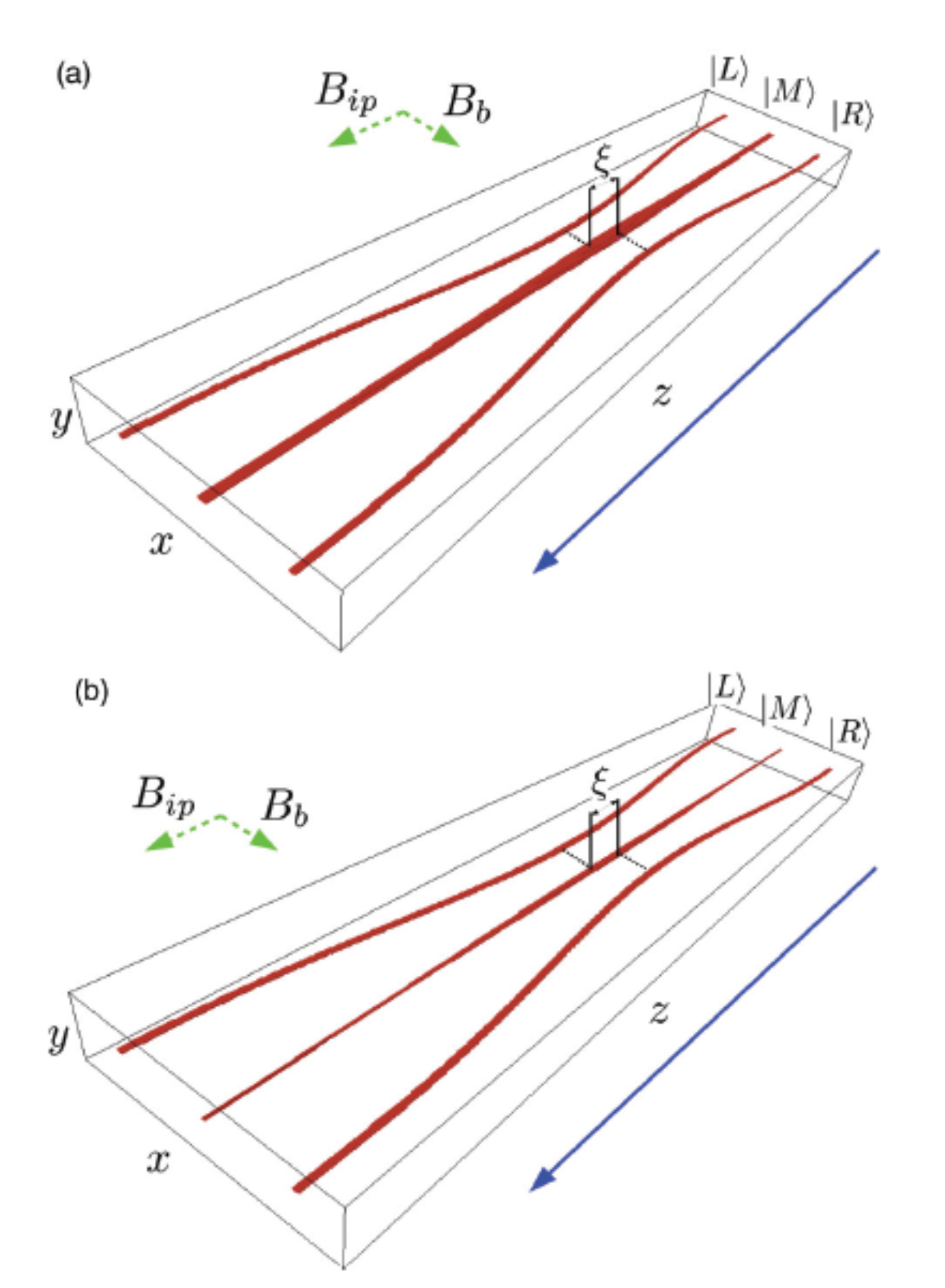}}}
\caption{
Isosurfaces of the waveguides created on an atom chip with the direction of propagation indicated by the blue solid arrow (for clarity the longitudinal harmonic potential is omitted in this plot).
In (a) all the wires carry the same current, and in (b) the current in the middle wire is reduced.
For full parameters, see the original reference. From Ref.~\cite{morgan_coherent_2013}.
}
\label{Fig:AtomChipPotential}
\end{figure}

An example of a waveguide potential for $^6$Li atoms and for experimentally realistic
parameters is shown in \fref{Fig:AtomChipPotential}.
If an atom is injected in the left waveguide, and moves in the positive $z$ direction, these waveguides provide the desired counterintuitive tunnel coupling needed to transfer it to the right waveguide. 
However, this system also suffers from the effect that the overlap between neighboring waveguides leads to a deformation of the individual potentials (see \fref{Fig:AtomChipPotential}(a)), and consequently to a breakdown of resonance.
Similarly to the optical traps discussed above, a reduction of the current in the middle wire allows to compensate for this (see \fref{Fig:AtomChipPotential}(b)).

If one considers a wavepacket traveling in the waveguides, a couple of additional points need to be considered: the wavepacket must be launched with an initial velocity, and its dispersion in the longitudinal direction needs to be compensated.
Both of these tasks can be dealt with by adding a harmonic potential along the $z$ direction centered at the middle of the chip. 
This will allow to launch the wavepacket and also lead to a refocusing at the classical turning point on the other side of the chip.
In \cite{morgan_coherent_2013} it was shown that the process is adiabatic if its total time is taken to be much larger than the inverse of the transverse trapping frequencies of the individual waveguides.
Choosing realistic parameters for the atom chip (waveguide length in the $z$ direction 1 mm, initial separation between waveguides 7 $\mu$m, transverse trapping frequencies $2\pi\times 5$ kHz, frequency of the longitudinal harmonic potential $2\pi\times 5$ Hz), this time was shown to be 0.1 s.
Even though the bend in the waveguide couples the longitudinal and transversal directions, the large difference in trapping frequencies ensures that this coupling is small.

Finally, the bend in the wires will also lead to a potential from the currents in the $z$-direction (note that this is absent in the optical waveguide realization, discussed below), which requires the atom to have enough kinetic energy to overcome it.
This has direct consequences for being able to fulfill the adiabaticity condition.
However, this local potential maximum can be reduced by increasing the length of the
atom chip ($z$-direction) and therefore reducing the curvature of the wires.

Full 3D simulations with experimentally realistic parameters were carried out in  \cite{morgan_coherent_2013} and confirmed that such systems allow to achieve SAP with very high fidelities for appropriate tunings of the currents in the individual waveguides.

\subsubsection{Radio-frequency traps}

A system that is more robust against distortion from the overlap of two neighbouring potentials than optical traps can be constructed by radio-frequency (rf) traps, which rely on coupling magnetic sublevels in the presence of an inhomogeneous magnetic field \cite{Zobay_RF_Traps_2001,Schumm:05,Lesanovsky:07,Zimmermann:06}.
This coupling leads to avoided crossings in the energy spectrum and therefore to minima and maxima in the potential landscape the atoms see.
As the couplings are spatial resonances, changes in the eigenspectrum only alter the potentials locally and do not affect the overlap with neighbouring traps.

To produce an rf potential with three minima along the $x$-direction, it is necessary to employ six different radio frequencies.
Moving the traps can be achieved by changing the individual rf frequencies that are associated with each trap.
An exact sequence that allows to realize the SAP transfer was given in \cite{morgan_coherent_2011}, where it was also shown that high transfer fidelities can be achieved without the need for any additional compensation potentials.

The use of rf traps also allows to design a realistic setup to extend SAP to non-linear systems, for example Bose--Einstein condensates (BECs) (see \sref{sec:BEC}).
Here the Hamiltonian in the three-level approximation has a non-linear term in the diagonal which is as a function of the particle numbers in each trap.
As these numbers are a function of time, the non-linear terms will change during the SAP process and therefore modify the resonances between the traps.
A straightforward way to compensate for this is to allow for the trapping frequencies to be functions of time as well \cite{morgan_coherent_2011}.
Starting with the BEC in the left trap, $N |a_L|^2$ will decrease during the process, while $N |a_R|^2$ will increase.
Adjusting the trapping frequencies $\omega_L$ and $\omega_R$ can restore the resonance between the uncoupled traps by ensuring that $\hbar\omega_i+g |a_i|^2$ is approximately constant at all times.
However, in order to be able to make the three-level approximation, one must ensure that $g  |a_i|^2<\hbar\omega_i$ for all values of $g |a_i|^2$ and $\omega_i$. 
This means in practice that the process is limited to cold atomic clouds with small nonlinearities.
A detailed description of this process is given in \cite{morgan_coherent_2011}. Note that a different setup for SAP transfer of BECs using optical traps was described in \cite{graefe_mean-field_2006}.

\subsubsection{Speeding up adiabatic techniques}

As discussed above, the robustness of adiabatic processes such as SAP make them very useful tools to prepare and manipulate quantum states.
However, the price for this are the long time scales these processes require, i.e., the fact that the inverse of the total time has to be much larger than all characteristic frequencies of the system. 
At first look, this makes adiabatic processes uninteresting for quantum information processing because they not only require highly efficient protocols, but also fast transport processes between gate operations.
Slow processes can be problematic because decoherence and noise can affect the system, leading to final states with reduced fidelity.
Therefore, it is desirable to develop techniques which are fast and lead to high fidelities.

One such group of techniques are based on optimal control theory (OCT) algorithms.
These work by determining the time dependence of a number of control parameters which minimizes a cost function, e.g. the infidelity of a process, subject to some constraints such as the initial state or the maximum coupling strengths.
OCT has been applied to speed up many adiabatic processes~\cite{salamon_maximum_2009,hohenester_optimal_2007,murphy_high-fidelity_2009,rahmani_optimal_2011}, and in particular SAP transport  of a BEC in a realistic model of Gaussian-shaped optical microtraps~\cite{negretti_speeding_2013}.
In this work, the authors divided the entire SAP process into three steps and optimized them individually via the CRAB algorithm~\cite{caneva_chopped_2011,doria_optimal_2011}.
These three steps consisted of (i) the initial approach between all traps before tunneling set in, (ii) the actual SAP process, and (iii) the final separation of the traps once the SAP dynamics are finished.
Furthermore, the optimization algorithms allowed to take the atomic interactions into account and showed that a significant speed up could be achieved without compromising high fidelities.

Another group of techniques are shortcuts to adiabaticity (STA), which were first introduced in Ref.~\cite{chen_fast_2010} to describe different protocols that allow to speed up adiabatic processes~\cite{torrontegui_shortcuts_2013}.
These rely on a number of different approaches, such as counterdiabatic or transitionless tracking algorithms~\cite{demirplak_adiabatic_2003,demirplak_assisted_2005, demirplak_consistency_2008,berry_transitionless_2009}, the use of the Lewis--Riesenfeld invariants~\cite{Lewis_1969,chen_fast_2010,Muga_invariant_2010}, or fast-forward techniques~\cite{Masuda_fastforward_2010, Torrontegui_fastforward_2012}.
By today they have not yet been applied to SAP processes.
The combination of OCT with invariant-based engineering techniques is promising to be particularly successful, since the later allows by construction for a perfect fidelity, while OCT can help to select the ones that optimize some physically relevant variables~\cite{chen_optimal_2011}.

\section{Dark state adiabatic passage}
\label{sec:darkstate}

While until now SAP has been discussed in the context of the motion of particles around some defined spatial network, a similar formalism can be used to describe the transport of spins on a lattice \cite{petrosyan_coherent_2006,OEO+07,OSF+13}.
This protocol is usually called dark state adiabatic passage (DSAP) \cite{OEO+07}, and while much of DSAP is conceptually very similar to conventional SAP, the robustness of spins to decoherence, as compared with degrees of freedom such as charge mean that there are potentially practical advantages in solid-state implementations.
Furthermore, high-spin systems introduce additional richness to the properties of the adiabatic passage~\cite{romero-isart_efficient_2007}.

Following the notation in Ref.~\cite{BK14}, as this highlights the connections with the previous analyses, the Hamiltonian for a three-spin system with an excitation exchange coupling is
\begin{align}
\mathcal{H} = B \sum_{i=1}^3 J_{z,i} + \left[d_{12} (t) J_1^{+}J_2^- + d_{23} (t) J_2^{+}J_3^- + \text{h.c.}\right], \label{eq:DSAPHam}
\end{align}
where $B$ is the (possibly time varying) Zeeman energy associated with the magnetic field, $J_{z,i}$ is the spin projection operator along the $z$ axis for particle $i$, $J_i^+$ ($J_i^-$) is the spin raising (lowering) operator for particle $i$, and $d_{ij}(t)$ is the time-varying (gated) coupling energy between (nearest neighbour) particles $i$ and $j$.  This generic Hamiltonian takes into account dipole-dipole or exchange-based coupling schemes, and the mechanism for the time control of the $d_{ij}$ can vary with the particular implementation.
Extensions to longer chains are, for the most part, relatively straightforward.

\begin{figure}
\centerline{ \resizebox{0.95\columnwidth}{!}{\includegraphics{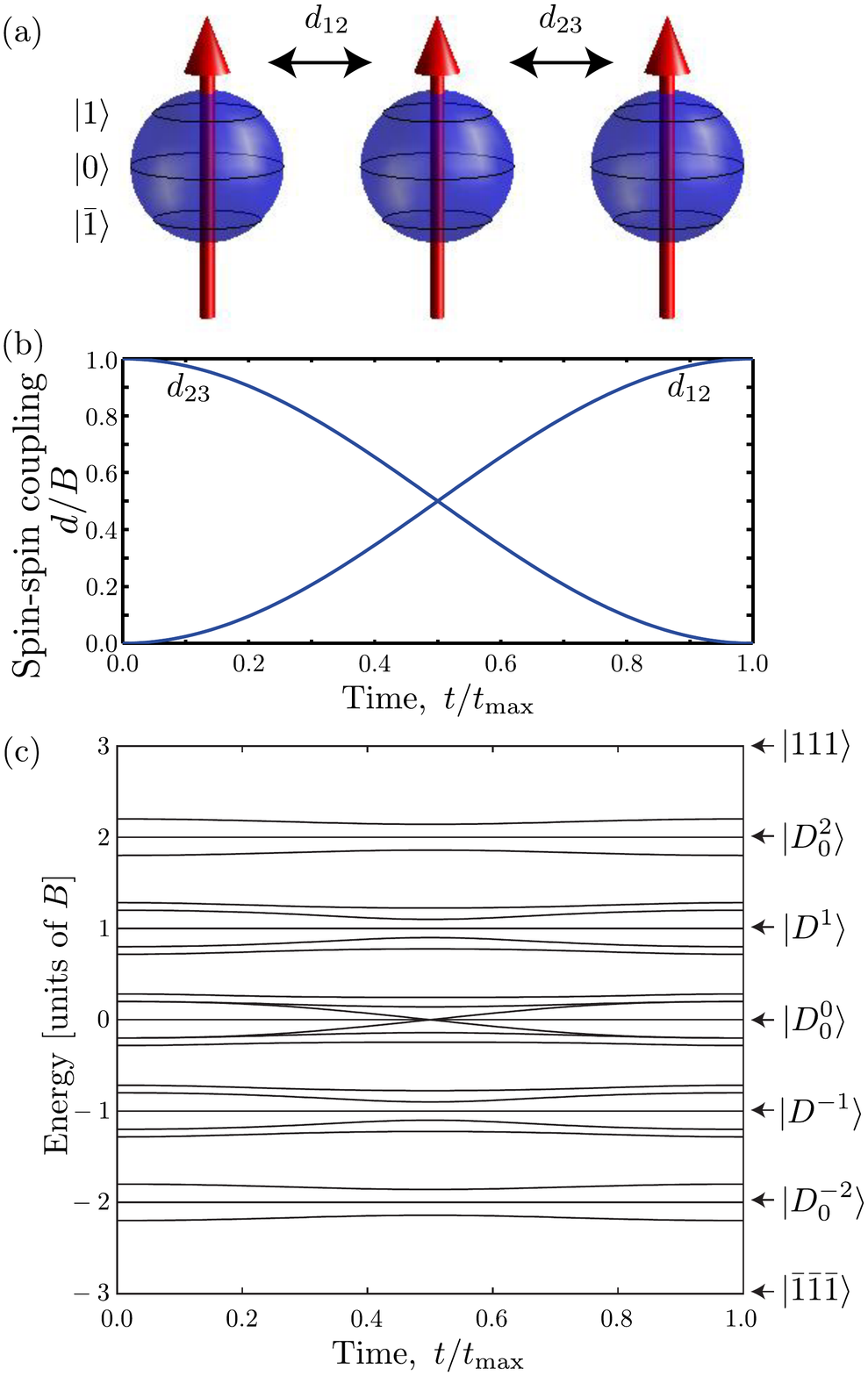}}}
\caption{(a) Schematic of a three-spin-1 system for DSAP with externally controlled spin-spin coupling. (b) Counterintuitive pulse sequence chosen to effect adiabatic passage. (c) Eigenspectrum through the adiabatic passage using a squared sinusoidal coupling with $d/B = 0.2$.  Note the separation of the eigenspectrum into distinct manifolds associated with the number of excitations in the system.
See text for details.
Reprinted from Ref.~\cite{BK14}.
}
\label{fig:DSAPfig1}
\end{figure}

\Fref{fig:DSAPfig1} shows a schematic of a three-spin-1 implementation, with counterintuitive pulse sequence of spin-spin coupling and eigenspectrum as a function of time through the protocol, assuming a squared sinusoidal coupling of the form
\begin{align}
d_{12}(t) =& d \sin^2\left(\pi t/2 t_{\max}\right), \\ 
d_{23}(t) =& d \cos^2\left(\pi t/2 t_{\max}\right).
\end{align} 
Here, $d$ is the maximum coupling, and the total time $t_{\max}$ is assumed long enough to ensure adiabatic evolution.  In the presence of a modest $B$ field, the eigenspectrum splits into manifolds defined by the number of excitations in the system, since the form of the coupling also conserves the total number of excitations.  We label the states $\ket{\bar{1}}$, $\ket{0}$ and $\ket{1}$, denoting the alignment of the spins with respect to the $z$ axis of the applied magnetic field.

Certain aspects of the eigenspectrum in \fref{fig:DSAPfig1}(c) are intuitively easy to understand with reference to conventional SAP.  The trivial manifolds with $\pm 3$ excitations, which consist of a single state, $\ket{111}$ and $\ket{\bar{1}\bar{1}\bar{1}}$ respectively, are not affected by the DSAP protocol as there is no possibility for excitation exchange.  The manifolds with $\pm2$ excitations, which are spanned by the states $\{\ket{0\bar{1}\bar{1}}, \ket{\bar{1}0\bar{1}}, \ket{\bar{1}\bar{1}0}\}$ for $-2$ excitations, and by $\{\ket{011}, \ket{101}, \ket{110}\}$ for 2 excitations allow for DSAP. In fact, these manifolds are identical to the spin-1/2 case with one excitation, and one can trivially generalize this result to any high-dimensional $m$-spin chain with $N$ particles, if the system is restricted to having either $-mN+1$  or $mN-1$ excitations.  In this way, DSAP in these manifolds is analogous to single particle or single hole SAP (see \sref{Sect:Holes}).

The remaining three manifolds also allow for DSAP, as they include 
a state which shows no change in energy throughout the adiabatic passage.
The states $\ket{D^{\pm1}}$ are adiabatically separated from the rest of the states in their manifold, but $\ket{D_0^0}$ crosses with other states in its manifold.
However, there is no possibility of coupling to the crossing states as there is no matrix element for interaction.
The dark states marked in \fref{fig:DSAPfig1}(c) are~\cite{BK14}
\begin{align}
\ket{D_0^2} &= \frac{d_{23}\ket{011} - d_{12}\ket{110}}{\sqrt{d_{12}^2 + d_{23}^2}}, \\
\ket{D_0^1} &= \frac{d_{23}^2 \ket{\bar{1}11} - d_{12}d_{23}\ket{010} + d_{12}^2\ket{11\bar{1}}}{\sqrt{d_{12}^4 - d_{12}^2d_{23}^2 + d_{23}^4}}, \\
\ket{D^0_0} &= \frac{d_{23}\left(\ket{01\bar{1}} - \ket{0\bar{1}1}\right) - d_{12}\left(\ket{1\bar{1}0} - \ket{\bar{1}10}\right)}{\sqrt{2}\sqrt{d_{12}^2 + d_{23}^2}}, \\
\ket{D_0^{-1}} &= \frac{d_{23}^2 \ket{1\bar{1}\bar{1}} - d_{12}d_{23}\ket{0\bar{1}0} + d_{12}^2\ket{\bar{1}\bar{1}1}}{\sqrt{d_{12}^4 - d_{12}^2d_{23}^2 + d_{23}^4}}, \\
\ket{D_0^{-2}} &= \frac{d_{23}\ket{0\bar{1}\bar{1}} - d_{12}\ket{\bar{1}\bar{1}0}}{\sqrt{d_{12}^2 + d_{23}^2}}.
\end{align}

Because of the presence of multiple dark states, it is possible to transfer an arbitrary qutrit using DSAP from spin 1 to spin 3, provided the other two spins are in the well-defined states $\ket{11}$, $\ket{\bar{1}\bar{1}}$ or $\left(\ket{11} - \ket{\bar{1}\bar{1}}\right)/\sqrt{2}$.  If the other two spins are not in these states, then noisy evolution is observed, which is similar to having extra particles in the canonical SAP schemes discussed above.

The properties of DSAP with higher spins and for longer chains are not yet fully known. One can speculate that qudit transport for the trivial cases of all of the other qudits in either the maximal or minimal spin orientation will be very similar.
However it is not clear what the set of all allowed configurations for the non-transported qudits is. Nor it is obvious what the properties of multidimensional orientations, as discussed in \sref{sec:beyond} will be.  These are expected to be fruitful areas for future research.

\section{Spatial adiabatic passage of light and sound}
\label{sec:classicalSAP}

Analogies between matter waves and light propagation have been very successful in the past years~\cite{garanovich_light_2012} and SAP processes have been studied to coherently control the propagation of light and sound in evanescently-coupled waveguides.
Moreover, optical-waveguide systems have allowed for the first experimental realizations of SAP, and enabled a simple and direct visualization of these coherent transport processes. 

Optical waveguides typically consist of a core of higher refractive index surrounded by lower refractive index material, so that light is transversally confined within the waveguide due to total internal reflection~\cite{chen_foundations_2007, katsunari_okamoto_fundamentals_2006}. Light beams travelling through a waveguide are in a superposition of the allowed solutions of Maxwell'€™s equations, i.e. the modes of the waveguide. These modes mainly propagate through the core of the waveguide, but also exist in the surrounding material of lower refractive index, 
where the electromagnetic field is called evanescent, as its amplitude decays exponentially outside the core of the waveguide. If two waveguides are close enough, they can be coupled through these evanescent fields. 

The scenario is similar for sound waves travelling through linear defects in sonic crystals, which work as waveguides~\cite{kafesaki_frequency_2000, khelif_transmittivity_2002, khelif_transmission_2003, miyashita_full_2002, khelif_trapping_2003, khelif_guiding_2004, vasseur_absolute_2008}. The linear defect creates allowed bands within the bandgap of the sonic crystal, which are the propagating modes of the sound waves. Sound travels mostly within the defect, but there are also evanescent fields surrounding the linear defect. If two linear defects are close enough, the evanescent field can couple them~\cite{sun_analyses_2005}.

Electromagnetic or sound waves propagating in a system of coupled waveguides are usually described in terms of a set of coupled-mode equations,
where the dynamics take place along the propagation distance.
Similarly to \sref{sec:general}, to perform a SAP process of light or sound, an eigenvector of the waveguide structure (also called a supermode) must be followed.
This supermode is a superposition of the individual modes of the waveguides.
The separation and shape of the waveguides can be used to control the strength of the couplings and/or the propagation constants of the modes along the system.

\subsection{Two coupled waveguides}

Wave propagation in a system of two coupled single-mode waveguides can be described by the following coupled-mode equations
\begin{equation}
i\frac{d}{dz}\left(\begin{array}{c}a_{L}\\a_{R}\end{array}\right)= \left( \begin{array}{cc}
\beta_L (z) & -\frac{\Omega_{LR (z)}}{2}\\
-\frac{\Omega_{LR} (z)}{2} &\beta_R (z)
\end{array} \right)\left(\begin{array}{c}a_{L}\\a_{R}\end{array}\right).
\label{sc_coupled}
\end{equation}
Here, $z$ is the propagation direction,
$a_i$ are the mode amplitudes (with $i=L,R$ representing the left and right waveguides, respectively),
$\beta_i$ are the propagation constants of each waveguide individual mode, 
and $\Omega_{LR}$ is the coupling coefficient between the two waveguides.
For simplicity we have considered that the coupling from the left to the right waveguide and from the right to the left waveguide is equal.
In the case of evanescent coupling, the dependence of $\Omega_{LR}$ on the separation between the two waveguides, $d(z)$, is given by
 \begin{equation}
 \Omega_{LR}(z) = \Omega_0\exp\left(-d(z)/l\right),
 \end{equation}
where $\Omega_0$ and $l$ are constant at a given frequency.
The propagation constants $\beta_i$ are related to the geometry of the system and the material used for the individual waveguides.
For the case of light propagation, these dependences can be written as an effective refractive index for each waveguide mode, $n^\textrm{eff}_i$, i.e., $\beta_i=k_0 n^\textrm{eff}_i$, where $k_0$ is the propagation constant in free space.

\begin{figure}
\centerline{ \resizebox{0.9\columnwidth}{!}{\includegraphics{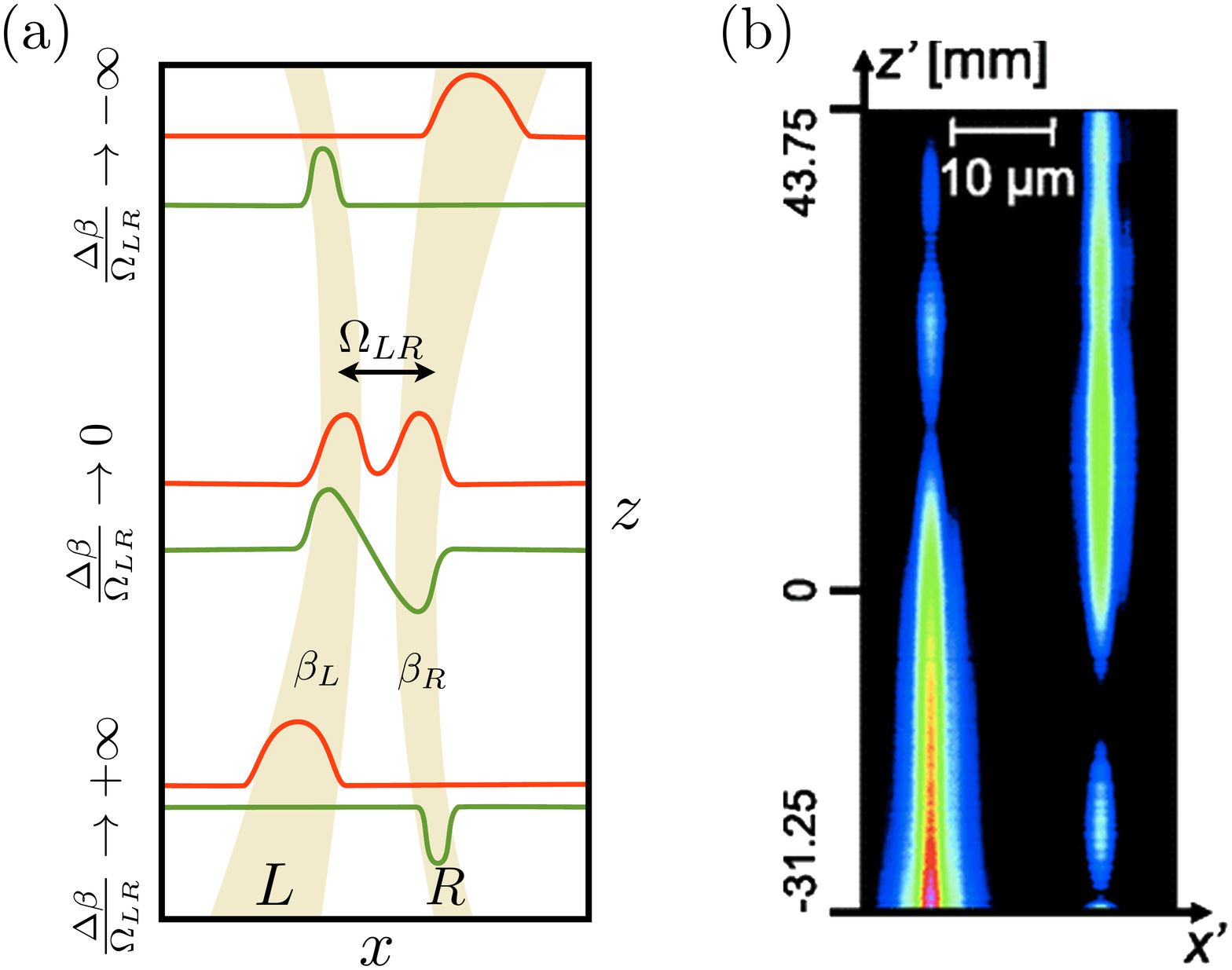}}}
\caption{(a) Schematic of a system of two coupled single-mode waveguides with the supermodes $S(z)$ (in red) and $A(z)$ (in green) at three different positions corresponding, respectively, to $\theta = 0$, $\pi/4$, and $\pi/2$.
(b) 
Experimental observation of a RAP process mimicking a Landau--Zener tunneling process in two coupled optical waveguides.
From \cite{dreisow_direct_2009}.
}
\label{light_fig1}
\end{figure}

The diagonalization of \eref{sc_coupled} gives the two supermodes of the structure, $S$ and $A$, as they correspond, in the case of the two waveguides having the same propagation constant, to the symmetric and antisymmetric superposition of the individual modes.
They can be expressed as 
\begin{align}
S(z)&=\left (\begin{array}{c}\cos\theta\\\sin\theta\end{array}\right),
\label{sc_sim} \\
A(z)&=\left (\begin{array}{c}\sin\theta\\-\cos\theta\end{array}\right),
\label{sc_antisim}
\end{align}
where the mixing angle $\theta$ is defined by 
\begin{equation}
\tan{2\theta}=\frac{\Omega_{LR}}{\Delta \beta},
\label{sc_mixing}
\end{equation}
with $\Delta \beta=\beta_R-\beta_L$.
The propagation constants of the supermodes are
\begin{equation}
\beta_{S, A}=\frac{\beta_L+\beta_R\mp\sqrt{\Omega_{LR}^2+{\Delta \beta}^2}}{2}.
\label{sc_kmodes}
\end{equation}
One can see from \etoeref{sc_sim}{sc_mixing} that when $|\Delta \beta| \gg \Omega_{LR}$, $\theta$ is either $0$ or $\pi/2$, depending on the sign of $\Delta \beta$.
In that case, each supermode stays localized in a different waveguide.
Moreover, when $|\Delta \beta| \ll \Omega_{LR}$ (the two propagation constants are very similar), $\theta=\pi/4$, and the supermodes spread equally between the two waveguides (see schematic example in \fref{light_fig1}(a)).
Therefore, by initially exciting one of the supermodes and adiabatically varying the propagation constants and the coupling, it is possible to coherently control the wave propagation.
A particular case of this would be a SAP transfer between the two waveguides following one of the supermodes, analogous to a quantum-optical RAP process~\cite{allen_optical_1987, vitanov_laser-induced_2001}.

Two-waveguide couplers performing SAP were extensively investigated
initially
for microwaves~\cite{cook_tapered_1955, fox_wave_1955, louisell_analysis_1955} and later on for the optical regime~\cite{wilson_improved_1973, smith_coupling_1975, smith_analytic_1976, silberberg_digital_1987, rowland_tapered_1991, ramadan_adiabatic_1998, chen_optimal_2007}.
Light propagation in coupled waveguides has also been considered as a classical system that allows simulating quantum dynamics.
One of the first presented analogies, discussed an optical realization of Landau--Zener dynamics in a two-level quantum system~\cite{longhi_landauzener_2005}.
This particular case, which can be viewed as a RAP scheme for which the coupling is considered constant, was experimentally realized in \cite{dreisow_direct_2009}, see \fref{light_fig1}(b).

\subsection{Three coupled waveguides}
We now explore a system of three single-mode coupled waveguides, for which the wave propagation is given by the coupled-mode equations \cite{chen_foundations_2007, katsunari_okamoto_fundamentals_2006}
\begin{align}
&i\frac{d}{dz}\left(\begin{array}{c}a_L\\a_M\\a_R\end{array}\right)= \notag \\
&\left( \begin{array}{ccc}
\beta_{L} & -\frac{\Omega_{LM}(z)}{2} & 0\\
-\frac{\Omega_{LM}(z)}{2} & \beta_{M} & -\frac{\Omega_{RM}(z)}{2}\\
0 & -\frac{\Omega_{RM}(z)}{2} & \beta_{R}
\end{array} \right)\left(\begin{array}{c}a_L\\a_M\\a_R\end{array}\right).
\label{cap1_coup}
\end{align}
Here, $a_i$ represent the mode amplitudes with $i=L,M,R$ accounting for the left, middle and right waveguides, $\Omega_{LM}$ ($\Omega_{RM}$) is the coupling coefficient between the left (right) and the middle waveguides, and $\beta_i$ are the propagation constants of each waveguide mode.
Following most of the studied cases, one considers that the waveguides on the left and on the right are not directly coupled
and takes the waveguides to be identical and constant along $z$ such that $\beta_L=\beta_M=\beta_R$.

\begin{figure}
\centerline{ \resizebox{0.85\columnwidth}{!}{\includegraphics{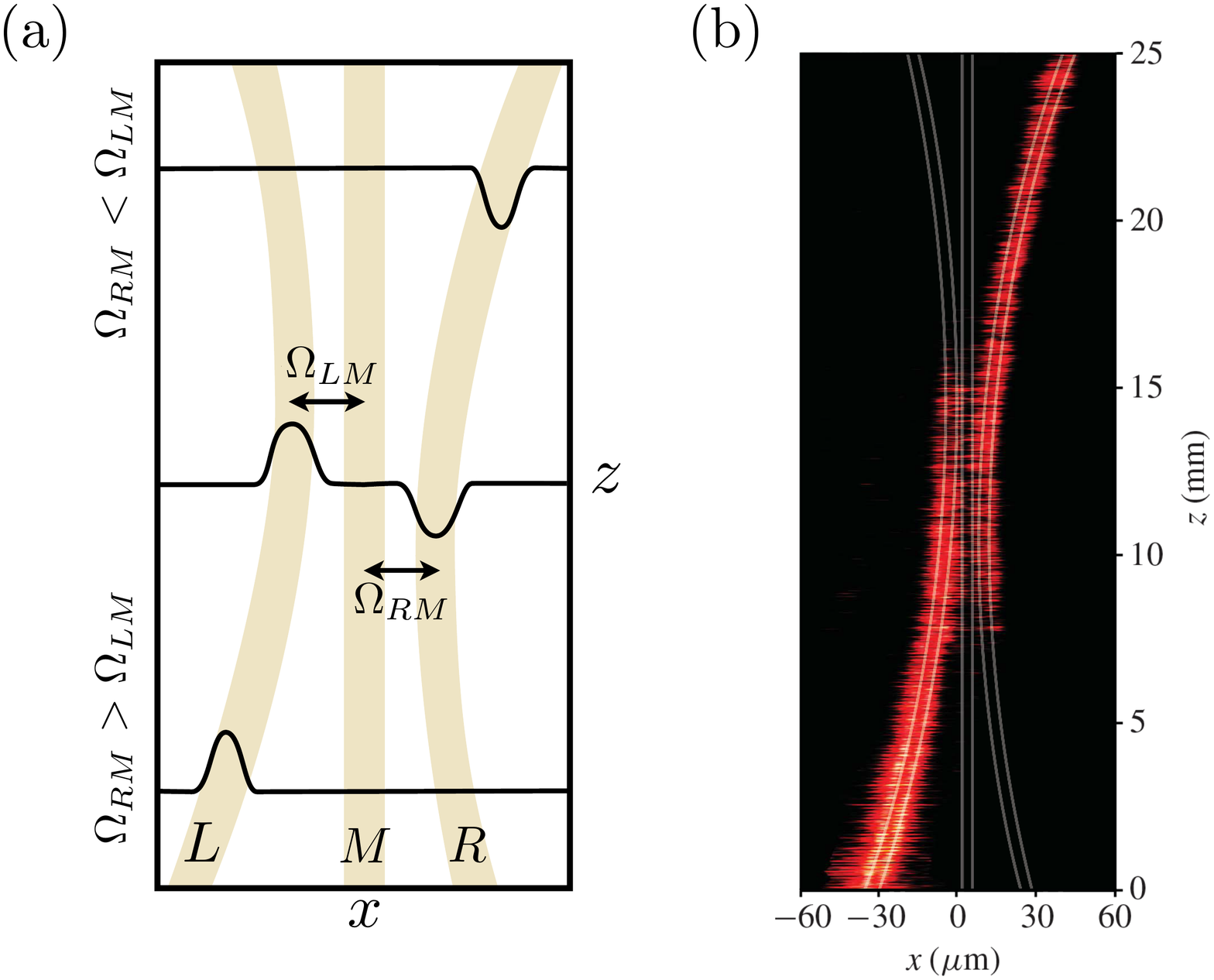}}}
\caption{
(a) Schematic of a system of three single-mode coupled waveguides with the dark state supermode (in black) at three different positions corresponding, respectively, to $\Theta = 0$, $\pi/4$, and $\pi/2$. (b) Top view experimental image of a triple-waveguide system performing SAP, where the transfer of light between the outermost waveguides takes place with almost no intensity in the central waveguide.
The shape of the waveguides is depicted on top of the experimental image.
From \cite{menchon-enrich_adiabatic_2012}.
}
\label{light_fig2}
\end{figure}

One of the supermodes obtained by diagonalizing \eref{cap1_coup} is
\begin{equation}
D(\Theta)=\left(\begin{array}{c}\cos\Theta\\0\\-\sin\Theta\end{array}\right),
\label{cap1_eigenvector}
\end{equation}
where the mixing angle is given by $\tan \Theta = \Omega_{LM} / \Omega_{RM}$.
This supermode is analogous to the dark state previously discussed for matter-wave systems, whose adiabatic following leads to a complete and robust transfer of a classical wave between the two outermost waveguides.
Analogously to \sref{formalism_SAP_three_well}, this is achieved by considering that the left waveguide of the triple-waveguide system is excited at the input port ($\Theta=0$), and that the spatial configuration of the waveguides forces the couplings to follow a counterintuitive sequence along $z$, as shown in \fref{light_fig2}(a).
If this modification is performed adiabatically, $D(\Theta)$ transfers the light to the right output port ($\Theta=\pi/2$),
without any light intensity in the middle waveguide.
This constitutes an example of SAP-based transport analogous to quantum-optical STIRAP~\cite{bergmann_coherent_1998}.
If the couplings are such that $\Theta$ evolves from $0$ to $\pi/4$,
$D(\Theta)$ involves the left waveguide at the input and 
an equal coherent splitting of the beam between the two outer waveguides at the output, analogously to the so-called fractional STIRAP~\cite{fractionalSTIRAP}.

Transport via SAP between the outermost waveguides of an optical three-waveguide system as discussed above was first proposed in \cite{paspalakis_adiabatic_2006, longhi_adiabatic_2006}, 
and experimentally observed shortly thereafter~\cite{longhi_coherent_2007,lahini_effect_2008}.
More recent experiments with CMOS-compatible technology were carried out in~\cite{menchon-enrich_adiabatic_2012,menchon-enrich_light_2013}, see \fref{light_fig2}(b).
All these experiments constituted the first implementations of SAP techniques in a three-mode system, and clearly demonstrated their robustness.
Moreover, they also showed the ability of optical waveguides to simulate and test certain aspects of quantum dynamics.

Other schemes for achieving adiabatic transport in systems of three waveguides have been proposed,
one suggests a new geometry consisting of three straight but non-parallel waveguides~\cite{salandrino_analysis_2009}, 
and one based on Raman chirped adiabatic passage (RCAP) \cite{Chelkowski_RCAP_1997}, where the following of the supermode is achieved by varying the propagation constants of the waveguides~\cite{longhi_photonic_2007}.

\subsection{Non-linearities and absorption} 

While the above considerations assumed the propagation dynamics to be linear, when intense light propagates through a medium, there is a change in its index of refraction $n$.
The first correction, the Kerr effect, yields $n = n_0 + n_2 I$, where $n_0$ is the index of refraction at low intensities, $n_2$ is the nonlinear coefficient, and $I$ is the light intensity.
The effects of such nonlinearities on SAP were studied in \cite{lahini_effect_2008}, by including them into the coupled-mode equations through a modification of the propagation constants as $\beta_i \rightarrow \beta_i + \Gamma |a_i|^2$. Here $\Gamma = \omega_0n_2/cA_\textrm{eff}$, with $\omega_0$ being the optical angular frequency, $c$ the speed of light, and $A_\textrm{eff}$ the common effective area of the waveguide modes~\cite{eisenberg_discrete_1998}.
The inclusion of the Kerr non-linearities into the description of the optical waveguides is analogous to the non-linearities in Bose--Einstein condensates, see \sref{sec:BEC}.

The effect of the nonlinearity on SAP in a system of two coupled waveguides
was analyzed theoretically and experimentally in \cite{barak_autoresonant_2009}, reporting the first observation of optical autoresonance:
at a sharp threshold of nonlinearity
the phase difference of the wave between the two waveguides stays locked.
For three coupled waveguides, it has been shown, theoretically and experimentally, that the transfer of light between the outermost waveguides depends critically on the excitation power~\cite{lahini_effect_2008}.
The role of the nonlinearity in each of the waveguides was also studied in \cite{kazazis_effects_2010}.

The effect of decay or absorption on SAP processes in systems of three coupled optical waveguides has also been studied in the literature.
The adiabatic transfer was shown to break down even for small values of the absorption in the initial or target waveguide \cite{graefe_breakdown_2013}, while it remains robust against absorption in the central waveguide~\cite{chung_broadband_2012}.

\subsection{Beyond three coupled waveguides, long range coupling, and SAP-based applications}
\label{LightSound:Beyond3LA}

\begin{figure}
\centerline{ \resizebox{0.8\columnwidth}{!}{\includegraphics{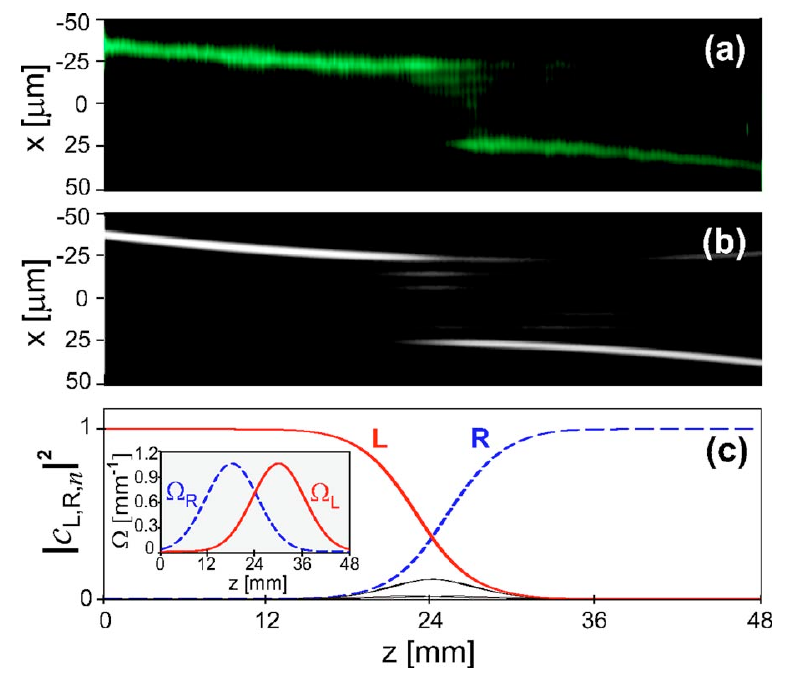}}}
\caption{Light transfer via a dressed state in the waveguide array with $N=5$. (a) Measured fluorescence pattern. (b) Numerically computed light propagation in the structure. (c) Evolution of fractional mode power in the waveguides as predicted by coupled mode equations. The inset in (c) shows the behavior of coupling rates. From \cite{valle_adiabatic_2008}.}
\label{valle_adiabatic_2008_fig}
\end{figure}

Multilevel structures are also of significant interest, as they allow SAP light transfer between the outermost waveguides of an array with an odd number (larger than three) of optical waveguides by following a supermode of the system \cite{longhi_optical_2006}.
This has been experimental demonstrated~\cite{valle_adiabatic_2008,ciret_broadband_2013}(see ~\fref{valle_adiabatic_2008_fig}), and it was shown that the transfer works for a broad range of wavelengths.
Different configurations of multiple waveguides have also been suggested to function as  multiple optical beam splitters \cite{rangelov_achromatic_2012, hristova_adiabatic_2013}.
Additionally, studies of the robustness of SAP processes in the presence of asymmetry variations in the waveguides were carried out in~\cite{chung_broadband_2012}.

Optical analogues of Rabi oscillations and STIRAP transfer via the continuum were proposed for waveguide-based photonic structures~\cite{longhi_transfer_2008} and experimentally demonstrated~\cite{dreisow_adiabatic_2009}.
In these systems, the continuum is approximated by a central array of waveguides that forms a plane perpendicular to the plane that contains the outermost waveguides used for the SAP process.
Similarly, other works study the wave optics analogue of coherent population trapping of bound states coupled to a common continuum using systems of multiple waveguides~\cite{longhi_optical_2009, longhi_optical_2009-1}.

The evanescent coupling inherent in waveguide approaches restricts strong coupling to nearest neighbours.
This means that breaking free from effectively 1D topologies without introducing significant cross talk is problematic (although a notable exception is presented in \cite{rangelov_achromatic_2012}).
One method for avoiding the restriction of nearest-neighbour coupling is to use a long-range `bus' mode, which for example, can be a thin shallow ridge optical waveguides on a slab mode~\cite{hope_long-range_2013}.
Thin shallow ridge optical waveguides are similar to conventional rib waveguides, although the height of the rib is smaller compared to typical rib designs and perturbations in the geometry of the rib can cause the modes to be radiative \cite{WPM+07}.
Although such losses are usually considered detrimental, this loss is coherent and can couple to other slab modes, allowing for long range coupling.
Furthermore, the waveguides themselves can be spaced far apart so that the evanescent coupling between them is negligible.
Hence by controlling the distance between the waveguides, it is possible to perform SAP processes across arbitrary distances within the slab.
However, as the slab is increased in size, the spectrum of the bus modes becomes denser, which increases unwanted off-resonant coupling to slab modes with different symmetry from the main (resonant) slab mode.

The long-range bus allows particular freedom to explore complex device geometries.
The implementation of a full non-deterministic CNOT gate \cite{RLB+02} was proposed in \cite{HNM+15}, by utilizing the Unanyan, Shore and Bergmann approach to realizing the 50:50 and 1/3:2/3 gates \cite{unanyan_laser-driven_1999}. 
This also allowed for a realization of the Hong--Ou--Mandel effect \cite{HOM87}.
While an adiabatic implementation of a beam splitter immediately implies that the two-photon input must demonstrate the Hong--Ou--Mandel effect (based on symmetry considerations), the actual adiabatic pathway is highly non-trivial involving coupling across ten states of four waveguide modes.

\begin{figure}
\centerline{ \resizebox{1\columnwidth}{!}{\includegraphics{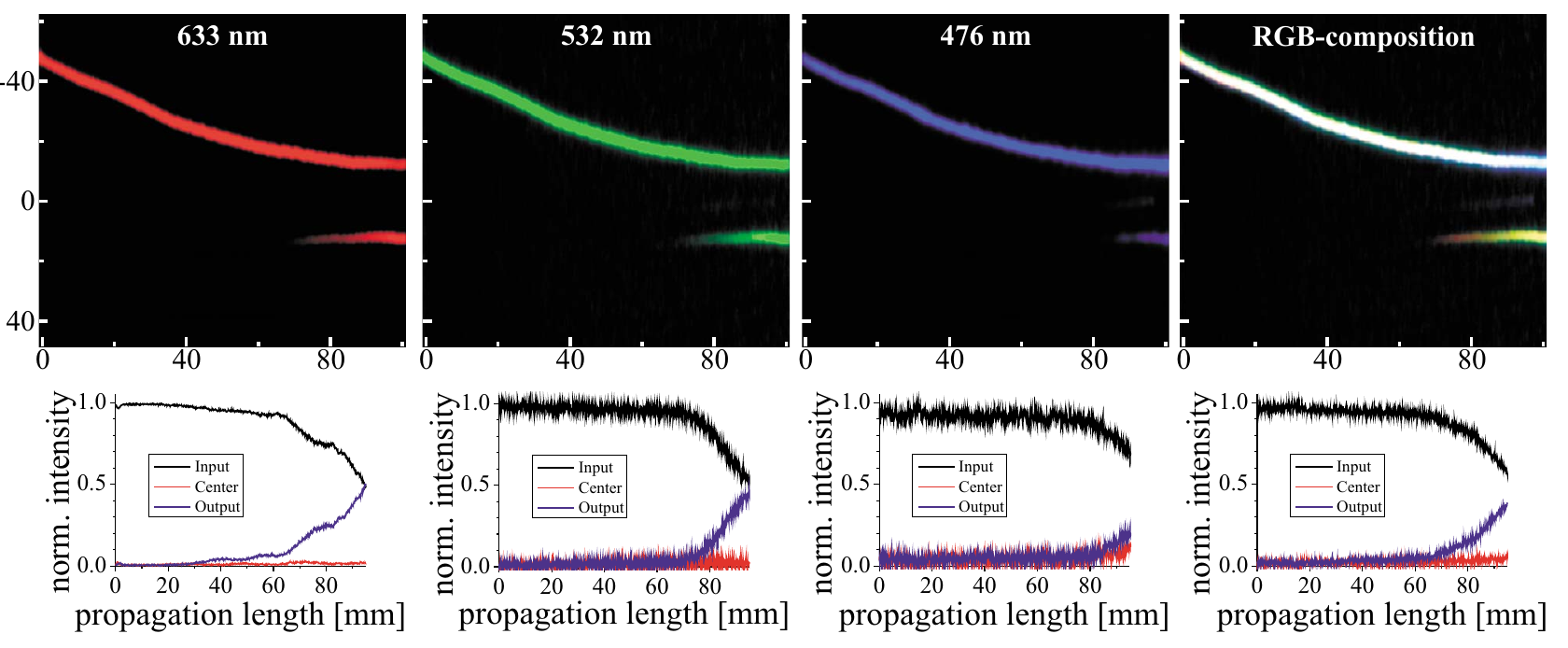}}}
\caption{Multicolor fluorescence measurements, showing light transfer in the fractional STIRAP device. Upper row: fluorescence images for three wavelengths and merged RGB-image. Lower row: measured intensities of input, center, and output waveguides, and the incoherent sum of 633 nm, 532 nm, and 476 nm (fourth column). From \cite{dreisow_polychromatic_2009}.}
\label{poli}
\end{figure}

Several applications based on SAP of light waves have been proposed.
For instance, it was experimentally demonstrated that a spatial analog of the fractional STIRAP technique works as a broad-band polychromatic beam splitter~\cite{dreisow_polychromatic_2009}, see \fref{poli}.
Other examples of optical beam splitters based on SAP are given in~\cite{rangelov_achromatic_2012, hristova_adiabatic_2013}.

A multiple waveguide system for interaction-free measurements based on the modification of the supermodes of the system due to the effect of a measured object on the couplings between the waveguides was suggested in \cite{hill_parallel_2011}.
Unlike the original proposal for quantum bomb detection \cite{EV93}, which utilized quantum mechanical interference, the adiabatic passage version performs sensing by having the bomb shift states out of the null space, leading to a different final outcome and preventing population from reaching the bomb.

\begin{figure}
\centerline{ \resizebox{0.75\columnwidth}{!}{\includegraphics{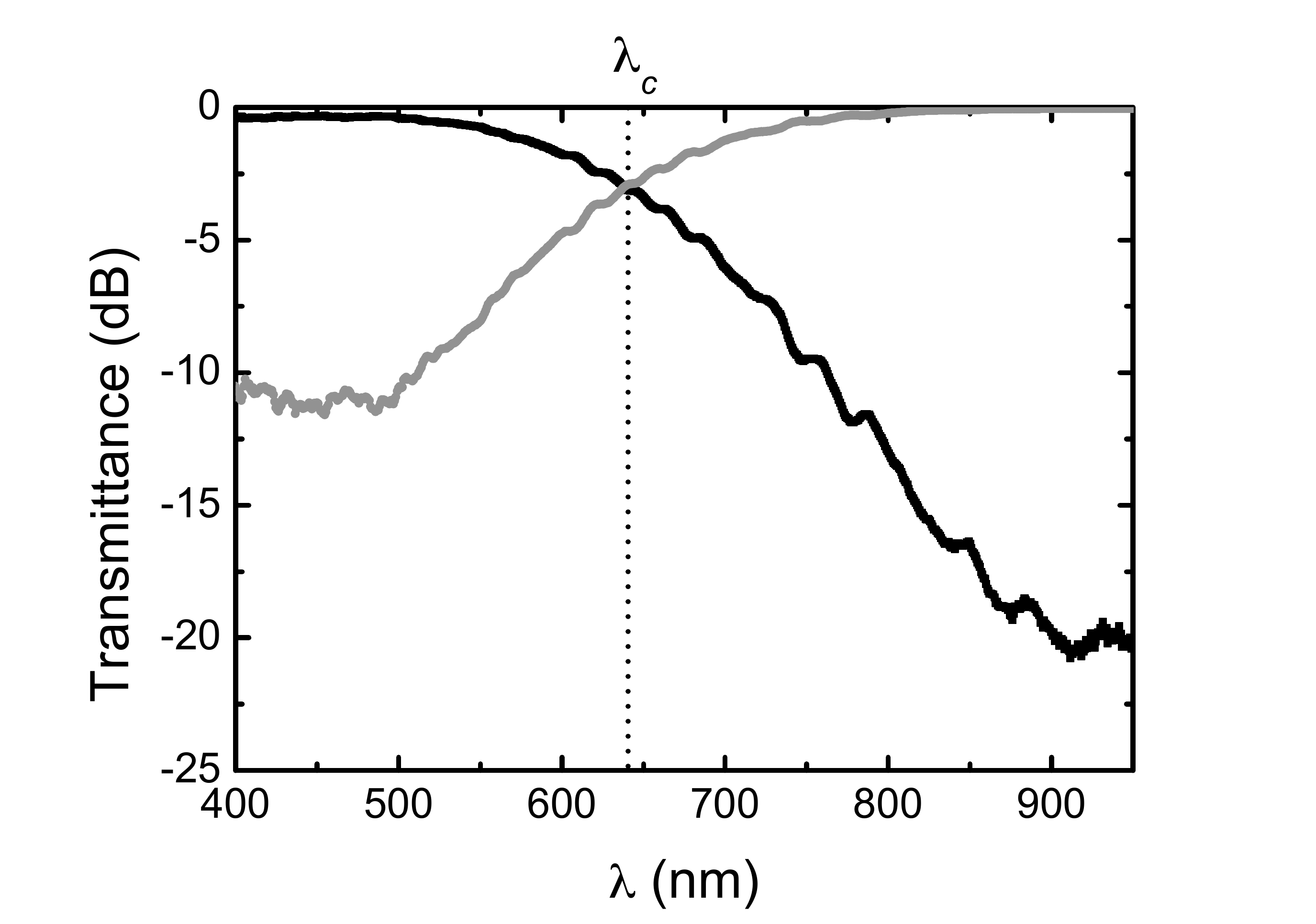}}}
\caption{Simultaneous high- and low-pass spectral filter based on SAP. Experimentally measured intensities are represented as the transmittance at the outermost waveguide (grey) and the sum of the middle and the initial waveguide (black) outputs relative to the total output intensity, as a function of the wavelength. From \cite{menchon-enrich_light_2013}.}
\label{filtre_light}
\end{figure}

\begin{figure}
\centerline{ \resizebox{0.7\columnwidth}{!}{\includegraphics{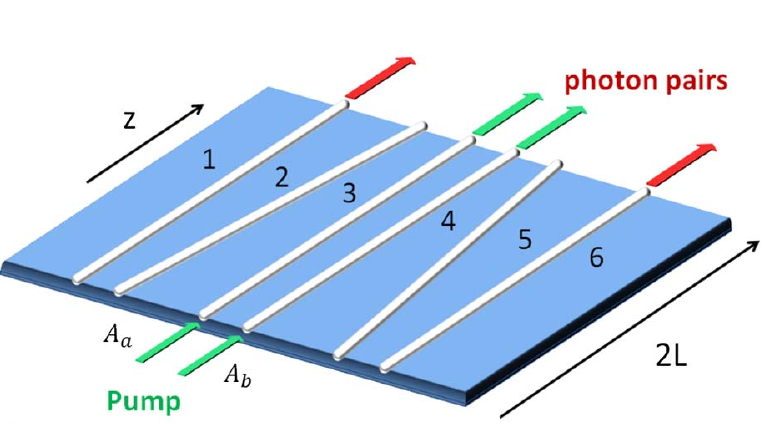}}}
\caption{
Scheme of an adiabatically coupled waveguiding structure for separating down-converted photons (red) from the pump beam (green).
Pump photons are coupled to the central waveguides 3 and 4, and the Bell states are formed in the edge waveguides 1 and 6 that are separated from the pumped waveguides. From \cite{wu_photon_2014}.
}
\label{photon_pair}
\end{figure}

SAP of light can also be used for spectral filtering  as the evanescent coupling depends on the light wavelength.
Because the coupling between waveguides is stronger for longer wavelengths, these are capable of adiabatically following a dark-state supermode and are transferred between the outermost waveguides of a triple-waveguide system.
Short wavelengths, on the other hand, are unable to adiabatically follow the supermode, and propagate towards the outputs of the middle and initial waveguides, see \fref{filtre_light}.
Such a simultaneous high- and low-pass spectral filter was demonstrated experimentally  with identical coupled TIR silicon oxide waveguides~\cite{menchon-enrich_light_2013},
and shown to represent an alternative to other integrated filtering devices, such as interference filters or absorbance-based filters.
As the waveguides are fully CMOS-compatible, they can be monolithically combined with other photonic and electronic elements into photonic integrated circuits, allowing SAP processes to become promising candidates to control the flow of light. 
In a similar manner, SAP processes have been proposed to robustly spatially separate spontaneous parametric down-converted photon pairs from the pumping photons~\cite{wu_photon_2014}, see \fref{photon_pair}.
This consitutes a novel integrated scheme for the generation of Bell states. 

Another example of a SAP-based application is a polarization rotor \cite{xiong_integrated_2013}, which can also be integrated in photonic circuits.
This system requires two waveguides, a signal and an ancillary one, and three modes:
two orthogonal polarization modes in the signal waveguide and an additional one in the ancillary waveguide. 
A dark-state supermode can then be modified and followed by controlling the distance between the two waveguides and the width of the signal waveguide.
This allows for a highly efficient adiabatic conversion between the two orthogonal modes in the signal waveguide. 

\subsection{SAP of sound waves}

As we previously commented, systems of linear defects in sonic crystals (acting as waveguides for sound waves) are suitable for SAP processes.
As an example, a system with two linear defects was considered in~\cite{menchon-enrich_spatial_2014} and two frequency ranges were studied.
For the first one, two supermodes of the system coexist and the sonic crystal was described using coupled-mode equations equivalent to \eref{sc_coupled}.
In this regime, by spatially modifying the linear defects, it is possible to design a broad-band beam splitter, based on the following of either of the two supermodes.
Exciting a superposition of the two supermodes of the system, the same structure can also work as a phase difference analyzer, i.e., a device that allows to infer the phase difference of two beams at the input ports by measuring the intensity at the output ports.

\begin{figure}
\centerline{ \resizebox{1.1\columnwidth}{!}{\includegraphics{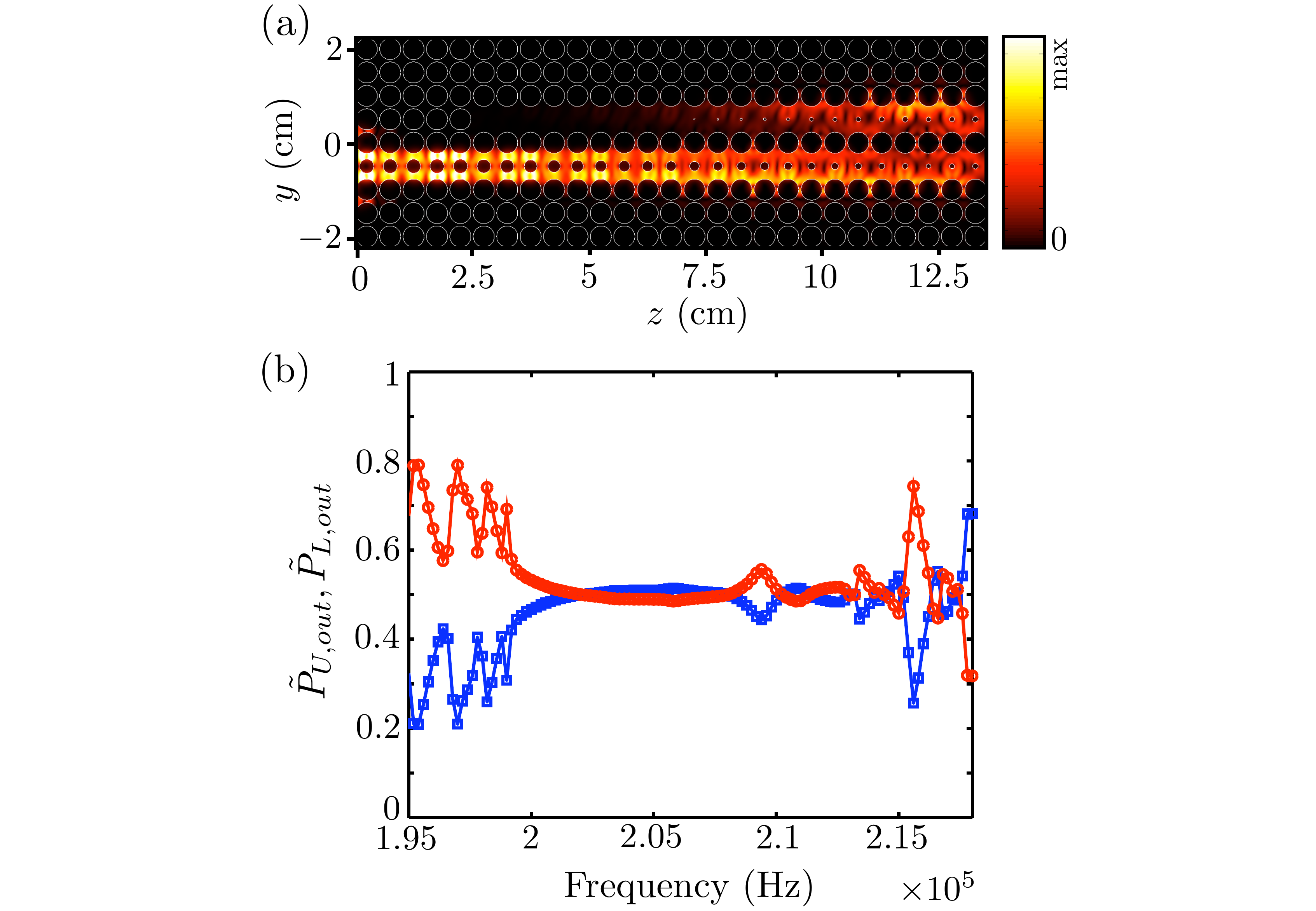}}}
\caption{
Coherent beam splitter in a sonic crystal with two linear defects.
(a) Numerical simulation of the intensity of a harmonic sound wave propagating with a frequency of $2.08\times 10^5$ Hz along the structure. 
(b) Power at the upper, $\tilde{P}_{U,out}$ (red dots), and lower, $\tilde{P}_{L,out}$ (blue squares), output ports, normalized with respect to the total output power, as a function of the frequency.
From \cite{menchon-enrich_spatial_2014}.
}
\label{sound_split}
\end{figure}

In the second studied frequency range, only one relevant supermode exists, making it
possible to reduce the length of the designed structures since there are no diabatic transitions to any other allowed supermode.
A broad-band beam splitter (see \fref{sound_split}) and a coupler were proposed, with much shorter lengths as compared to the devices in the first range of frequencies~\cite{menchon-enrich_spatial_2014}.

\section{Final remarks}
\label{sec:conclusions}

Since the first observation of dark states in the `riga nera' experiments during the mid 70s~\cite{alzetta_experimental_1976,alzetta_nonabsorption_1979,arimondo_nonabsorbing_1976}, they have been the basis of some of the most fruitful quantum-optical techniques: coherent population trapping (CPT)~\cite{CPT_Arimondo_1996}, electromagnetically induced transparency (EIT)~\cite{EIT_1,EIT_2}, and stimulated Raman adiabatic passage (STIRAP)~\cite{gaubatz_population_1990,bergmann_coherent_1998}.
These techniques, where two transitions of three internal levels of an atom (or molecule) are coupled with two laser fields, have given rise to a large number of applications routinely used in modern physics:
velocity selective laser cooling~\cite{CPT_A1} and atomic clocks~\cite{CPT_A2} based on CPT, slow/superluminal light~\cite{EIT_A1} and quantum memories \cite{EIT_A2} based on EIT, and single-photon guns~\cite{STIRAP_A1} and holonomic quantum computation~\cite{STIRAP_A2} based on STIRAP, to name only a few.

About one decade ago, spatial adiabatic passage (SAP) in one-dimensional triple-well systems, the matter-wave analogue of the STIRAP technique was proposed~\cite{eckert_three-level_2004,greentree_coherent_2004}. 
In it, the tunneling interaction plays the role of the laser fields and spatially localized states take the place of the internal states.
In the first two seminal papers this was shown to allow for a high fidelity and robust quantum transport, which led to subsequent works that showed that SAP techniques hold a large potential beyond the STIRAP analogy.
The fact that SAP consists of adiabatically following a spatially delocalized dark state allows for the possibility to perform matter-wave interferometry to measure non-uniform fields, to extend it to higher dimensions involving, for instance, orbital angular momentum states or to consider coupling geometries that are often forbidden in STIRAP.
Another extension is dark state adiabatic passage where a quantum spin state is transported along a chain of particles.

It is also worth noting that STIRAP schemes using the bright eigenstates exist~\cite{gaubatz_population_1990,vitanov_analytic_1997,klein_robust_1999}, which are challenging to realize in quantum-optical systems due to the presence of spontaneous emission.
However, as this problem does not exist in spatial systems, they are promising candidates for future extensions of SAP.

Despite the many theoretical investigations on matter-wave SAP, it has not yet been observed. Experimental difficulties stem from a number of facts.
First, a single atom or, alternatively, a BEC, needs to be prepared in the ground state of one of the three wells only.
Moreover, the tunneling amplitudes between left-middle and middle-right wells need to be independently controllable in a way that avoids excitations of unwanted vibrational states.
While this is experimentally demanding, no fundamental obstacles to observe matter-wave SAP exist, and significant experimental effort to observe SAP in real physical systems is currently ongoing.

Finally, SAP can also be applied to other systems, most notably to light beams propagating in evanescently-coupled optical waveguides.
Therefore, these systems can be used to simulate some fundamental aspects of quantum mechanics, while at the same time, SAP allows for the development of efficient photonic devices.
In fact, SAP processes have been already observed for light beams.

Given the potential and the fast development of the research in spatial dark states in recent years, it is highly likely that they will play a significant role in many applications related to quantum transport, quantum metrology, spintronics, quantum information science and condensed matter physics in the near future.

\section*{Acknowledgements}

This work was supported by the Okinawa Institute of Science and Technology Graduate University.  A.D.G. acknowledges the ARC for financial support (Grant No. DP130104381). J.M and V.A. acknowledge financial support through the Spanish and Catalan contracts FIS2014-57460-P and SGR2014-1639.
J.M. also acknowledges financial support from JSPS Research Fellowship S-15025.

\bibliographystyle{unsrt}
\bibliography{SAP_REV}

\begin{thebibliography}{100}

\bibitem{micheli_single_2004}
A.~Micheli, A.~J. Daley, D.~Jaksch, and P.~Zoller.
\newblock Single atom transistor in a 1{D} optical lattice.
\newblock {\em Physical Review Letters}, 93(14):140408, October 2004.

\bibitem{ruschhaupt_atom_2004}
A.~Ruschhaupt and J.~G. Muga.
\newblock Atom diode: {A} laser device for a unidirectional transmission of
  ground-state atoms.
\newblock {\em Physical Review A}, 70(6):061604, December 2004.

\bibitem{stickney_transistorlike_2007}
James~A. Stickney, Dana~Z. Anderson, and Alex~A. Zozulya.
\newblock Transistorlike behavior of a {B}ose--{E}instein condensate in a
  triple-well potential.
\newblock {\em Physical Review A}, 75(1):013608, January 2007.

\bibitem{thorn_experimental_2008}
Jeremy~J. Thorn, Elizabeth~A. Schoene, Tao Li, and Daniel~A. Steck.
\newblock Experimental realization of an optical one-way barrier for neutral
  atoms.
\newblock {\em Physical Review Letters}, 100(24):240407, June 2008.

\bibitem{price_single-photon_2008}
Gabriel~N. Price, S.~Travis Bannerman, Kirsten Viering, Edvardas Narevicius,
  and Mark~G. Raizen.
\newblock Single-photon atomic cooling.
\newblock {\em Physical Review Letters}, 100(9):093004, March 2008.

\bibitem{pepino_atomtronic_2009}
R.~A. Pepino, J.~Cooper, D.~Z. Anderson, and M.~J. Holland.
\newblock Atomtronic circuits of diodes and transistors.
\newblock {\em Physical Review Letters}, 103(14):140405, September 2009.

\bibitem{benseny_atomtronics_2010}
A.~Benseny, S.~Fern\'andez-Vidal, J.~Bagud\`a, R.~Corbal\'an,
  A.~Pic\'on, L.~Roso, G.~Birkl, and J.~Mompart.
\newblock Atomtronics with holes: Coherent transport of an empty site in a
  triple-well potential.
\newblock {\em Physical Review A}, 82(1):013604, July 2010.

\bibitem{zozulya_principles_2013}
Alex~A. Zozulya and Dana~Z. Anderson.
\newblock Principles of an atomtronic battery.
\newblock {\em Physical Review A}, 88(4):043641, October 2013.

\bibitem{ryu_experimental_2013}
C.~Ryu, P.~W. Blackburn, A.~A. Blinova, and M.~G. Boshier.
\newblock Experimental realization of {J}osephson junctions for an atom
  {SQUID}.
\newblock {\em Physical Review Letters}, 111(20):205301, November 2013.

\bibitem{wright_driving_2013}
K.~C. Wright, R.~B. Blakestad, C.~J. Lobb, W.~D. Phillips, and G.~K. Campbell.
\newblock Driving phase slips in a superfluid atom circuit with a rotating weak
  link.
\newblock {\em Physical Review Letters}, 110(2):025302, January 2013.

\bibitem{jendrzejewski_resistive_2014}
F.~Jendrzejewski, S.~Eckel, N.~Murray, C.~Lanier, M.~Edwards, C. J. Lobb, and
  G. K. Campbell.
\newblock Resistive flow in a weakly interacting {B}ose--{E}instein condensate.
\newblock {\em Physical Review Letters}, 113(4):045305, July 2014.

\bibitem{born_beweis_1928}
M.~Born and V.~Fock.
\newblock Beweis des {A}diabatensatzes.
\newblock {\em Zeitschrift f{\"u}r Physik}, 51(3-4):165--180, March 1928.

\bibitem{messiah_quantum_1976}
Albert Messiah.
\newblock {\em Quantum {M}echanics}, volume~2.
\newblock North-Holland, Amsterdam, 1976.

\bibitem{allen_optical_1987}
L.~Allen and J.~H. Eberly.
\newblock {\em Optical Resonance and Two-level Atoms}.
\newblock Dover Publications, New York, 1987.

\bibitem{vitanov_laser-induced_2001}
Nikolay~V Vitanov, Thomas Halfmann, Bruce~W Shore, and Klaas Bergmann.
\newblock Laser-induced population transfer by adiabatic passage techniques.
\newblock {\em Annual Review of Physical Chemistry}, 52(1):763--809, 2001.

\bibitem{bergmann_coherent_1998}
K.~Bergmann, H.~Theuer, and B.~W. Shore.
\newblock Coherent population transfer among quantum states of atoms and
  molecules.
\newblock {\em Reviews of Modern Physics}, 70(3):1003--1025, July 1998.

\bibitem{fewell_coherent_1997}
M.~P. Fewell, B.~W. Shore, and K.~Bergmann.
\newblock Coherent population transfer among three states: Full algebraic
  solutions and the relevance of non adiabatic processes to transfer by delayed
  pulses.
\newblock {\em Australian Journal of Physics}, 50(2):281--308, January 1997.

\bibitem{alzetta_experimental_1976}
G.~Alzetta, A.~Gozzini, L.~Moi, and G.~Orriols.
\newblock An experimental method for the observation of r.f. transitions and
  laser beat resonances in oriented na vapour.
\newblock {\em Il Nuovo Cimento B Series 11}, 36(1):5--20, November 1976.

\bibitem{arimondo_nonabsorbing_1976}
E.~Arimondo and G.~Orriols.
\newblock Nonabsorbing atomic coherences by coherent two-photon transitions in
  a three-level optical pumping.
\newblock {\em Lettere Al Nuovo Cimento Series 2}, 17(10):333--338, November
  1976.

\bibitem{gray_coherent_1978}
H.~R. Gray, R.~M. Whitley, and C.~R. Stroud, Jr.
\newblock Coherent trapping of atomic populations.
\newblock {\em Optics Letters}, 3(6):218, December 1978.

\bibitem{alzetta_nonabsorption_1979}
G.~Alzetta, L.~Moi, and G.~Orriols.
\newblock Nonabsorption hyperfine resonances in a sodium vapour irradiated by a
  multimode dye-laser.
\newblock {\em Il Nuovo Cimento B}, 52(2):209--218, August 1979.

\bibitem{shore_pre-history_2013}
Bruce~W Shore.
\newblock Pre-history of the concepts underlying stimulated {R}aman adiabatic
  passage ({STIRAP}).
\newblock {\em acta physica slovaca}, 63(6):361--481, 2013.

\bibitem{bergmann_perspective_2015}
Klaas Bergmann, Nikolay~V. Vitanov, and Bruce~W. Shore.
\newblock Perspective: Stimulated {R}aman adiabatic passage: The status after
  25 years.
\newblock {\em The Journal of Chemical Physics}, 142(17):--, 2015.

\bibitem{renzoni_charge_2001}
F.~Renzoni and T.~Brandes.
\newblock Charge transport through quantum dots via time-varying tunnel
  coupling.
\newblock {\em Physical Review B}, 64(24):245301, November 2001.

\bibitem{eckert_three-level_2004}
K.~Eckert, M.~Lewenstein, R.~Corbal\'an, G.~Birkl, W.~Ertmer, and J.~Mompart.
\newblock Three-level atom optics via the tunneling interaction.
\newblock {\em Physical Review A}, 70(2):023606, August 2004.

\bibitem{greentree_coherent_2004}
Andrew~D. Greentree, Jared~H. Cole, A.~R. Hamilton, and Lloyd C.~L. Hollenberg.
\newblock Coherent electronic transfer in quantum dot systems using adiabatic
  passage.
\newblock {\em Physical Review B}, 70(23):235317, December 2004.

\bibitem{busch_quantum_2007}
Th. Busch, K.~Deasy, and S.~Nic Chormaic.
\newblock Quantum state preparation using multi-level-atom optics.
\newblock {\em Journal of Physics: Conference Series}, 84(1):012002, October
  2007.

\bibitem{loiko_filtering_2011}
Yu. Loiko, V.~Ahufinger, R.~Corbal\'an, G.~Birkl, and J.~Mompart.
\newblock Filtering of matter-wave vibrational states via spatial adiabatic
  passage.
\newblock {\em Physical Review A}, 83(3):033629, March 2011.

\bibitem{loiko_coherent_2014}
Yury Loiko, Veronica Ahufinger, Ricard Menchon-Enrich, Gerhard Birkl, and Jordi
  Mompart.
\newblock Coherent injecting, extracting, and velocity filtering of neutral
  atoms in a ring trap via spatial adiabatic passage.
\newblock {\em The European Physical Journal D}, 68(6):1--5, June 2014.

\bibitem{jong_interferometry_2010}
Lenneke~M. Jong and Andrew~D. Greentree.
\newblock Interferometry using spatial adiabatic passage in quantum dot
  networks.
\newblock {\em Physical Review B}, 81(3):035311, January 2010.

\bibitem{menchon-enrich_single-atom_2014}
R.~Menchon-Enrich, S.~McEndoo, Th. Busch, V.~Ahufinger, and J.~Mompart.
\newblock Single-atom interferometer based on two-dimensional spatial adiabatic
  passage.
\newblock {\em Physical Review A}, 89:053611, May 2014.

\bibitem{lu_coherent_2011}
Gengbiao Lu, Wenhua Hai, and Qiongtao Xie.
\newblock Coherent control of atomic tunneling in a driven triple well.
\newblock {\em Physical Review A}, 83(1):013407, January 2011.

\bibitem{menchon-enrich_tunneling-induced_2014}
R.~Menchon-Enrich, S.~McEndoo, J.~Mompart, V.~Ahufinger, and Th. Busch.
\newblock Tunneling-induced angular momentum for single cold atoms.
\newblock {\em Physical Review A}, 89(1):013626, January 2014.

\bibitem{longhi_adiabatic_2006}
Stefano Longhi.
\newblock Adiabatic passage of light in coupled optical waveguides.
\newblock {\em Physical Review E}, 73(2):026607, February 2006.

\bibitem{longhi_coherent_2007}
S.~Longhi, G.~Della~Valle, M.~Ornigotti, and P.~Laporta.
\newblock Coherent tunneling by adiabatic passage in an optical waveguide
  system.
\newblock {\em Physical Review B}, 76(20):201101, November 2007.

\bibitem{longhi_optical_2006}
S.~Longhi.
\newblock Optical realization of multilevel adiabatic population transfer in
  curved waveguide arrays.
\newblock {\em Physics Letters A}, 359(2):166--170, November 2006.

\bibitem{valle_adiabatic_2008}
G.~Della Valle, M.~Ornigotti, T.~Toney Fernandez, P.~Laporta, S.~Longhi,
  A.~Coppa, and V.~Foglietti.
\newblock Adiabatic light transfer via dressed states in optical waveguide
  arrays.
\newblock {\em Applied Physics Letters}, 92(1):011106, January 2008.

\bibitem{rangelov_achromatic_2012}
Andon~A. Rangelov and Nikolay~V. Vitanov.
\newblock Achromatic multiple beam splitting by adiabatic passage in optical
  waveguides.
\newblock {\em Physical Review A}, 85(5):055803, May 2012.

\bibitem{ciret_broadband_2013}
Charles Ciret, Virginie Coda, Andon~A. Rangelov, Dragomir~N. Neshev, and
  Germano Montemezzani.
\newblock Broadband adiabatic light transfer in optically induced waveguide
  arrays.
\newblock {\em Physical Review A}, 87(1):013806, January 2013.

\bibitem{longhi_transfer_2008}
S.~Longhi.
\newblock Transfer of light waves in optical waveguides via a continuum.
\newblock {\em Physical Review A}, 78(1):013815, July 2008.

\bibitem{dreisow_adiabatic_2009}
F.~Dreisow, A.~Szameit, M.~Heinrich, R.~Keil, S.~Nolte, A.~T\"unnermann, and
  S.~Longhi.
\newblock Adiabatic transfer of light via a continuum in optical waveguides.
\newblock {\em Optics Letters}, 34(16):2405--2407, August 2009.

\bibitem{barak_autoresonant_2009}
Assaf Barak, Yuval Lamhot, Lazar Friedland, and Mordechai Segev.
\newblock Autoresonant dynamics of optical guided waves.
\newblock {\em Physical Review Letters}, 103(12):123901, September 2009.

\bibitem{graefe_breakdown_2013}
Eva-Maria Graefe, Alexei~A. Mailybaev, and Nimrod Moiseyev.
\newblock Breakdown of adiabatic transfer of light in waveguides in the
  presence of absorption.
\newblock {\em Physical Review A}, 88(3):033842, September 2013.

\bibitem{dreisow_polychromatic_2009}
F.~Dreisow, M.~Ornigotti, A.~Szameit, M.~Heinrich, R.~Keil, S.~Nolte,
  A.~T\"unnermann, and S.~Longhi.
\newblock Polychromatic beam splitting by fractional stimulated {R}aman
  adiabatic passage.
\newblock {\em Applied Physics Letters}, 95(26):261102--261102--3, December
  2009.

\bibitem{chung_broadband_2012}
Kelvin Chung, Timothy~J. Karle, Masum Rab, Andrew~D. Greentree, and Snjezana
  Tomljenovic-Hanic.
\newblock Broadband and robust optical waveguide devices using coherent
  tunnelling adiabatic passage.
\newblock {\em Optics Express}, 20(21):23108--23116, October 2012.

\bibitem{hristova_adiabatic_2013}
H.~S. Hristova, A.~A. Rangelov, S.~Gu\'erin, and N.~V. Vitanov.
\newblock Adiabatic evolution of light in an array of parallel curved optical
  waveguides.
\newblock {\em Physical Review A}, 88(1):013808, July 2013.

\bibitem{menchon-enrich_light_2013}
Ricard Menchon-Enrich, Andreu Llobera, Jordi Vila-Planas, V\'ictor~J. Cadarso,
  Jordi Mompart, and Veronica Ahufinger.
\newblock Light spectral filtering based on spatial adiabatic passage.
\newblock {\em Light: Science \& Applications}, 2(8):e90, August 2013.

\bibitem{hill_parallel_2011}
Charles~D. Hill, Andrew~D. Greentree, and Lloyd C.~L. Hollenberg.
\newblock Parallel interaction-free measurement using spatial adiabatic
  passage.
\newblock {\em New Journal of Physics}, 13(12):125002, December 2011.

\bibitem{hope_long-range_2013}
A.~P. Hope, T.~G. Nguyen, A.~D. Greentree, and A.~Mitchell.
\newblock Long-range coupling of silicon photonic waveguides using lateral
  leakage and adiabatic passage.
\newblock {\em Optics Express}, 21(19):22705--22716, September 2013.

\bibitem{xiong_integrated_2013}
Xiao Xiong, Chang-Ling Zou, Xi-Feng Ren, and Guang-Can Guo.
\newblock Integrated polarization rotator/converter by stimulated {R}aman
  adiabatic passage.
\newblock {\em Optics express}, 21(14):17097--17107, July 2013.

\bibitem{wu_photon_2014}
Che~Wen Wu, Alexander~S. Solntsev, Dragomir~N. Neshev, and Andrey~A.
  Sukhorukov.
\newblock Photon pair generation and pump filtering in nonlinear adiabatic
  waveguiding structures.
\newblock {\em Optics Letters}, 39(4):953--956, Feb 2014.

\bibitem{menchon-enrich_spatial_2014}
R.~Menchon-Enrich, J.~Mompart, and V.~Ahufinger.
\newblock Spatial adiabatic passage processes in sonic crystals with linear
  defects.
\newblock {\em Physical Review B}, 89(9):094304, March 2014.

\bibitem{eckert_quantum_2002}
K.~Eckert, J.~Mompart, X.~X. Yi, J.~Schliemann, D.~Bru\ss{}, G.~Birkl, and
  M.~Lewenstein.
\newblock Quantum computing in optical microtraps based on the motional states
  of neutral atoms.
\newblock {\em Physical Review A}, 66:042317, Oct 2002.

\bibitem{holstein_mobilities_1952}
T.~Holstein.
\newblock Mobilities of positive ions in their parent gases.
\newblock {\em The Journal of Physical Chemistry}, 56(7):832--836, 1952.

\bibitem{herring_critique_1962}
Conyers Herring.
\newblock Critique of the {H}eitler--{L}ondon method of calculating spin
  couplings at large distances.
\newblock {\em Review of Modern Physics}, 34:631--645, Oct 1962.

\bibitem{unanyan_laser_1997}
R.~G. Unanyan, L.~P. Yatsenko, K~Bergmann, and B.~W. Shore.
\newblock Laser-induced adiabatic atomic reorientation with control of diabatic
  losses.
\newblock {\em Optics Communications}, 139:48--54, 1997.

\bibitem{fleischhauer_coherent_1999}
M.~Fleischhauer, R.~Unanyan, L.~P., B.~W. Shore, and K~Bergmann.
\newblock Coherent population transfer beyond the adiabatic limit: Generalized
  matched pulses and higher-order trapping states.
\newblock {\em Physical Review A}, 59:3751--3760, 1999.

\bibitem{giannelli_three_2014}
Luigi Giannelli and Ennio Arimondo.
\newblock Three-level superadiabatic quantum driving.
\newblock {\em Physical Review A}, 89:033419, Mar 2014.

\bibitem{kral_cyclic_2001}
P~Kr\'{a}l and M~Shapiro.
\newblock Cyclic population transfer in quantum systems with broken symmetry.
\newblock {\em Physical Review Letters}, 87:183002, 2001.

\bibitem{kral_two-step_2003}
P~Kr\'{a}l, I~Thanopulos, M~Shapiro, and D.~Cohen.
\newblock Two-step enantio-selective optical switch.
\newblock {\em Physical Review Letters}, 87:033001, 2003.

\bibitem{GRB+88}
U.~Gaubatz, P.~Rudecki, M.~Becker, S.~Schiemann, M.~K{\"u}lz, and K.~Bergmann.
\newblock Population switching between vibrational levels in molecular beams.
\newblock {\em Chemical Physics Letters}, 149(5 - 6):463 -- 468, 1988.

\bibitem{CH90}
C.~E. Carroll and F.~T. Hioe.
\newblock Analytic solutions for three-state systems with overlapping pulses.
\newblock {\em Physical Review A}, 42:1522--1531, Aug 1990.

\bibitem{LS96}
Timo~A. Laine and Stig Stenholm.
\newblock Adiabatic processes in three-level systems.
\newblock {\em Physical Review A}, 53:2501--2512, Apr 1996.

\bibitem{DGH07}
Simon~J. Devitt, Andrew~D. Greentree, and Lloyd~C.L. Hollenberg.
\newblock Information free quantum bus for generating stabiliser states.
\newblock {\em Quantum Information Processing}, 6(4):229--242, 2007.

\bibitem{rab_spatial_2008}
M.~Rab, J.~H. Cole, N.~G. Parker, A.~D. Greentree, L.~C.~L. Hollenberg, and
  A.~M. Martin.
\newblock Spatial coherent transport of interacting dilute {B}ose gases.
\newblock {\em Physical Review A}, 77(6):061602, June 2008.

\bibitem{vaitkus_digital_2013}
Jesse~A. Vaitkus and Andrew~D. Greentree.
\newblock Digital three-state adiabatic passage.
\newblock {\em Physical Review A}, 87(6):063820, June 2013.

\bibitem{SMM07}
E.~A. Shapiro, V.~Milner, C.~Menzel-Jones, and M.~Shapiro.
\newblock Piecewise adiabatic passage with a series of femtosecond pulses.
\newblock {\em Physical Review Letters}, 99:033002, Jul 2007.

\bibitem{SMS09}
Evgeny~A. Shapiro, Valery Milner, and Moshe Shapiro.
\newblock Complete transfer of populations from a single state to a preselected
  superposition of states using piecewise adiabatic passage: Theory.
\newblock {\em Physical Review A}, 79:023422, Feb 2009.

\bibitem{opatrny_conditions_2009}
Tom\'{a}\v{s} Opatrn\'{y} and Kunal~K. Das.
\newblock Conditions for vanishing central-well population in triple-well
  adiabatic transport.
\newblock {\em Physical Review A}, 79(1):012113, January 2009.

\bibitem{eckert_three_2006}
K.~Eckert, J.~Mompart, R.~Corbal\'an, M.~Lewenstein, and G.~Birkl.
\newblock Three level atom optics in dipole traps and waveguides.
\newblock {\em Optics Communications}, 264(2):264--270, August 2006.

\bibitem{razavy_quantum_2003}
M.~Razavy.
\newblock {\em Quantum theory of tunneling}.
\newblock World Scientific Publishing Co. Pte. Ltd., 2003.

\bibitem{cole_spatial_2008}
J.~H. Cole, A.~D. Greentree, L.~C.~L. Hollenberg, and S.~Das~Sarma.
\newblock Spatial adiabatic passage in a realistic triple well structure.
\newblock {\em Physical Review B}, 77(23):235418, June 2008.

\bibitem{vitanov_stimulated_2006}
N.~V. Vitanov and B.~W. Shore.
\newblock Stimulated {R}aman adiabatic passage in a two-state system.
\newblock {\em Physical Review A}, 73:053402, 2006.

\bibitem{ottaviani_adiabatic_2010}
C.~Ottaviani, V.~Ahufinger, R.~Corbal\'an, and J.~Mompart.
\newblock Adiabatic splitting, transport, and self-trapping of a
  {B}ose--{E}instein condensate in a double-well potential.
\newblock {\em Physical Review A}, 81(4):043621, April 2010.

\bibitem{benseny_need_2012}
A.~Benseny, Joan Bagud\`a, X.~Oriols, and J.~Mompart.
\newblock Need for relativistic corrections in the analysis of spatial
  adiabatic passage of matter waves.
\newblock {\em Physical Review A}, 85(5):053619, May 2012.

\bibitem{benseny_atomtronics:_2012}
Albert Benseny, Joan Bagud{\`a}, Xavier Oriols, Gerhard Birkl, and Jordi
  Mompart.
\newblock Atomtronics: Coherent control of atomic flow via adiabatic passage.
\newblock In Xavier Oriols and Jordi Mompart, editors, {\em Applied Bohmian
  mechanics: From nanoscale systems to cosmology}. Pan Stanford Publishing,
  Singapore, June 2012.

\bibitem{bohm_suggested_1952-1}
David Bohm.
\newblock A suggested interpretation of the quantum theory in terms of
  ``hidden'' variables. {I}.
\newblock {\em Physical Review}, 85:166--179, Jan 1952.

\bibitem{bohm_suggested_1952-2}
David Bohm.
\newblock A suggested interpretation of the quantum theory in terms of
  ``hidden'' variables. {II}.
\newblock {\em Physical Review}, 85:180--193, Jan 1952.

\bibitem{benseny_applied_2014}
Albert Benseny, Guillermo Albareda, \'Angel~S. Sanz, Jordi Mompart, and Xavier
  Oriols.
\newblock Applied {B}ohmian mechanics.
\newblock {\em The European Physical Journal D}, 68(10), 2014.

\bibitem{leavens_are_1998}
{C.R.} Leavens and R.~Sala~Mayato.
\newblock Are predicted superluminal tunneling times an artifact of using the
  nonrelativistic {S}chr{\"o}dinger equation?
\newblock {\em Annalen der Physik}, 7(7-8):662--670, 1998.

\bibitem{struyve_uniqueness_2004}
W.~Struyve, W.~De Baere, J.~De Neve, and S.~De Weirdt.
\newblock On the uniqueness of paths for spin-0 and spin-1 quantum mechanics.
\newblock {\em Physics Letters A}, 322(1–2):84 -- 95, 2004.

\bibitem{gajdacz_transparent_2011}
Miroslav Gajdacz, Tom\'{a}\v{s} Opatrn\'{y}, and Kunal~K. Das.
\newblock Transparent nonlocal species-selective transport in an optical
  superlattice containing two interacting atom species.
\newblock {\em Physical Review A}, 83(3):033623, March 2011.

\bibitem{mcendoo_phase_2010}
S.~McEndoo, S.~Croke, J.~Brophy, and Th. Busch.
\newblock Phase evolution in spatial dark states.
\newblock {\em Physical Review A}, 81(4):043640, April 2010.

\bibitem{shore_multilevel_1991}
B.~W. Shore, K.~Bergmann, J.~Oreg, and S.~Rosenwaks.
\newblock Multilevel adiabatic population transfer.
\newblock {\em Physical Review A}, 44(11):7442--7447, December 1991.

\bibitem{jong_coherent_2009}
L.~M. Jong, A.~D. Greentree, V.~I. Conrad, L.~C.~L. Hollenberg, and D.~N.
  Jamieson.
\newblock Coherent tunneling adiabatic passage with the alternating coupling
  scheme.
\newblock {\em Nanotechnology}, 20(40):405402, October 2009.

\bibitem{bradly_coherent_2012}
C.~J. Bradly, M.~Rab, A.~D. Greentree, and A.~M. Martin.
\newblock Coherent tunneling via adiabatic passage in a three-well
  {B}ose--{H}ubbard system.
\newblock {\em Physical Review A}, 85(5):053609, May 2012.

\bibitem{osullivan_using_2010}
B.~O'Sullivan, P.~Morrissey, T.~Morgan, and Th~Busch.
\newblock Using adiabatic coupling techniques in atom-chip waveguide
  structures.
\newblock {\em Physica Scripta}, 2010(T140):014029, September 2010.

\bibitem{morgan_coherent_2013}
T.~Morgan, L.~J. O'Riordan, N.~Crowley, B.~O'Sullivan, and Th. Busch.
\newblock Coherent transport by adiabatic passage on atom chips.
\newblock {\em Physical Review A}, 88(5):053618, November 2013.

\bibitem{morgan_coherent_2011}
T.~Morgan, B.~O'Sullivan, and Th. Busch.
\newblock Coherent adiabatic transport of atoms in radio-frequency traps.
\newblock {\em Physical Review A}, 83(5):053620, May 2011.

\bibitem{kreutzmann_coherence_2004}
H.~Kreutzmann, U.~V. Poulsen, M.~Lewenstein, R.~Dumke, W.~Ertmer, G.~Birkl, and
  A.~Sanpera.
\newblock Coherence properties of guided-atom interferometers.
\newblock {\em Physical Review Letters}, 92(16):163201, April 2004.

\bibitem{LD98}
Daniel Loss and David~P. DiVincenzo.
\newblock Quantum computation with quantum dots.
\newblock {\em Physical Review A}, 57:120--126, Jan 1998.

\bibitem{Kan98}
B.~E. Kane.
\newblock A silicon-based nuclear spin quantum computer.
\newblock {\em Nature}, 393(6681):133--137, 05 1998.

\bibitem{HDW+04}
L.~C.~L. Hollenberg, A.~S. Dzurak, C.~Wellard, A.~R. Hamilton, D.~J. Reilly,
  G.~J. Milburn, and R.~G. Clark.
\newblock Charge-based quantum computing using single donors in semiconductors.
\newblock {\em Physical Review B}, 69:113301, Mar 2004.

\bibitem{DiV00}
David~P. DiVincenzo.
\newblock The physical implementation of quantum computation.
\newblock {\em Fortschritte der Physik}, 48(9-11):771--783, 2000.

\bibitem{COI+03}
D.~Copsey, M.~Oskin, F.~Impens, T.~Metodiev, A.~Cross, F.T. Chong, I.L. Chuang,
  and J.~Kubiatowicz.
\newblock Toward a scalable, silicon-based quantum computing architecture.
\newblock {\em Selected Topics in Quantum Electronics, IEEE Journal of},
  9(6):1552--1569, Nov 2003.

\bibitem{SB04}
Jens Siewert and Tobias Brandes.
\newblock Applications of adiabatic passage in solid-state devices.
\newblock In Bernhard Kramer, editor, {\em Advances in Solid State Physics},
  volume~44 of {\em Advances in Solid State Physics}, pages 181--189. Springer
  Berlin Heidelberg, 2004.

\bibitem{SBF06}
Jens Siewert, Tobias Brandes, and Giuseppe Falci.
\newblock Adiabatic passage with superconducting nanocircuits.
\newblock {\em Optics Communications}, 264(2):435 -- 440, 2006.
\newblock Quantum Control of Light and Matter In honor of the 70th birthday of
  Bruce Shore.

\bibitem{TRM2000}
S.~Tsujino, M.~R\"{u}fenacht, P.~Miranda, S.J. Allen, P.~Tamborenea,
  W.~Schoenfeld, G.~Herold, G.~Lupke, T.~Lundstrom, P.~Petroff, H.~Metiu, and
  D.~Moses.
\newblock Quantum control of electron transfer.
\newblock {\em Physica Status Solidi (b)}, 221(1):391--396, 2000.

\bibitem{RTA2000}
M.~R\"{u}fenacht, S.~Tsujino, S.J. Allen, W.~Schoenfeld, and P.~Petroff.
\newblock Coherent transfer and electron teleportation in semiconductor double
  quantum well.
\newblock {\em Physica Status Solidi (b)}, 221(1):407--411, 2000.

\bibitem{BRB01}
T.~Brandes, F.~Renzoni, and R.~H. Blick.
\newblock Adiabatic steering and determination of dephasing rates in double-dot
  qubits.
\newblock {\em Physical Review B}, 64:035319, Jun 2001.

\bibitem{PH07}
Thorsten Peters and Thomas Halfmann.
\newblock Stimulated {R}aman adiabatic passage via the ionization continuum in
  helium: Experiment and theory.
\newblock {\em Optics Communications}, 271(2):475 -- 486, 2007.

\bibitem{MT97}
Vladimir~S. Malinovsky and David~J. Tannor.
\newblock Simple and robust extension of the stimulated {R}aman adiabatic
  passage technique to $n$-level systems.
\newblock {\em Physical Review A}, 56:4929--4937, Dec 1997.

\bibitem{HGF+06}
L.~C.~L. Hollenberg, A.~D. Greentree, A.~G. Fowler, and C.~J. Wellard.
\newblock Two-dimensional architectures for donor-based quantum computing.
\newblock {\em Physical Review B}, 74:045311, Jul 2006.

\bibitem{GCH+05}
A.D. Greentree, J.H. Cole, A.R. Hamilton, and L.C.L. Hollenberg.
\newblock Scaling of coherent tunneling adiabatic passage in solid-state
  coherent quantum systems.
\newblock {\em Proceedings of SPIE - The International Society for Optical
  Engineering}, 5650:72--80, 2005.

\bibitem{petrosyan_coherent_2006}
D.~Petrosyan and P.~Lambropoulos.
\newblock Coherent population transfer in a chain of tunnel coupled quantum
  dots.
\newblock {\em Optics Communications}, 264(2):419--425, August 2006.

\bibitem{VGA+10}
Jessica A.~Van Donkelaar, Andrew~D. Greentree, Andrew D.~C. Alves, Lenneke~M.
  Jong, Lloyd C.~L. Hollenberg, and David~N. Jamieson.
\newblock Top-down pathways to devices with few and single atoms placed to high
  precision.
\newblock {\em New Journal of Physics}, 12(6):065016, 2010.

\bibitem{rahman_atomistic_2009}
Rajib Rahman, Seung~H. Park, Jared~H. Cole, Andrew~D. Greentree, Richard~P.
  Muller, Gerhard Klimeck, and Lloyd C.~L. Hollenberg.
\newblock Atomistic simulations of adiabatic coherent electron transport in
  triple donor systems.
\newblock {\em Physical Review B}, 80(3):035302, July 2009.

\bibitem{rahman_coherent_2010}
Rajib Rahman, Richard~P. Muller, James~E. Levy, Malcolm~S. Carroll, Gerhard
  Klimeck, Andrew~D. Greentree, and Lloyd C.~L. Hollenberg.
\newblock Coherent electron transport by adiabatic passage in an imperfect
  donor chain.
\newblock {\em Physical Review B}, 82(15):155315, October 2010.

\bibitem{huneke_steady-state_2013}
Jan Huneke, Gloria Platero, and Sigmund Kohler.
\newblock Steady-state coherent transfer by adiabatic passage.
\newblock {\em Physical Review Letters}, 110(3):036802, January 2013.

\bibitem{FMM+15}
E.~Ferraro, M.~De~Michielis, M.~Fanciulli, and E.~Prati.
\newblock Coherent tunneling by adiabatic passage of an exchange-only spin
  qubit in a double quantum dot chain.
\newblock {\em Physical Review B}, 91:075435, Feb 2015.

\bibitem{SGG+07}
D.~Schr\"oer, A.~D. Greentree, L.~Gaudreau, K.~Eberl, L.~C.~L. Hollenberg,
  J.~P. Kotthaus, and S.~Ludwig.
\newblock Electrostatically defined serial triple quantum dot charged with few
  electrons.
\newblock {\em Physical Review B}, 76:075306, Aug 2007.

\bibitem{BBR+2013}
F.~R. Braakman, P.~Barthelemy, C.~Reichl, W.~Wegscheider, and L.~M.~K.
  Vandersypen.
\newblock Long-distance coherent coupling in a quantum dot array.
\newblock {\em Nature Nanotechnology}, 8(6):432--437, 06 2013.

\bibitem{PTS+2015}
G.~Poulin-Lamarre, J.~Thorgrimson, S.~A. Studenikin, G.~C. Aers, A.~Kam,
  P.~Zawadzki, Z.~R. Wasilewski, and A.~S. Sachrajda.
\newblock Three-spin coherent oscillations and interference.
\newblock {\em Physical Review B}, 91:125417, Mar 2015.

\bibitem{kamleitner_adiabatic_2008}
I.~Kamleitner, J.~Cresser, and J.~Twamley.
\newblock Adiabatic information transport in the presence of decoherence.
\newblock {\em Physical Review A}, 77(3):032331, March 2008.

\bibitem{rech_effect_2011}
J\'er\^ome Rech and Stefan Kehrein.
\newblock Effect of measurement backaction on adiabatic coherent electron
  transport.
\newblock {\em Physical Review Letters}, 106(13):136808, March 2011.

\bibitem{vogt_influence_2012}
Nicolas Vogt, Jared~H. Cole, Michael Marthaler, and Gerd Sch\"on.
\newblock Influence of two-level fluctuators on adiabatic passage techniques.
\newblock {\em Physical Review B}, 85(17):174515, May 2012.

\bibitem{greentree_quantum-information_2006}
Andrew~D. Greentree, Simon~J. Devitt, and Lloyd C.~L. Hollenberg.
\newblock Quantum-information transport to multiple receivers.
\newblock {\em Physical Review A}, 73(3):032319, March 2006.

\bibitem{WC15}
Xin Wei and Mei-Feng Chen.
\newblock Preparation of multi-qubit {W} states in multiple resonators coupled
  by a superconducting qubit via adiabatic passage.
\newblock {\em Quantum Information Processing}, 14(7):2419--2433, 2015.

\bibitem{unanyan_laser-driven_1999}
R.~G. Unanyan, B.~W. Shore, and K.~Bergmann.
\newblock Laser-driven population transfer in four-level atoms: Consequences of
  non-abelian geometrical adiabatic phase factors.
\newblock {\em Physical Review A}, 59:2910--2919, 1999.

\bibitem{KR02}
Z.~Kis and F.~Renzoni.
\newblock Qubit rotation by stimulated {R}aman adiabatic passage.
\newblock {\em Physical Review A}, 65:032318, Feb 2002.

\bibitem{FSF03}
Lara Faoro, Jens Siewert, and Rosario Fazio.
\newblock Non-abelian holonomies, charge pumping, and quantum computation with
  josephson junctions.
\newblock {\em Physical Review Letters}, 90:028301, Jan 2003.

\bibitem{RLA99}
Ferruccio Renzoni, Albrecht Lindner, and Ennio Arimondo.
\newblock Coherent population trapping in open systems: A
  coupled/noncoupled-state analysis.
\newblock {\em Physical Review A}, 60:450--455, Jul 1999.

\bibitem{HNM+15}
Anthony~P Hope, Thach~G Nguyen, Arnan Mitchell, and Andrew~D Greentree.
\newblock Adiabatic two-photon quantum gate operations using a long-range
  photonic bus.
\newblock {\em Journal of Physics B: Atomic, Molecular and Optical Physics},
  48(5):055503, 2015.

\bibitem{MS83}
James~R. Morris and Bruce~W. Shore.
\newblock Reduction of degenerate two-level excitation to independent two-state
  systems.
\newblock {\em Physical Review A}, 27:906--912, Feb 1983.

\bibitem{RVS06}
A.~A. Rangelov, N.~V. Vitanov, and B.~W. Shore.
\newblock Extension of the {M}orris--{S}hore transformation to multilevel
  ladders.
\newblock {\em Physical Review A}, 74:053402, Nov 2006.

\bibitem{kestner_proposed_2011}
J.~P. Kestner and S.~Das~Sarma.
\newblock Proposed spin-qubit controlled-not gate robust against noisy
  coupling.
\newblock {\em Physical Review A}, 84(1):012315, July 2011.

\bibitem{Boy73}
Timothy~H. Boyer.
\newblock Classical electromagnetic deflections and lag effects associated with
  quantum interference pattern shifts: Considerations related to the
  aharonov-bohm effect.
\newblock {\em Physical Review D}, 8:1679--1693, Sep 1973.

\bibitem{pitaevskii_Bose_2003}
L.~Pitaevskii and S.~Stringari.
\newblock {\em Bose-Einstein condensation}.
\newblock Clarendon Press, Oxford, 2003.

\bibitem{graefe_mean-field_2006}
E.~M. Graefe, H.~J. Korsch, and D.~Witthaut.
\newblock Mean-field dynamics of a {B}ose--{E}instein condensate in a
  time-dependent triple-well trap: Nonlinear eigenstates, {L}andau--{Z}ener
  models, and stimulated {R}aman adiabatic passage.
\newblock {\em Physical Review A}, 73(1):013617, January 2006.

\bibitem{nesterenko_stirap_2009}
V.~O. Nesterenko, A.~N. Novikov, F.~F. de~Souza Cruz, and E.~L. Lapolli.
\newblock {STIRAP} transport of {B}ose--{E}instein condensate in triple-well
  trap.
\newblock {\em Laser Physics}, 19(4):616--624, April 2009.
\newblock {WOS}:000265043500014.

\bibitem{rab_interferometry_2012}
M.~Rab, A.~L.~C. Hayward, J.~H. Cole, A.~D. Greentree, and A.~M. Martin.
\newblock Interferometry using adiabatic passage in dilute-gas
  {B}ose--{E}instein condensates.
\newblock {\em Physical Review A}, 86(6):063605, December 2012.

\bibitem{nesterenko_adiabatic_2009}
V.~O. Nesterenko, A.~N. Novikov, A.~Yu Cherny, F.~F. de~Souza Cruz, and
  E.~Suraud.
\newblock An adiabatic transport of {B}ose--{E}instein condensates in
  double-well traps.
\newblock {\em Journal of Physics B: Atomic, Molecular and Optical Physics},
  42(23):235303, December 2009.

\bibitem{nesterenko_adiabatic_2010}
V.~O. Nesterenko, A.~N. Novikov, and E.~Suraud.
\newblock Adiabatic transport of {B}ose--{E}instein condensates in a
  double-well trap: Case of weak nonlinearity.
\newblock {\em Laser Physics}, 20(5):1149--1155, May 2010.

\bibitem{longhi_coherent_2014}
S.~{Longhi}.
\newblock {Coherent transfer by adiabatic passage in two-dimensional lattices}.
\newblock {\em Annals of Physics}, 348:161--175, September 2014.

\bibitem{birkl_atom_2001}
G~Birkl, {F.B.J} Buchkremer, R~Dumke, and W~Ertmer.
\newblock Atom optics with microfabricated optical elements.
\newblock {\em Optics Communications}, 191(1{\textendash}2):67--81, May 2001.

\bibitem{dumke_micro-optical_2002}
R.~Dumke, M.~Volk, T.~M{\"u}ther, F.~B.~J. Buchkremer, G.~Birkl, and W.~Ertmer.
\newblock Micro-optical realization of arrays of selectively addressable dipole
  traps: A scalable configuration for quantum computation with atomic qubits.
\newblock {\em Physical Review Letters}, 89(9):097903, August 2002.

\bibitem{AtomChipReview}
J\'ozsef Fort\'agh and Claus Zimmermann.
\newblock Magnetic microtraps for ultracold atoms.
\newblock {\em Review of Modern Physics}, 79:235--289, Feb 2007.

\bibitem{Zobay_RF_Traps_2001}
O.~Zobay and B.M. Garraway.
\newblock Two-dimensional atom trapping in field-induced adiabatic potentials.
\newblock {\em Physical Review Letters}, 86:1195, Feb 2001.

\bibitem{Schumm:05}
T.~Schumm, S.~Hofferberth, L.M. Andersson, S.~Wildermuth, S.~Groth,
  I.~Bar-Joseph, J.~Schmiedmayer, and P.~Kr\"uger.
\newblock Matter-wave interferometry in a double well on an atom chip.
\newblock {\em Nature Physics}, 1:57--62, 2005.

\bibitem{Lesanovsky:07}
S.~Hofferberth, B.~Fischer, T.~Schumm, J.~Schmiedmayer, and I.~Lesanovsky.
\newblock Ultracold atoms in radio-frequency dressed potentials beyond the
  rotating-wave approximation.
\newblock {\em Physical Review A}, 76:013401, Jul 2007.

\bibitem{Zimmermann:06}
Ph.~W. Courteille, B.~Deh, J.~Fort\'agh, A.~G\"unther, S.~Kraft, C.~Marzok,
  S.~Slama, and C.~Zimmermann.
\newblock Highly versatile atomic micro traps generated by multifrequency
  magnetic field modulation.
\newblock {\em Journal of Physics B: Atomic, Molecular and Optical Physics},
  39(5):1055, 2006.

\bibitem{salamon_maximum_2009}
Peter Salamon, Karl~Heinz Hoffmann, Yair Rezek, and Ronnie Kosloff.
\newblock Maximum work in minimum time from a conservative quantum system.
\newblock {\em Physical Chemistry Chemical Physics}, 11(7):1027--1032, February
  2009.

\bibitem{hohenester_optimal_2007}
Ulrich Hohenester, Per~Kristian Rekdal, Alfio Borz{\`i}, and J{\"o}rg
  Schmiedmayer.
\newblock Optimal quantum control of {B}ose--{E}instein condensates in magnetic
  microtraps.
\newblock {\em Physical Review A}, 75(2):023602, February 2007.

\bibitem{murphy_high-fidelity_2009}
M.~Murphy, L.~Jiang, N.~Khaneja, and T.~Calarco.
\newblock High-fidelity fast quantum transport with imperfect controls.
\newblock {\em Physical Review A}, 79(2):020301, February 2009.

\bibitem{rahmani_optimal_2011}
Armin Rahmani and Claudio Chamon.
\newblock Optimal control for unitary preparation of many-body states:
  Application to luttinger liquids.
\newblock {\em Physical Review Letters}, 107(1):016402, July 2011.

\bibitem{negretti_speeding_2013}
Antonio Negretti, Albert Benseny, Jordi Mompart, and Tommaso Calarco.
\newblock Speeding up the spatial adiabatic passage of matter waves in optical
  microtraps by optimal control.
\newblock {\em Quantum Information Processing}, 12(3):1439--1467, March 2013.

\bibitem{caneva_chopped_2011}
Tommaso Caneva, Tommaso Calarco, and Simone Montangero.
\newblock Chopped random-basis quantum optimization.
\newblock {\em Physical Review A}, 84:022326, Aug 2011.

\bibitem{doria_optimal_2011}
Patrick Doria, Tommaso Calarco, and Simone Montangero.
\newblock Optimal control technique for many-body quantum dynamics.
\newblock {\em Physical Review Letters}, 106:190501, May 2011.

\bibitem{chen_fast_2010}
Xi~Chen, A.~Ruschhaupt, S.~Schmidt, A.~{{del Campo}}, D.~Gu{\'e}ry-Odelin, and
  J.~G. Muga.
\newblock Fast optimal frictionless atom cooling in harmonic traps: Shortcut to
  adiabaticity.
\newblock {\em Physical Review Letters}, 104(6):063002, February 2010.

\bibitem{torrontegui_shortcuts_2013}
Erik Torrontegui, Sara Ib{\'a}{\~n}ez, Sofia Mart\'inez-Garaot, Michele
  Modugno, Adolfo {del Campo}, David Guery-Odelin, Andreas Ruschhaupt, Xi~Chen,
  and Juan Gonzalo~Muga.
\newblock Shortcuts to adiabaticity.
\newblock In {\em Advances in Atomic, Molecular, and Optical Physics},
  volume~62, page 117. Elsevier Academic Press Inc, San Diego, 2013.

\bibitem{demirplak_adiabatic_2003}
Mustafa Demirplak and Stuart~A. Rice.
\newblock Adiabatic population transfer with control fields.
\newblock {\em The Journal of Physical Chemistry A}, 107(46):9937--9945,
  November 2003.

\bibitem{demirplak_assisted_2005}
Mustafa Demirplak and Stuart~A Rice.
\newblock Assisted adiabatic passage revisited.
\newblock {\em The Journal of Physical Chemistry B}, 109(14):6838--6844, April
  2005.

\bibitem{demirplak_consistency_2008}
Mustafa Demirplak and Stuart~A. Rice.
\newblock On the consistency, extremal, and global properties of
  counterdiabatic fields.
\newblock {\em The Journal of Chemical Physics}, 129(15):154111, October 2008.

\bibitem{berry_transitionless_2009}
M.~V. Berry.
\newblock Transitionless quantum driving.
\newblock {\em Journal of Physics A: Mathematical and Theoretical},
  42(36):365303, September 2009.

\bibitem{Lewis_1969}
H.~R. Lewis and W.~B. Riesenfeld.
\newblock An exact quantum theory of the time‐dependent harmonic oscillator
  and of a charged particle in a time‐dependent electromagnetic field.
\newblock {\em Journal of Mathematical Physics}, 10(8):1458--1473, 1969.

\bibitem{Muga_invariant_2010}
J~G Muga, X~Chen, S~Ib{\'a}{\~n}ez, I~Lizuain, and A~Ruschhaupt.
\newblock Transitionless quantum drivings for the harmonic oscillator.
\newblock {\em Journal of Physics B: Atomic, Molecular and Optical Physics},
  43(8):085509, 2010.

\bibitem{Masuda_fastforward_2010}
Shumpei Masuda and Katsuhiro Nakamura.
\newblock Fast-forward of adiabatic dynamics in quantum mechanics.
\newblock {\em Proceedings of the Royal Society of London A: Mathematical,
  Physical and Engineering Sciences}, 466(2116):1135--1154, 2010.

\bibitem{Torrontegui_fastforward_2012}
E.~Torrontegui, S.~Mart\'{i}nez-Garaot, A.~Ruschhaupt, and J.~G. Muga.
\newblock Shortcuts to adiabaticity: Fast-forward approach.
\newblock {\em Physical Review A}, 86:013601, Jul 2012.

\bibitem{chen_optimal_2011}
Xi~Chen, E.~Torrontegui, Dionisis Stefanatos, Jr-Shin Li, and J.~G. Muga.
\newblock Optimal trajectories for efficient atomic transport without final
  excitation.
\newblock {\em Physical Review A}, 84(4):043415, October 2011.

\bibitem{OEO+07}
Toshio Ohshima, Artur Ekert, Daniel K.~L. Oi, Dagomir Kaslizowski, and L.~C.
  Kwek.
\newblock Robust state stansfer and rotation through a spin chain via dark
  passage.
\newblock {\em arXiv:quant-ph/0702019}, 2007.

\bibitem{OSF+13}
Sangchul Oh, Yun-Pil Shim, Jianjia Fei, Mark Friesen, and Xuedong Hu.
\newblock Resonant adiabatic passage with three qubits.
\newblock {\em Physical Review A}, 87:022332, Feb 2013.

\bibitem{romero-isart_efficient_2007}
K~Eckert, O~Romero-Isart, and A~Sanpera.
\newblock Efficient quantum state transfer in spin chains via adiabatic
  passage.
\newblock {\em New Journal of Physics}, 9(5):155, 2007.

\bibitem{BK14}
Andrew~D. Greentree and Belita Koiller.
\newblock Dark-state adiabatic passage with spin-one particles.
\newblock {\em Physical Review A}, 90:012319, Jul 2014.

\bibitem{garanovich_light_2012}
Ivan~L. Garanovich, Stefano Longhi, Andrey~A. Sukhorukov, and Yuri~S. Kivshar.
\newblock Light propagation and localization in modulated photonic lattices and
  waveguides.
\newblock {\em Physics Reports}, 518(1{\textendash}2):1--79, September 2012.

\bibitem{chen_foundations_2007}
Chin-Lin Chen.
\newblock {\em Foundations for guided-wave optics}.
\newblock Wiley Interscience, Hoboken, 2007.

\bibitem{katsunari_okamoto_fundamentals_2006}
Katsunari Okamoto.
\newblock {\em Fundamentals of Optical Waveguides}.
\newblock Elsevier, Burlingtong, 2nd edition, 2006.

\bibitem{kafesaki_frequency_2000}
M.~Kafesaki, M.~M. Sigalas, and N.~Garc{\'i}a.
\newblock Frequency modulation in the transmittivity of wave guides in
  elastic-wave band-gap materials.
\newblock {\em Physical Review Letters}, 85(19):4044--4047, November 2000.

\bibitem{khelif_transmittivity_2002}
A.~Khelif, B.~Djafari-Rouhani, J.~O. Vasseur, P.~A. Deymier, Ph. Lambin, and
  L.~Dobrzynski.
\newblock Transmittivity through straight and stublike waveguides in a
  two-dimensional phononic crystal.
\newblock {\em Physical Review B}, 65(17):174308, May 2002.

\bibitem{khelif_transmission_2003}
A.~Khelif, B.~Djafari-Rouhani, J.~O. Vasseur, and P.~A. Deymier.
\newblock Transmission and dispersion relations of perfect and
  defect-containing waveguide structures in phononic band gap materials.
\newblock {\em Physical Review B}, 68(2):024302, July 2003.

\bibitem{miyashita_full_2002}
Toyokatsu Miyashita.
\newblock Full band gaps of sonic crystals made of acrylic cylinders in air\
  {{\textemdash}Numerical} and experimental investigations{\textemdash}.
\newblock {\em Japanese Journal of Applied Physics}, 41(Part 1, No.
  {5B}):3170--3175, 2002.

\bibitem{khelif_trapping_2003}
A.~Khelif, A.~Choujaa, B.~Djafari-Rouhani, M.~Wilm, S.~Ballandras, and
  V.~Laude.
\newblock Trapping and guiding of acoustic waves by defect modes in a
  full-band-gap ultrasonic crystal.
\newblock {\em Physical Review B}, 68(21):214301, December 2003.

\bibitem{khelif_guiding_2004}
A.~Khelif, A.~Choujaa, S.~Benchabane, B.~Djafari-Rouhani, and V.~Laude.
\newblock Guiding and bending of acoustic waves in highly confined phononic
  crystal waveguides.
\newblock {\em Applied Physics Letters}, 84(22):4400--4402, May 2004.

\bibitem{vasseur_absolute_2008}
J.~O. Vasseur, P.~A. Deymier, B.~Djafari-Rouhani, Y.~Pennec, and A-C.
  Hladky-Hennion.
\newblock Absolute forbidden bands and waveguiding in two-dimensional phononic
  crystal plates.
\newblock {\em Physical Review B}, 77(8):085415, February 2008.

\bibitem{sun_analyses_2005}
Jia-Hong Sun and Tsung-Tsong Wu.
\newblock Analyses of mode coupling in joined parallel phononic crystal
  waveguides.
\newblock {\em Physical Review B}, 71(17):174303, May 2005.

\bibitem{cook_tapered_1955}
J.~S. Cook.
\newblock Tapered velocity couplers.
\newblock {\em Bell System Technical Journal}, 34(4):807--822, July 1955.

\bibitem{fox_wave_1955}
A.~G. Fox.
\newblock Wave coupling by warped normal modes.
\newblock {\em Bell System Technical Journal}, 34(4):823, 1955.

\bibitem{louisell_analysis_1955}
W.~H. Louisell.
\newblock Analysis of the single tapered mode coupler.
\newblock {\em Bell System Technical Journal}, 34(4):853, 1955.

\bibitem{wilson_improved_1973}
M.G.F. Wilson and G.A. Teh.
\newblock Improved tolerance in optical directional couplers.
\newblock {\em Electronics Letters}, 9(19):453--455, 1973.

\bibitem{smith_coupling_1975}
R.B. Smith.
\newblock Coupling efficiency of the tapered coupler.
\newblock {\em Electronics Letters}, 11(10):204, 1975.

\bibitem{smith_analytic_1976}
Robert~B. Smith.
\newblock Analytic solutions for linearly tapered directional couplers.
\newblock {\em Journal of the Optical Society of America}, 66(9):882--892,
  September 1976.

\bibitem{silberberg_digital_1987}
Y.~Silberberg, P.~Perlmutter, and J.~E. Baran.
\newblock Digital optical switch.
\newblock {\em Applied Physics Letters}, 51(16):1230--1232, October 1987.

\bibitem{rowland_tapered_1991}
D.R. Rowland, Y.~Chen, and Allan~W. Snyder.
\newblock Tapered mismatched couplers.
\newblock {\em Journal of Lightwave Technology}, 9(5):567--570, 1991.

\bibitem{ramadan_adiabatic_1998}
T.A. Ramadan, Robert Scarmozzino, and R.M. Osgood.
\newblock Adiabatic couplers: design rules and optimization.
\newblock {\em Journal of Lightwave Technology}, 16(2):277--283, 1998.

\bibitem{chen_optimal_2007}
Chi-Feng Chen, Yun-Sheng Ku, and Tsu-Te Kung.
\newblock Optimal design of coupling waveguide structure for adiabatic optical
  directional full couplers weighted by sin-square and raised-cosine functions.
\newblock {\em Optics Communications}, 280(1):79--86, December 2007.

\bibitem{longhi_landauzener_2005}
Stefano Longhi.
\newblock {{L}andau--{Z}ener} dynamics in a curved optical directional coupler.
\newblock {\em Journal of Optics B: Quantum and Semiclassical Optics}, 7(6):L9,
  June 2005.

\bibitem{dreisow_direct_2009}
F.~Dreisow, A.~Szameit, M.~Heinrich, S.~Nolte, A.~T\"unnermann, M.~Ornigotti,
  and S.~Longhi.
\newblock Direct observation of {L}andau--{Z}ener tunneling in a curved optical
  waveguide coupler.
\newblock {\em Physical Review A}, 79(5):055802, May 2009.

\bibitem{fractionalSTIRAP}
N.V. Vitanov, K.-A. Suominen, and Shore B.W.
\newblock Creation of coherent atomic superpositions by fractional stimulated
  {R}aman adiabatic passage.
\newblock {\em Journal of Physics B: Atomic, Molecular and Optical Physics},
  32(18):4535, 1999.

\bibitem{menchon-enrich_adiabatic_2012}
R.~Menchon-Enrich, A.~Llobera, V.J. Cadarso, J.~Mompart, and V.~Ahufinger.
\newblock Adiabatic passage of light in {CMOS}-compatible silicon oxide
  integrated rib waveguides.
\newblock {\em {IEEE} Photonics Technology Letters}, 24(7):536--538, April
  2012.

\bibitem{paspalakis_adiabatic_2006}
Emmanuel Paspalakis.
\newblock Adiabatic three-waveguide directional coupler.
\newblock {\em Optics Communications}, 258(1):30--34, February 2006.

\bibitem{lahini_effect_2008}
Y.~Lahini, F.~Pozzi, M.~Sorel, R.~Morandotti, D.~N. Christodoulides, and
  Y.~Silberberg.
\newblock Effect of nonlinearity on adiabatic evolution of light.
\newblock {\em Physical Review Letters}, 101(19):193901, November 2008.

\bibitem{salandrino_analysis_2009}
Alessandro Salandrino, Konstantinos Makris, Demetrios~N. Christodoulides, Yoav
  Lahini, Yaron Silberberg, and Roberto Morandotti.
\newblock Analysis of a three-core adiabatic directional coupler.
\newblock {\em Optics Communications}, 282(23):4524--4526, December 2009.

\bibitem{Chelkowski_RCAP_1997}
S.~Chelkowski and A.~D. Bandrauk.
\newblock Raman chirped adiabatic passage: a new method for selective
  excitation of high vibrational states.
\newblock {\em Journal of Raman Spectroscopy}, 28(6):459--466, 1997.

\bibitem{longhi_photonic_2007}
Stefano Longhi.
\newblock Photonic transport via chirped adiabatic passage in optical
  waveguides.
\newblock {\em Journal of Physics B: Atomic, Molecular and Optical Physics},
  40(12):F189, June 2007.

\bibitem{eisenberg_discrete_1998}
H.~S. Eisenberg, Y.~Silberberg, R.~Morandotti, A.~R. Boyd, and J.~S. Aitchison.
\newblock Discrete spatial optical solitons in waveguide arrays.
\newblock {\em Physical Review Letters}, 81(16):3383, October 1998.

\bibitem{kazazis_effects_2010}
S.~Kazazis and E.~Paspalakis.
\newblock Effects of nonlinearity in asymmetric adiabatic three-waveguide
  directional couplers.
\newblock {\em Journal of Modern Optics}, 57(21):2123--2129, 2010.

\bibitem{longhi_optical_2009}
Stefano Longhi.
\newblock Optical analogue of coherent population trapping via a continuum in
  optical waveguide arrays.
\newblock {\em Journal of Modern Optics}, 56(6):729--737, 2009.

\bibitem{longhi_optical_2009-1}
Stefano Longhi.
\newblock Optical analog of population trapping in the continuum: Classical and
  quantum interference effects.
\newblock {\em Physical Review A}, 79(2):023811, February 2009.

\bibitem{WPM+07}
M.A. Webster, R.M. Pafchek, A.~Mitchell, and T.L. Koch.
\newblock Width dependence of inherent tm-mode lateral leakage loss in
  silicon-on-insulator ridge waveguides.
\newblock {\em Photonics Technology Letters, IEEE}, 19(6):429--431, March 2007.

\bibitem{RLB+02}
T.~C. Ralph, N.~K. Langford, T.~B. Bell, and A.~G. White.
\newblock Linear optical controlled-not gate in the coincidence basis.
\newblock {\em Physical Review A}, 65:062324, Jun 2002.

\bibitem{HOM87}
C.~K. Hong, Z.~Y. Ou, and L.~Mandel.
\newblock Measurement of subpicosecond time intervals between two photons by
  interference.
\newblock {\em Physical Review Letters}, 59:2044--2046, Nov 1987.

\bibitem{EV93}
A.~C. Elitzur and L.~Vaidman.
\newblock Quantum mechanical interaction-free measurements.
\newblock {\em Foundations of Physics}, 23(7):987--997, 1993.

\bibitem{CPT_Arimondo_1996}
E.~Arimondo.
\newblock {\em Progress in {O}ptics}, volume~35.
\newblock Elsevier, Amsterdam, 1996.

\bibitem{EIT_1}
K.-J. Boller, A.~Imamo\ifmmode~\breve{g}\else \u{g}\fi{}lu, and S.~E. Harris.
\newblock Observation of electromagnetically induced transparency.
\newblock {\em Physical Review Letters}, 66:2593--2596, May 1991.

\bibitem{EIT_2}
Michael Fleischhauer, Atac Imamoglu, and Jonathan~P. Marangos.
\newblock Electromagnetically induced transparency: Optics in coherent media.
\newblock {\em Review of Modern Physics}, 77:633--673, Jul 2005.

\bibitem{gaubatz_population_1990}
U.~Gaubatz, P.~Rudecki, S.~Schiemann, and K.~Bergmann.
\newblock Population transfer between molecular vibrational levels by
  stimulated raman scattering with partially overlapping laser fields. a new
  concept and experimental results.
\newblock {\em The Journal of Chemical Physics}, 92(9):5363--5376, 1990.

\bibitem{CPT_A1}
A.~Aspect, E.~Arimondo, R.~Kaiser, N.~Vansteenkiste, and C.~Cohen-Tannoudji.
\newblock Laser cooling below the one-photon recoil energy by
  velocity-selective coherent population trapping.
\newblock {\em Physical Review Letters}, 61:826--829, Aug 1988.

\bibitem{CPT_A2}
S.~Knappe, P.~Schwindt, V.~Shah, L.~Hollberg, J.~Kitching, L.~Liew, and
  J.~Moreland.
\newblock A chip-scale atomic clock based on $^{87}${R}b with improved
  frequency stability.
\newblock {\em Optics Express}, 13(4):1249--1253, Feb 2005.

\bibitem{EIT_A1}
Lene~Vestergaard Hau, S.~E. Harris, Zachary Dutton, and Cyrus~H. Behroozi.
\newblock Light speed reduction to 17 metres per second in an ultracold atomic
  gas.
\newblock {\em Nature}, 397(6720):594--598, 02 1999.

\bibitem{EIT_A2}
Rui Zhang, Sean~R. Garner, and Lene~Vestergaard Hau.
\newblock Creation of long-term coherent optical memory via controlled
  nonlinear interactions in {B}ose--{E}instein condensates.
\newblock {\em Physical Review Letters}, 103:233602, Dec 2009.

\bibitem{STIRAP_A1}
Axel Kuhn, Markus Hennrich, and Gerhard Rempe.
\newblock Deterministic single-photon source for distributed quantum
  networking.
\newblock {\em Physical Review Letters}, 89:067901, Jul 2002.

\bibitem{STIRAP_A2}
Ditte M\o{}ller, Lars~Bojer Madsen, and Klaus M\o{}lmer.
\newblock Geometric phase gates based on stimulated {R}aman adiabatic passage
  in tripod systems.
\newblock {\em Physical Review A}, 75:062302, Jun 2007.

\bibitem{vitanov_analytic_1997}
N.~V. Vitanov and S.~Stenholm.
\newblock Analytic properties and effective two-level problems in stimulated
  raman adiabatic passage.
\newblock {\em Physical Review A}, 55:648--660, Jan 1997.

\bibitem{klein_robust_1999}
Jens Klein, Fabian Beil, and Thomas Halfmann.
\newblock Robust population transfer by stimulated raman adiabatic passage in a
  {Pr}$^{3+}$:{Y}$_{2}${SiO}$_{5}$ crystal.
\newblock {\em Physical Review Letters}, 99:113003, Sep 2007.

\end{thebibliography}

\end{document}